\documentclass[fleqn,usenatbib]{mnras}

\usepackage{newtxtext,newtxmath}

\usepackage[T1]{fontenc}

\DeclareRobustCommand{\VAN}[3]{#2}
\let\VANthebibliography\thebibliography
\def\thebibliography{\DeclareRobustCommand{\VAN}[3]{##3}\VANthebibliography}

\usepackage{graphicx}	
\usepackage{amsmath}	
\usepackage{threeparttable}  
\usepackage{gensymb}

\newcommand{\Msol}{M$_{\odot}$}                           

\newcommand{\hii}{H\,{\sc ii}}

\newcommand{\oiii}{[O\,{\sc iii}]}
\newcommand{\oii}{[O\,{\sc ii}]}
\newcommand{\oi}{[O\,{\sc i}]}
\newcommand{\nii}{[N\,{\sc ii}]}
\newcommand{\sii}{[S\,{\sc ii}]}

\newcommand{\lin}{$\lambda$}

\def\usfr{M$_{\odot}$yr$^{-1}$}
\def\ulum{erg $\rm s^{-1}$}

\def\uvel{\mbox{km $\rm s^{-1}$}}

\def\hdelta {H$\delta$}

\def\hbeta {H$\beta$}
\def\halpha {H$\alpha$}

\def\Av{$A_\mathrm{v}$}	
\def\Te{$T_e$}	
\def\ne{$n_e$}

\title[Ionized gas in NGC~4258]{Ionized gas in NGC~4258: Exploring the AGN - Star formation connection}

\author[SIGNALS et. al...]{D. Fernández-Arenas$^{1}$\thanks{E-mail: arenas@chft.hawaii.edu}, L. Rousseau-Nepton$^{1, 2, 3}$,  C. Robert$^{4,5}$, L. Drissen$^{4,5}$, R. P. Martin$^{6}$, P. Amram$^{7}$, \newauthor B. Epinat$^{1,7}$, Duarte Puertas$^{4,5,8,9}$, R. Garner, III$^{10}$,  G. Savard$^{4,5}$, S. Vicens-Mouret$^{4,5}$,  É. Massé$^{4}$. \\
$^{1}$Canada-France-Hawaii Telescope, Kamuela, HI, USA\\
$^{2}$Department of Astronomy and Astrophysics, University of Toronto, 50 St George St, Toronto ON M5S 3H4, Canada\\
$^{3}$Dunlap Institute for Astronomy and Astrophysics, 50 St George St, Toronto ON M5S 3H4, Canada\\
$^{4}$D\'epartement de Physique, de G\'enie Physique et d'Optique, Universit\'e Laval, Canada\\ $^{5}$Centre de Recherche en Astrophysique du Qu\'ebec (CRAQ), Qu\'ebec, QC, G1V 0A6, Canada\\
$^{6}$Dept. Physics \& Astronomy, University of Hawaii at Hilo, Hilo, HI, USA\\
$^{7}$ Aix Marseille Univ, CNRS, CNES, LAM, Marseille, France\\
$^{8}$Departamento de F\'isica Te\'orica y del Cosmos Universidad de Granada, 18071 Granada, Spain \\ 
$^{9}$Instituto Universitario Carlos I de F\'isica Te\'orica y Computacional, Universidad de Granada, 18071 Granada, Spain\\
$^{10}$George P. and Cynthia W. Mitchell Institute for Fundamental Physics \& Astronomy, Texas A\&M University, College Station, TX 77843-4242, USA
}

\date{Accepted XXX. Received YYY; in original form ZZZ}

\begin{document}
\label{firstpage}
\pagerange{\pageref{firstpage}--\pageref{lastpage}}
\maketitle

\begin{abstract}
NGC 4258 is a prime target for studying feedback in Low-Luminosity Active Galactic Nuclei (LLAGNs) due to its proximity and comprehensive multi-wavelength coverage. Using new Integral Field Spectroscopy (IFS) data from SITELLE at the Canada–France–Hawaii Telescope, we analysed the galaxy's nebular emission lines. Our study focused on spatially resolved line ratios and Baldwin–Phillips–Terlevich diagrams, revealing that the "anomalous spiral arms" exhibit intense interactions between the jet and interstellar medium (ISM) extending up to 6 kpc with velocity dispersions peak at 200-250 km/s in these regions, contrasting with star-forming areas showing lower values around of 30-50 km/s. Analysis of covering fractions indicates heightened AGN ionization cones aligned with the radio jet, alongside evidence of shock quenching observed in the lower "anomalous arc". Conversely, jet-induced compression may stimulate star formation in other areas. We derived a galaxy-wide star formation rate of  $\sim3 M_{\odot}\mathrm{yr}^{-1}$, decreasing to $0.3 M_{\odot}\mathrm{yr}^{-1}$ within the central 3.4 kpc$^2$. SITELLE's broad field coverage elucidates the galaxy's structural details, confirming that low-power jets significantly influence the host galaxy across parsec and kpc scales. The velocity dispersion map reveals asymmetric or double-peaked emission lines, tracing jet-disk interactions likely responsible for the formation of anomalous arm features. Small-scale ionizing clusters were detected in regions with disrupted gas flows, possibly formed through tidal interactions or shock compression. NGC~4258 thus presents a compelling case for studying LLAGN-driven feedback, illustrating how optical IFS combined with multi-wavelength data clarifies the impact of outflows and shocks on nearby spiral galaxies, providing insights into how these processes shape star formation and ISM conditions.

\end{abstract}

\begin{keywords}
instrumentation: interferometers -galaxies: Seyfert -galaxies: ISM   - galaxies: NGC~4258
\end{keywords}


\section{Introduction}

The energy released by Active Galactic Nuclei (AGN), in the form of radiation and/or mechanical outflows, is believed to significantly influence the interstellar medium (ISM). Through various feedback mechanisms, including outflows that either trigger or quench star formation (SF), this energy plays a prominent role in shaping the host galaxy. Over the last two decades, the impact of AGN on galaxy evolution has been a major topic of discussion for understanding the connection between supermassive black holes (SMBHs) and their host galaxies \citep{Kormendy2013, Ding2020}.

Energy injected by the central AGN excites and ionizes gas, dissociates molecules, and ablates dust grains. Tracers of shocked gas, used to quantify this energy input, can track the ISM’s response to specific forms of feedback. Various energetic phenomena from AGNs—such as radiation, relativistic jets, and outflows—act as potential feedback channels that either expel or stabilize the ISM. At present, AGN jets appear to play a significant role in galaxy evolution, providing kinetic feedback to the host ISM and regulating star formation. Indeed, recent cosmological simulations suggest that the more common low-power jets may dominate sub-kiloparsec-scale feedback \citep[e.g.][]{Weinberger2017, Pillepich2018}. Nevertheless, AGN feedback remains controversial despite its broad acceptance as a crucial mechanism for regulating massive galaxy growth \citep{Husemann2018}.

To investigate the physical properties of AGN-driven outflows, spatially resolved spectroscopy is extensively used to map the distribution of warm ionized gas in luminous AGNs exhibiting signs of ionized outflows \citep[e.g.,][]{Harrison2014,Husemann2016,Villar2018,Kang2018,Mingozzi2019,Revalski2021,Ruschel2021}. However, there is still no clear consensus on how AGN activity influences host galaxies. In particular, key issues remain unsettled regarding the relationship between AGN properties and star formation \citep[e.g.,][]{Vayner2021,Vayner2023,Kakkad2023,Wylezalek2022,Cresci2023,Veilleux2023} and the distances over which AGN-driven processes can operate \citep[e.g.,][]{Greene2005,Woo2016,Perna2017,Wylezalek2020}.

Integral Field Spectroscopy (IFS) has yielded an unprecedented wealth of information on the spatial variation of ionized gas properties, the distribution of stellar populations, metallicity, and the kinematics of both gas and stars, as well as the interplay of different physical processes within individual galaxies \citep[e.g.,][]{Rosales2010, Croom2012, Sanchez2012, Blanc2013, Erroz2019, Poetrodjojo2019, Emsellem2022}. It has also provided valuable insights into the connection between star formation (SF) and AGN activity \citep[e.g.,][]{Husemann2017, Neumann2019, RobletoOrus2021, Agostino2021, Husemann2022, Smirnova2022, Molina2023}. However, only a small number of IFS studies have specifically focused on how AGN jets interact with both the cold molecular and ionized phases of the ISM \citep[e.g.,][]{Alatalo2011, Garcia2019, Alonso2019, FernandezOntiveros2020, Garcia2021, Mingozzi2019, Venturi2018, Venturi2021, Juneau2022}.

One common approach to characterizing the properties of ionized gas, investigating the nature of ionization sources, and identifying potential outflow locations is through the analysis of kinematic variations and the use of spatially resolved Baldwin–Phillips–Terlevich (BPT) diagrams \citep{Baldwin1981}. These diagrams enable us to distinguish the dominant ionizing contribution in each spaxel and locate it across the galaxy, making them invaluable for tracing the influence of AGN, shocks, and star-forming regions \citep[e.g.,][]{Lopez2019,Neumann2019,Agostino2021,RobletoOrus2021,Smirnova2022,Molina2023,Watts2024}. However, most of these studies have been conducted on kiloparsec scales, leaving uncertainty about the smallest spatial scales on which such feedback processes can operate \citep[e.g.][]{Husemann2018, Nandi2023}.

At the core of our study, we leveraged Integral Field Spectroscopy (IFS) with a large field of view (FoV) and parsec-scale sampling of the nearby galaxy NGC~4258. Using the SITELLE instrument at the Canada-France-Hawaii Telescope (CFHT), we mapped the spatial and kinematic distribution of ionized gas, enabling the construction of spatially and kinematically resolved Baldwin-Phillips-Terlevich (BPT) diagrams. These diagrams helped us distinguish between AGN activity, shocks, and star-forming regions. This approach also allowed us to investigate how the jet interacts with the interstellar medium (ISM) and study the so-called ``anomalous spiral arm", whose origin remains debated. Because NGC~4258 is a nearby galaxy that can be examined in detail, it offers strong evidence for the effects of AGN-driven feedback processes. In the following sections, we describe the main features of  NGC~4258 and discuss the relevance of this galaxy to star formation, AGN-ISM interactions, and feedback in the local Universe, with implications for interpreting observations of high-redshift galaxies.

In particular, the impact of AGN feedback on the surrounding ISM in NGC~4258 remains poorly constrained, and the relative contributions of star formation, AGN activity, and shocks to ionized gas excitation are not well quantified. Previous studies have focused primarily on the galaxy’s central region, leaving open questions about how these mechanisms vary across different locations and scales. One key question addressed by using IFS in NGC~4258 is: How do outflows from either AGN or star formation, including jets, interact with the ISM? Another central question is: What is the nature of the ionized gas in the ``anomalous spiral arms"? By employing spatially resolved emission-line diagnostics—particularly Baldwin–Phillips–Terlevich (BPT) diagrams—and velocity dispersion maps, we aim to explore the interplay between radiative feedback (AGN photoionization) and mechanical energy injection (shocks, jet-ISM interactions) in NGC~4258.

\subsection{The case of NGC~4258}

Studying the ionized gas in NGC~4258 (also known as Messier~106) is critical for understanding its ``anomalous spiral arms" and the feedback processes associated with Low-Luminosity Active Galactic Nuclei (LLAGNs). This section explores the key aspects of ionized gas in NGC~4258 and its implications for galaxy evolution.

NGC~4258 is noteworthy in the local Universe for its ``anomalous spiral arms" which differ from standard density-wave spiral structures \citep{ceccarelli2001jets}. Owing to its proximity, NGC~4258 has received considerable attention because of its complex structure and is an ideal candidate for investigating various astrophysical processes, such as accretion onto its central black hole, star formation in its spiral arms, and the interaction of ejected nuclear material with the ISM—particularly given its highly active SMBH. By examining properties like velocity, excitation, and density, we can understand how the jets interact with the surrounding material and shape the ``anomalous spiral arms," thereby providing insights into the mechanics of jet-driven feedback and its effect on galactic morphology.

Classified as a SABbc Seyfert~1.9 galaxy \citep{Ho1997} at a distance of $7.57\pm0.11$ Mpc \citep{Reid2019}, NGC~4258 hosts a central SMBH of \mbox{$4.7\times10^{7}$ \Msol} \citep{Drehmer2015}.  Although NGC~4258 looks like typical spiral galaxy, observations in radio, X-ray and \halpha\ images reveals an arc-like structure offset from the galaxy's plane, which formation could be the result of the interaction between outflows originated from the nuclear activity and the ISM. This object is well-known for its nearly edge-on molecular nuclear disc, located between 0.16 and 0.28 pc from the nucleus.  Its complex morphology features typical outer and distinctive inner spiral arms with an active star-forming process \citep{Roy1985,Cecil1992}, including jet emissions and active star-forming regions.  As a LLAGN, NGC~4258 provides a remarkable case study for examining the interplay between AGN-driven processes and SF.

Observational data on NGC~4258 are extensive, covering radio \citep[e.g.,][]{Sofue1989,Krause1990,Collison1995,Cox1996}, infrared \citep[e.g.,][]{Chary1997,Laine2010,Ogle2014,Menezes2018}, and X-ray \citep[e.g.,][]{Ptak1999,Vogler1999,Reynolds2000,Wilson2001,Yuan2002,Fruscione2005,Akyuz2013,Avdan2016,Avdan2020,Akyuz2020} observations. Spectroscopic studies in the optical range have focused on determining the abundance gradient through slit observations of the brightest \hii\ regions \cite[see][]{Bresolin2011,Diaz2000} or the kinematics of gas in small central region (<4 kpc) \citep{Cecil2000}. Observations using a Fabry-Perot interferometer have revealed intricate kinematics \citep{Roy1985,Cecil1992,Gach2002,Erroz2015} and only \citep{Drehmer2015} and \citep{Appleton2018} have been used IFS to trace the stellar kinematics and \halpha\ emission but only covering a projected area smaller than $\sim$30\arcsec$\times$30\arcsec.

\halpha\ images of NGC~4258 reveal two prominent elongated structures extending north and south of the nucleus. Commonly referred to as ``anomalous spiral arms," these features also appear in soft X-ray and radio continuum. Their origin remains debated: several studies propose that they are formed by collimated nuclear outflows driven by the AGN \citep{Ford1986,Martin1989,Makishima1994, Dutil1995,Wilkes1995,Cohen2018}. Overall, a unique combination of properties—such as proximity, a well-characterized nucleus, and a gas- and dust-rich disk—makes NGC~4258 an ideal subject for exploring the interplay between star formation (SF) and AGN activity; however, the potential impact of the jet on the warm ISM is not yet clear. Earlier investigations focused primarily on the nucleus or limited fields, thus failing to capture the jet’s broader effect on the ISM or disentangle how much of the emission arises from star formation versus AGN/shock ionization.

SITELLE data offer an excellent opportunity to investigate how a nuclear jet propagates through the disk of NGC~4258 at parsec-scale resolution using optical diagnostics. These observations extend beyond the central region to include the spiral arms and the so-called ``anomalous spiral arms." Thanks to the galaxy's proximity, we can conduct high-spatial and spectroscopic observations that shed light on interactions between its SMBH and the surrounding ISM across multiple scales. This study utilizes new data from the SITELLE instrument at CFHT to map ionized gas and emission-li"ne properties in NGC~4258 over a wide range of radii. The large field of view and favorable spatial resolution of SITELLE enable us to cover the galaxy’s center, its jets, and to extend analyses of kinematics and spectroscopic characteristics out to the disk periphery.

As part of the larger SIGNALS survey-which aims to systematically characterize ionized gas and star-formation processes across diverse galactic environments—NGC~4258 offers an instructive test case. Its well-known maser disk, pronounced jet, and disk–jet interactions provide clear, spatially resolvable signatures of mechanical and radiative feedback. Studying these phenomena at sub-kiloparsec and parsec scales helps SIGNALS develop robust analysis workflows (e.g., improved stellar correction or mixing-sequence) that can later be applied to more ``typical" spirals or less obvious AGN. Thus, NGC~4258 functions as a case test for the methodologies employed across the entire SIGNALS sample, bridging conventional star-forming regions and more extreme feedback scenarios. In particular, the galaxy’s unique structure—featuring anomalous spiral arms and notable jet activity—makes it a compelling laboratory for refining methods that disentangle multiple ionization sources and quantify their spatial distribution.

Current theories propose that AGN activity took place in NGC~4258, possibly accompanied by outflows oriented at an angle to the galactic disk \citep{Cecil2000}. Our aim is to investigate the ionizing mechanisms by examining morphology and kinematic properties at high resolution to track shocks and ionized gas flows. Future comparisons with James Webb Space Telescope data on stellar content will help confirm the location and nature of the ``anomalous spiral arm." The adopted parameters for NGC~4258 used throughout this study are listed in Table \ref{table_galaxy}.

The paper is organized as follows: In~\S \ref{data}, we describe the observations and the data–reduction procedures. In~\S \ref{fluxes}, we present the flux measurements, emission-line intensity maps, and the structural components previously reported in the literature. In~\S \ref{properties}, we discuss the flux maps; the galactocentric and azimuthal variations of the emission-line ratios at multiple spatial scales; the Baldwin–Phillips–Terlevich (BPT) diagnostic diagrams and their spatial and kinematic variations; the star-formation-rate estimates; and the emission-line profiles across the central region. In ~\S \ref{Discu}, we explore the physical implications for NGC~4258, examine the mechanisms that may drive the observed trends,  outline additional processes not treated in this work and offer guidance for interpreting the forthcoming \emph{James Webb Space Telescope} (JWST) observations of NGC~4258. Finally, in~\S \ref{sum_conclu},  we state the main conclusions and summarise our results.

\begin{table}

\caption{Key parameters of NGC~4258.}
\begin{threeparttable}  
\begin{tabular}{ll}\hline\hline
Parameter  & Value \\ \hline
R.A. (J2000) & 12h18m57s.50 $^\textrm a$\\
DEC. (J2000) & 47d18m14s3 $^\textrm a$ \\
Morphological type & SAB(s) bcb. $^\textrm b$\\
Distance          & $7.57\pm0.11$ $^\textrm c$ \\
$R_{25}$         & $9\arcmin.31$ $^\textrm b$\\
Inclination        & $72^{\circ}$ $^\textrm d$\\
Position angle of major axis & $150^{\circ}$ $^\textrm d$\\
M$^0_B$        & $-20.76$ $^\textrm b$ \\
V$_{\rm sys}${$^\textrm e$}       & $472\pm1$ \uvel \\\hline
\end{tabular}
 $^\textrm a$ \cite{Herrnstein2005}, $^\textrm b$ \cite{Vaucouleurs1991}, $^\textrm c$ \cite{Reid2019}, $^\textrm d$ \cite{Albada1980}, $^\textrm e$ In this study, the systemic velocity is derived by modelling the kinematics of the \halpha\ line, with fixed centre and inclination.
\end{threeparttable}    
\label{table_galaxy}
\end{table}%

\section{Observations and data reduction}\label{data}

NGC~4258 was observed with SITELLE at CFHT as part of the SIGNALS project \citep{Rousseau2019} during the 2022A semester campaign (program number 22AQ05). SITELLE is an Imaging Fourier Transform Spectrograph (IFTS) designed for the visible spectrum on the 3.6m Canada-France-Hawaii Telescope (CFHT), featuring two E2V detectors, each with 2048$\times$2064 pixels. The pixel size is 0.32\arcsec$\times$0.32\arcsec (corresponding to 11.5$\times$11.5 pc$^2$ at a distance of 7.57 kpc), and it has a FoV of \mbox{11\arcmin $\times$ 11\arcmin}. The resolution power is adjustable from 1 to 10,000. A single observation produces more than 4 million spectra in a specific filter \citep[see][for a detailed description of the instrument]{Grandmont2012,Drissen2019}.

Observations were centred at R.A. (J2000)= 12h18m57s.50 and DEC. (J2000)= 47d18m14s3 and three data cubes were obtained: SN1, SN2 and SN3 covering a spectral range of  3630\AA-3860\AA, 4840\AA–5120\AA\  and 6480\AA–6860\AA, respectively. The selected resolution power was $R\sim1000$ for SN1 and SN2 and $R\sim5000$ for SN3\footnote{For a complete description of filters available for SITELLE, see \url{https://www.cfht.hawaii.edu/Instruments/Sitelle/SITELLE_filters.php}}. The wavelength range of each filter bandpass allows the transmission of the main nebular lines \oii\lin3727, \hbeta, \oiii\lin4959 and \oiii\lin5007, \nii\lin6548, \halpha, \nii\lin6583, \sii\lin6716 and \sii\lin6731. The data reduction was performed with the \texttt{ORBS}\footnote{\texttt{ORBS}, Outil de Réduction Binoculaire pour SITELLE, is a data reduction software created to process data obtained with SITELLE, The code includes standard CCD corrections, astrometry calibration, fast Fourier transform, and finally flux and wavelength calibrations. Standard stars were observed for each filter as part of the usual observing procedures with SITELLE. For more information see \mbox{\protect\url{https://github.com/thomasorb/orbs}}.} code \citep{Martin2016}. A summary of the observations is presented in Table \ref{table_filters}.

\begin{table}
\caption{Filter setups for the observations of NGC~4258 with SITELLE}
\begin{threeparttable}  
\begin{tabular}{lccc}\hline\hline
 Filter  & SN1 &SN2 & SN3 \\\hline
 Wavelength coverage [\AA]  &  3650-3850  & 4800-5200 & 6510-6850\\
 Resolution R& 1000&1000&5000\\
 Number of steps&171&219&842\\
 Time per steps [sec]&59&45.5&13.3\\
 Total Exposure time[sec]&10089&9964&11199\\
 Image Quality [\arcsec]$^\textrm *$  &1.41&1.23&1.01\\\hline
\end{tabular}
$^\textrm *$Values for the image quality have been derived from the deep image and correspond to the average FWHM of unsaturated stars.
\end{threeparttable}    

\label{table_filters}
\end{table}

\section{Flux measurements}\label{fluxes}

In this section, we describe the methodology used to measure the flux from the emission lines for each spaxel. Additionally, we present the spatially resolved flux maps following additional data processing, which includes sky subtraction and correction for the stellar continuum as described below.

\subsection{Sky subtraction}\label{_sky_cor}

\begin{figure}
\centering
	\includegraphics[width=\columnwidth]{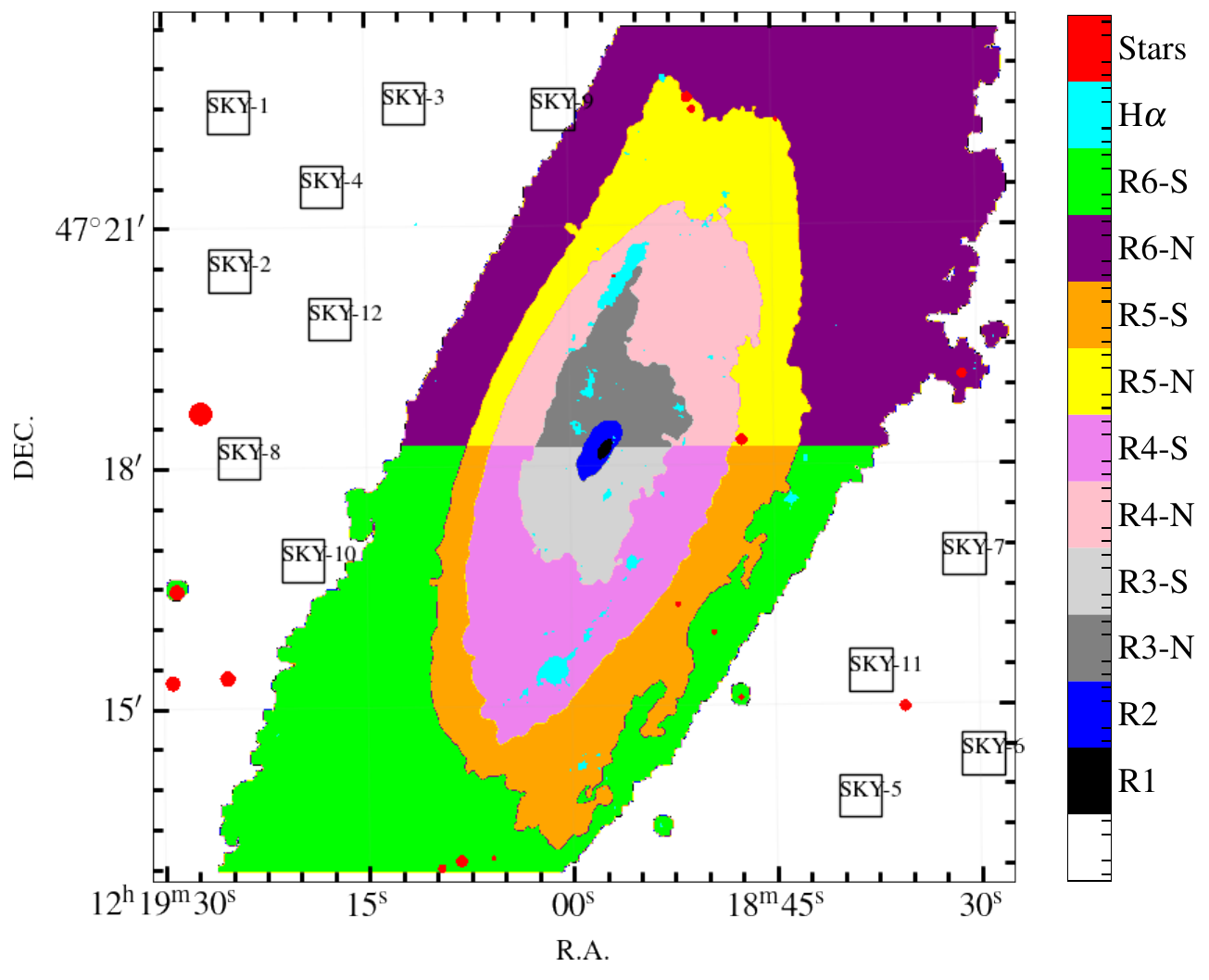}
    \caption{Mask selection for the sky emission subtraction and correction for the stellar continuum absorption for our observations of NGC~4258 with SITELLE. Black squares labelled from SKY-1 to -12 correspond to areas where the sky background extraction was performed to correct the spectra. Each square averages $10000$ spaxels covering each more than 1000 arcsec$^2$. The continuum areas were selected based on contours derived from the SN3 filter continuum, as indicated by the colour bar with the label ``R". For a more accurate characterization of the continuum, spaxels with high \halpha\ emission (S/N > 10, shown in cyan), foreground stars and background objects (marked in red) were excluded from the selection.}
    \label{_regions}
\end{figure}

\begin{figure}
\includegraphics[width=\columnwidth]{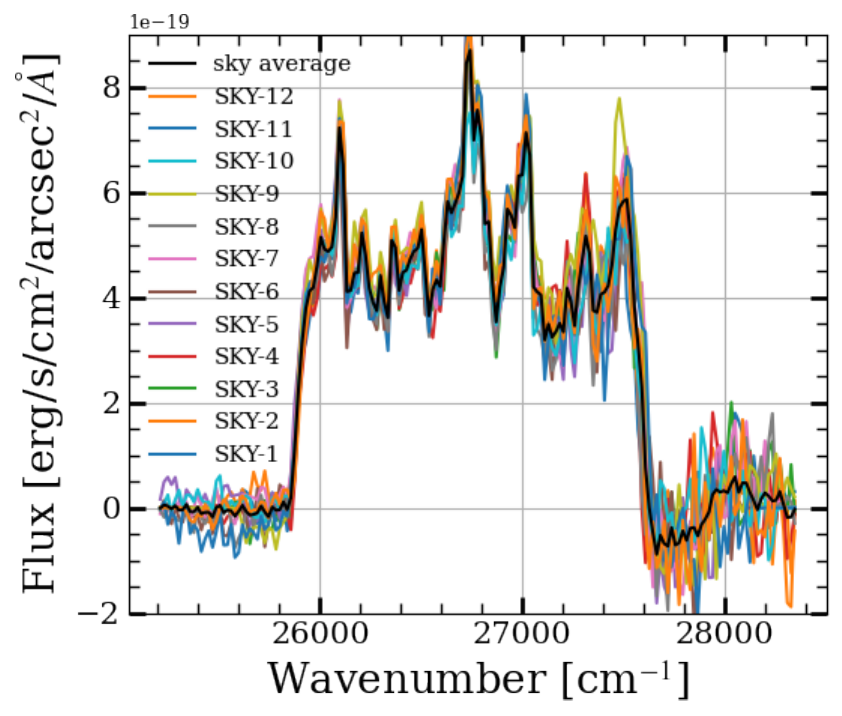}
\includegraphics[width=\columnwidth]{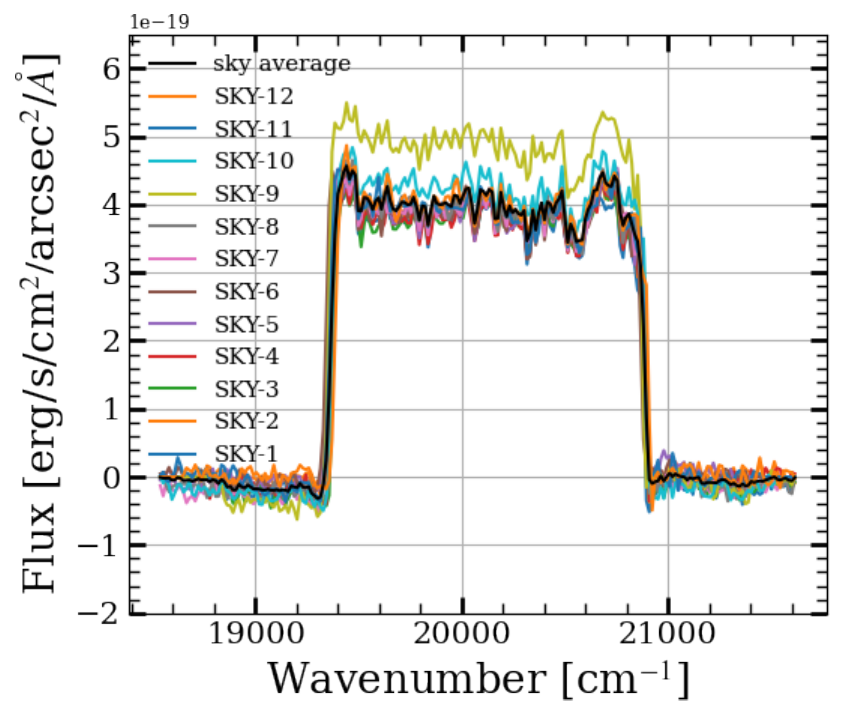}
\includegraphics[width=\columnwidth]{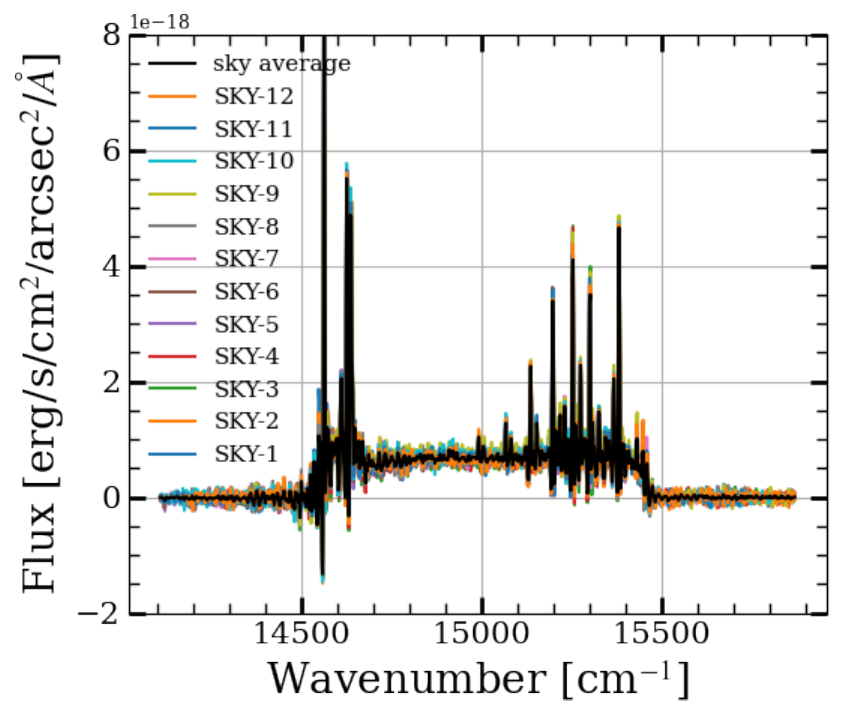}
    \caption{Spectra corresponding to the sky emission from the different regions marked by the black squares in Figure \ref{_regions}.From top to bottom: SN1, SN2, and SN3 filters. In all panels, the average spectrum in black is used for sky background subtraction. We observe more variations in the sky background in SN2, due to contamination from the galaxy disc in regions 9 and 10.}
    \label{sky_spec}
\end{figure}

As the final preprocessing step, the sky background is subtracted using a median sky spectrum obtained from 12 boxes of 100$\times$100 spaxels in regions located far away from the galaxy disc and rejecting spectra seen as outliers. Figure \ref{_regions} shows the regions in colours for the stellar correction (see next section), while areas designated for sky subtraction are marked with black squares. Figure \ref{sky_spec} shows the averaged spectrum for each region. The primary variation in the FoV is associated with the SN2 filter, wherein one outlier was identified and subsequently rejected due to potential contamination by the disk.

\subsection{Stellar absorption correction}\label{_stellar_cor}

Depending on the galaxy's mass and type, the measurement of emission lines in a galaxy can be significantly influenced by the absorption line features from the stellar populations and their distribution within the galaxy. After sky subtraction and before fitting the emission lines to isolate and study the ionized gas properties, we corrected the spectrum in each spaxel by subtracting an underlying stellar population. It's important to note that our aim is not to conduct an exhaustive analysis of the resolved stellar populations. Instead, our focus is on subtracting the stellar contribution from the observed spectra. Due to the absence of stellar absorption features in the filters used for the SIGNALS project, conducting a detailed analysis of stellar ages and metallicities would be beyond the scope of this study. Nevertheless, modelling the continuum stellar component remains crucial for mitigating the impact of absorption components on the Balmer emission lines, particularly for \hbeta.

The methodology, as previously presented by \cite{Rousseau2018}, utilizes a reference spectrum for subtracting the contribution of the stellar population. The reference spectrum is derived from a region devoid of emission lines and centred on the peak of the galaxy's continuum emission. After scaling this reference spectrum with the continuum in each spaxel and the observed local velocity, the subtraction is performed, resulting in a pure emission data cube.

In this study, we adopt a similar method; however, instead of using a global reference spectrum, we generate high signal-to-noise (S/N>200) spectra at various regions in the galaxy. This approach considers the potential variation in stellar populations across the galaxy. Figure \ref{_regions} illustrates the areas used to generate the continuum spectrum reference and subsequently remove the stellar continuum from the data cubes. Before combining the spectra, each one is shifted to the rest frame using the line-of-sight velocity obtained from an initial fit of the data cube.

Once the spectrum of each region is obtained, we apply the process described by Massé et al. (in preparation) to fit and remove the stellar continuum in each region. In summary, the code utilizes an adapted version of Penalized Pixel-Fitting \citep[\texttt{pPXF},][]{Emsellem2004, Cappellari2017}. This code aims to construct the stellar template that best reproduces the observed spectrum using absorption and emission lines, based on a linear combination of single stellar population spectra. For this purpose, we employ the MILES library \cite{Vazdekis2016}, which has a wavelength coverage of 3540-7409 Å and a spectral resolution of FWHM = 2.3 Å. This library's characteristics match very well with the properties of our SITELLE data, covering a wide range of metallicities ([M/H] from $-2.32$ to $+0.22$) and ages (0.06 to 18 Gyr) for the stellar populations. The code was adapted to consider the SITELLE’s sine cardinal instrumental profile (sinc), specifically related to the emission lines model.

To utilize \texttt{pPXF}, SITELLE spectra were converted from wavenumbers to wavelengths. The SN1 datacube was excluded due to its lower spectral resolution, narrow bandwidth, and the absence of strong absorption features. After testing, it was determined that SN1 did not provide any additional information for the procedure. Therefore, only SN2 and SN3 were employed for stellar continuum fitting. When combining the SN2 and SN3 spectra, interpolation was performed to match the spectral sampling of both bands and generate a unified spectrum at a constant $\Delta$$\lambda$ without compromising the spectral resolution of each filter.

\begin{figure*}
\centering
	\includegraphics[width=\columnwidth]{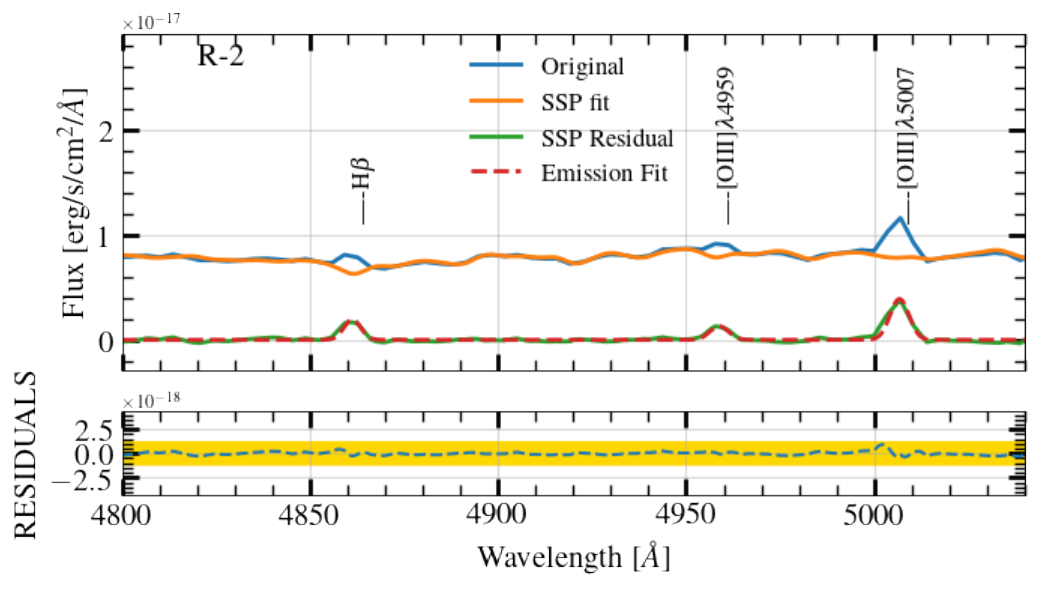}
 \includegraphics[width=\columnwidth]{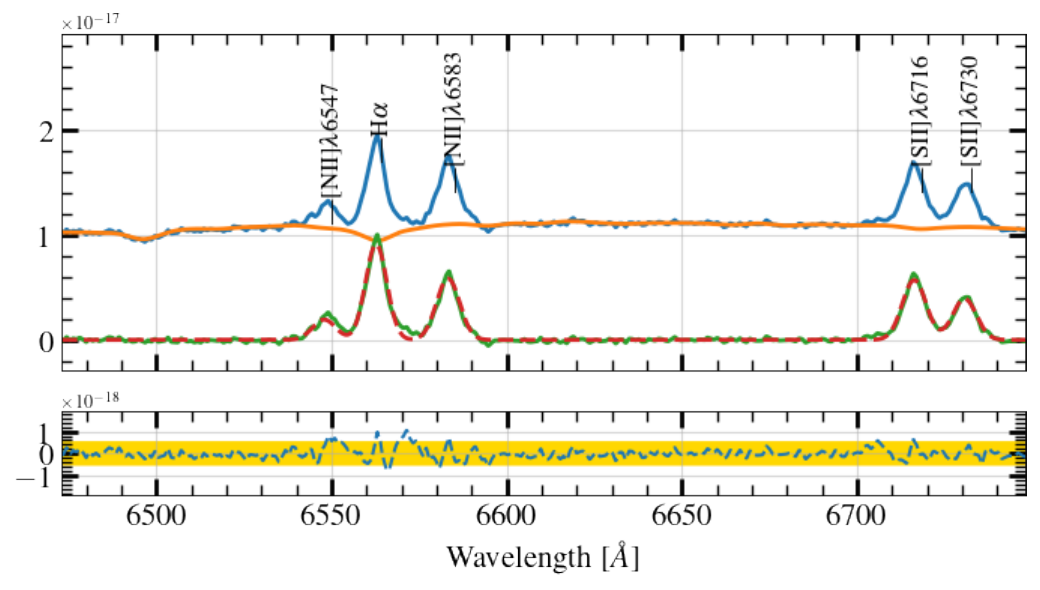}	
    \caption{Example of the spectrum extracted for region R2 (see Figure \ref{_regions}) using the SN2 and SN3 filters. Each spectrum in the region is corrected for sky emission, shifted, and then summed to enhance the S/N ratio. The blue line represents the final spectrum, the orange line depicts the fit of the stellar continuum, while the red dashed line corresponds to the emission line fit obtained using \texttt{pPXF}. The yellow band in the residuals is the standard deviation obtained from the continuum after masking emission lines.}
    \label{_stellar_spec_fit}
\end{figure*}

Details about the data spectral resolution were then conveyed to \texttt{pPXF}, enabling it to fit the stellar population spectra to our resolutions, which vary from filter to filter. An illustrative model obtained for one region in SN2 and SN3 filters is depicted in Figure \ref{_stellar_spec_fit}, with vertical dashed lines marking the emission lines. The residuals are consistent with the root mean square (r.m.s.) of the observed spectra, the yellow area in the bottom panels. The rest of the regions are shown in Figure \ref{_stellar_spec_fit_t}.

Once we obtain the spectra of the stellar component in each region, subtraction is performed after scaling the stellar population model to match the continuum level and the observed velocity of the spaxel. We compared the differences between the fluxes derived before and after applying the correction for stellar absorption and found an average ratio between the fluxes of the lines with and without this correction, corresponding to 1.2 and 1.1 for the \hbeta\ and \halpha\ lines, respectively. Other lines remain consistent with a ratio of 1.0, indicating they are not strongly affected by the stellar correction. The most significant impact of this correction is on the \hbeta\ line, where the uncorrected fluxes could be underestimated by on average approximately $\sim$20\%, taking into account the absorption features present in the stellar continuum.

\subsection{Flux measurements and spatially resolved structures}\label{emis_lines}

\begin{figure*}
\centering
	\includegraphics[width=\columnwidth]{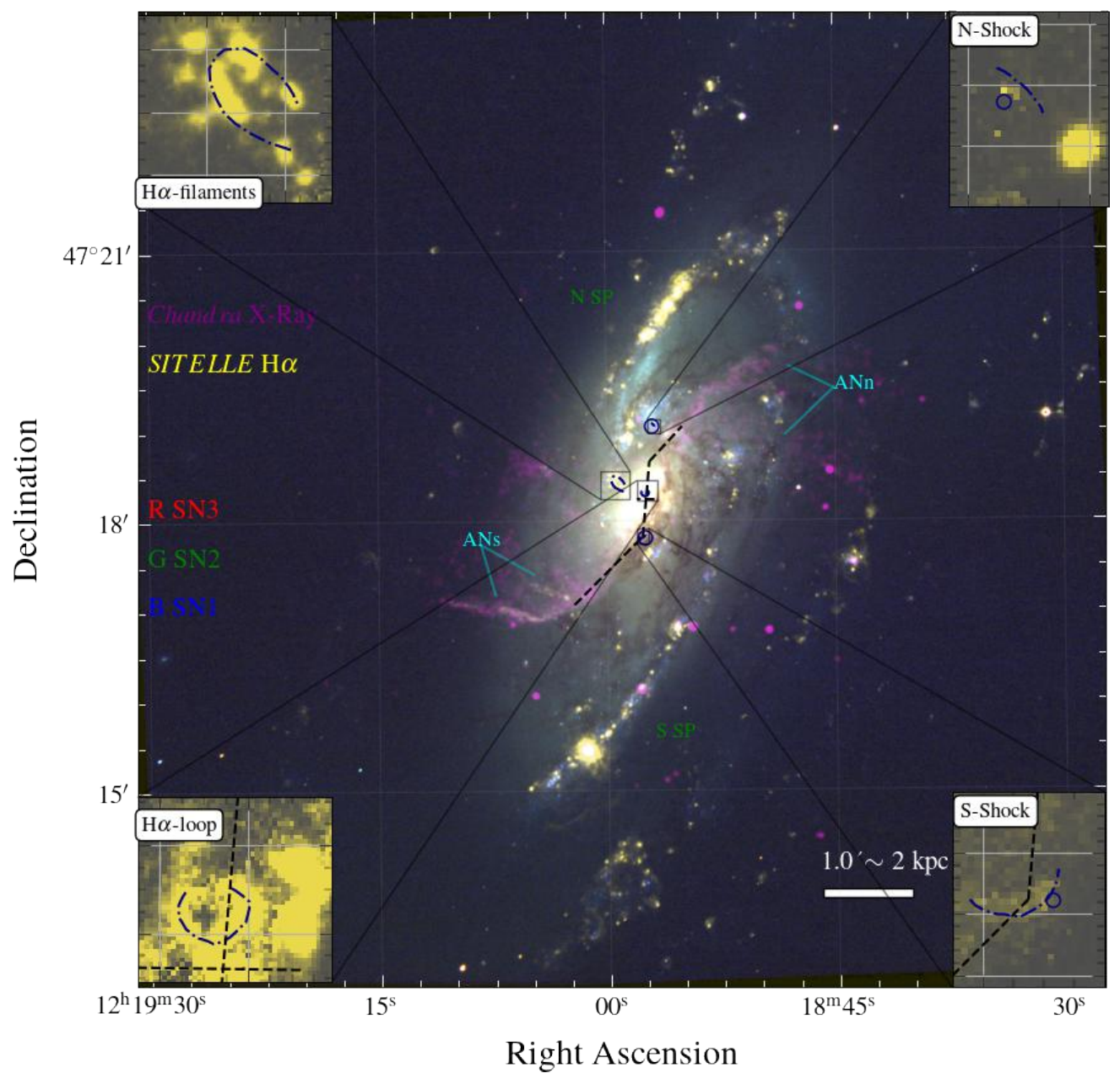}\includegraphics[width=\columnwidth]{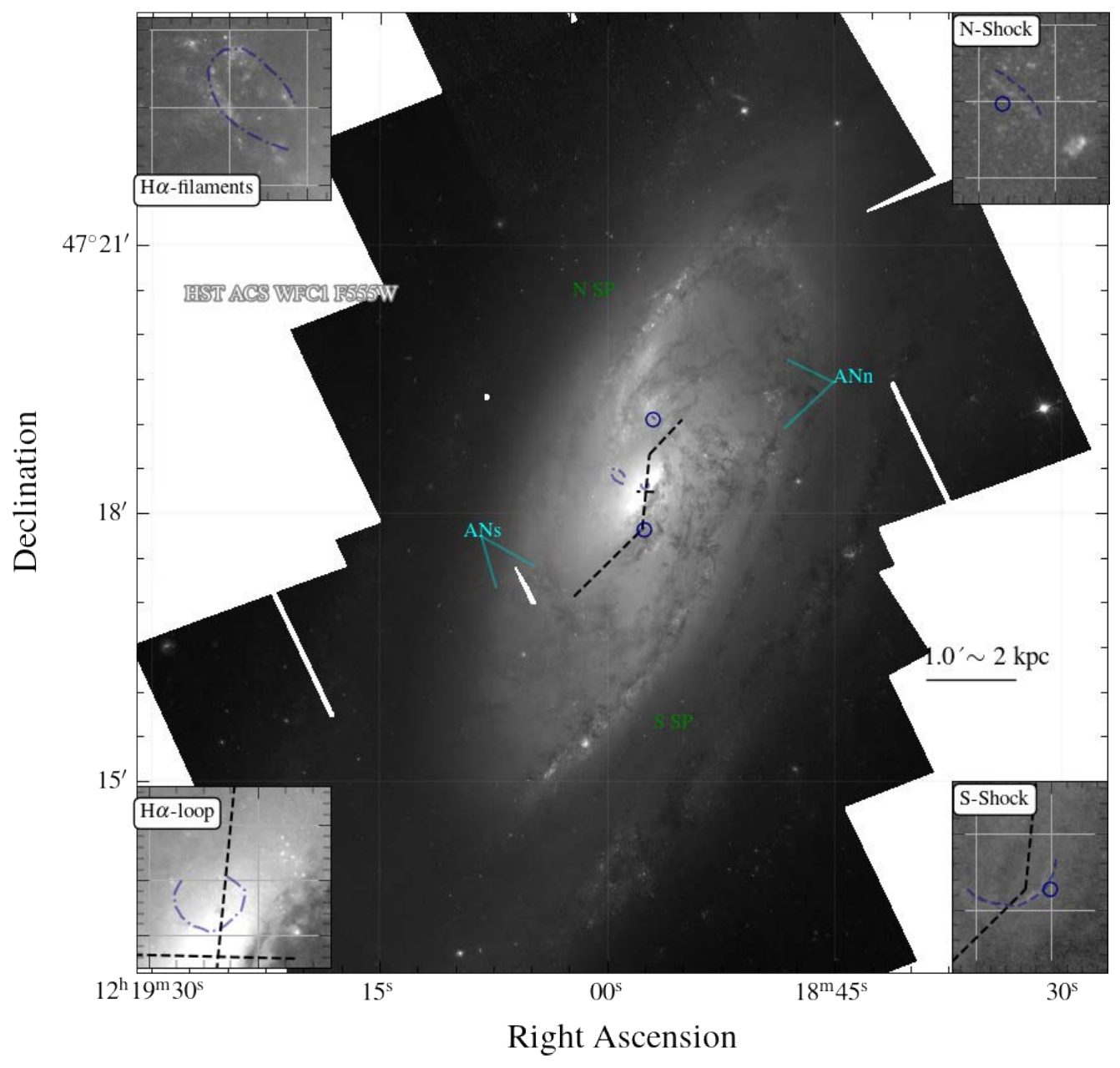}
		\includegraphics[width=\columnwidth]{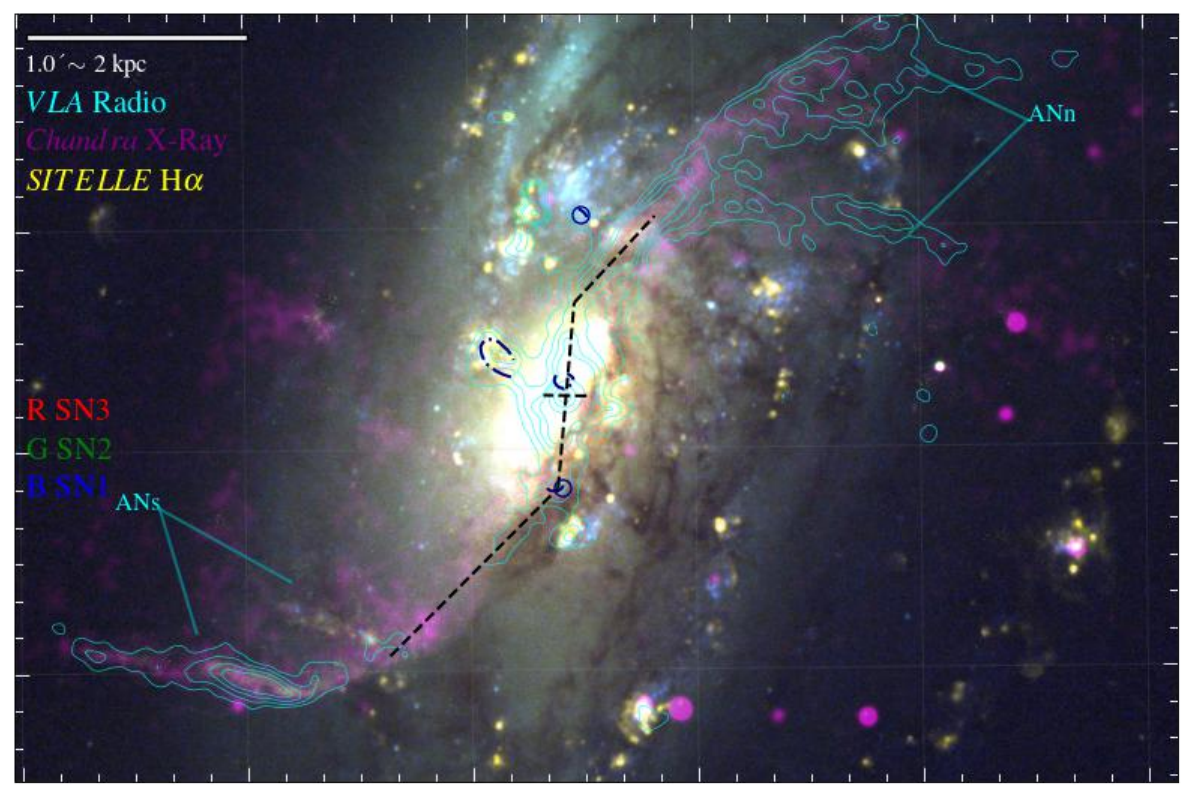}\includegraphics[width=\columnwidth]{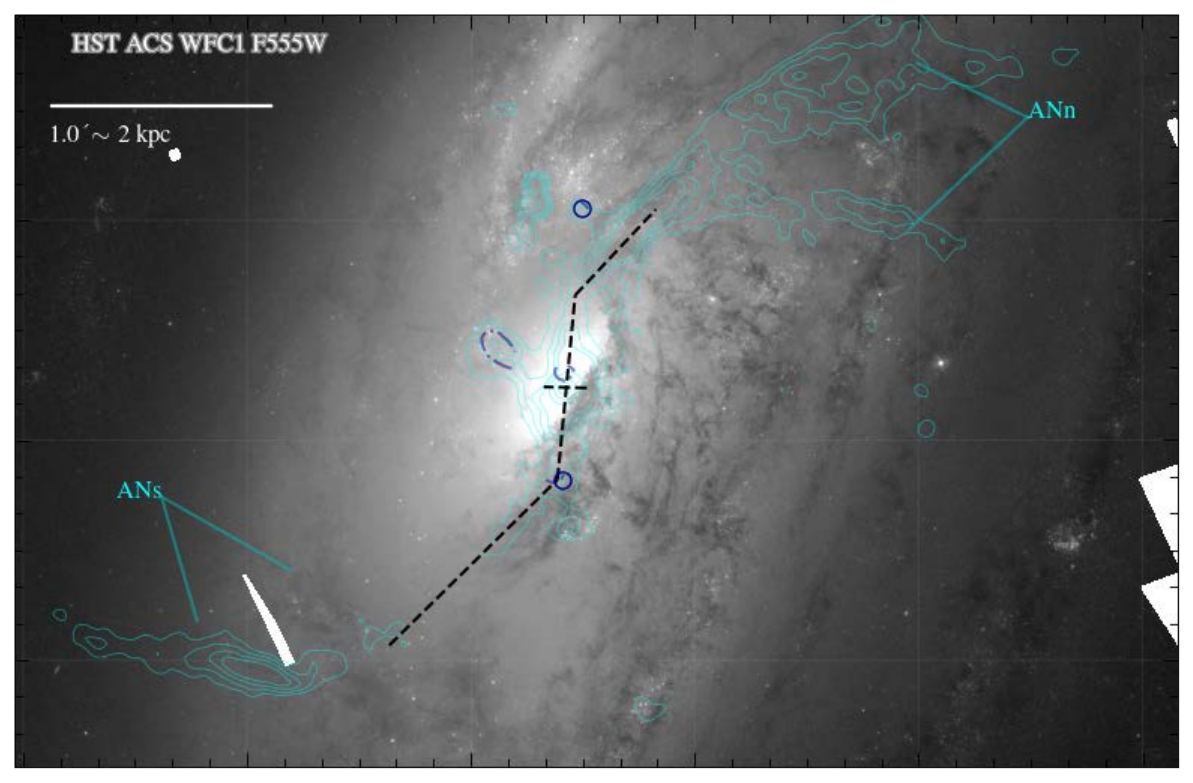}
    \caption{Left panels: The RGB image of NGC4258 is generated using three deep images obtained from the SN3 filter [6470\AA-6850\AA] (red), SN2 filter [4820\AA-5130\AA] (green), and SN1 filter [3630\AA-3860\AA] (blue). We highlight in yellow the  \halpha\ emission map and in purple the X-ray emission from ACIS-S instrument on $Chandra$ in the 0.3-8 keV band \protect\citep{Wilson2001}. The green labels of NSp and SSp indicate the typical north and south spiral arms in NGC~4258, while ANs and ANn correspond to the reported ``anomalous spiral arms" identified by \protect\cite{vanderKruit1972}. We identified the observed geometry of the jets and the north and south hot spots in radio and X-ray emission \protect\citep{Krause2004}. The \halpha\ filaments and shock arcs reported by \protect\cite{Appleton2018} are traced with blue dashed dot lines. We also highlighted the northern nuclear loop first detected by \protect\cite{Cecil2000} and \protect\cite{JVicente2010}. The top panels feature a zoomed-in view of these regions to enhance visualization. The left bottom panel displays a zoom to the central region with contours corresponding to the radio image at 1.49 GHz obtained with the VLA by \protect\cite{Cecil2000}. The right panels, both top and bottom, depict a mosaic obtained with the HST ACS WFC1 and F555W filter, available in the MAST archive. Bottom panels correspond to the same zoom in the central regions (right and left).}
    \label{RGB}
\end{figure*}

The left panels of Figure \ref{RGB} present the RGB image created by combining three deep images from the SN1, SN2 and SN3 filters, which result from summing all interferograms of each datacube after sky background subtraction. In this figure, we have highlighted the \halpha\ emission map in yellow and incorporated additional data from the literature. The X-ray emission from the $Chandra$ Telescope with ACIS-S instrument in the 0.3-8 keV band is depicted in purple \citep{Wilson2001}. The observed geometry of the jets, as well as the north and south hot spots in radio and X-ray emission, were identified by \cite{Krause2004}. The \halpha\ filaments and shock arcs presented by \cite{Appleton2018} are shown with blue dashed dot lines. In the bottom left panel, zoom into the central region of NGC~4258 is presented with contours representing the radio image at \mbox{1.49 GHz} obtained with the VLA by \cite{Cecil2000}.

In the same figure, we also provide a simple comparison of the data quality achievable with the SITELLE instrument by contrasting it with a mosaic constructed using the Hubble Space Telescope (HST)\footnote{The data can be downloaded via The Mikulski Archive for Space Telescopes (MAST), an astronomical data archive focused on the optical, ultraviolet, and near-infrared, \protect\url{https://archive.stsci.edu/} and the mosaic was reconstructed using  \texttt{reproject} python package.}. The HST mosaic comprises 16 pointings using the Advanced Camera for Surveys (ACS) with the Wide Field Channel (WFC1) and filter F555W (PI: Riess A., Proposal ID: 11570). Despite the lower spatial resolution sampling of SITELLE ($\sim$10 pc/pixel) compared to HST ($\sim$2 pc/pixel) at the distance of 7.57 Mpc, we can still distinguish numerous structures in Figure \ref{RGB}. This is particularly evident in obscured paths originating from dense sweeping dust lanes in the central region.

In our Figure \ref{RGB}, it is possible to identify some interesting morphological features previously reported in the literature in the central region of NGC~4258. The ``anomalous spiral arms" at north and south are very well defined from the \halpha\ map. We also note the spatial agreement between some emission point sources in X-ray and \halpha\ peaks associated with filaments and arcs, which could be related to supernova remnants or planetary nebulae. Other points in the X-ray emission are likely more associated with background sources than with the galaxy because no counterparts are observed in our optical images. In what follows we describe some of the main characteristics of the central region of NGC~4258.

\subsubsection{The jet and the ``anomalous spiral arms"}

The jets in NGC~4258 display a diffuse structure with bifurcation and strong bending at the extreme points approximately 5 kpc from the nucleus, clearly visible in our \halpha\ map. Since \cite{vanderKruit1972}, the ``anomalous spiral arms" of NGC~4258 have been extensively discussed in terms of matter ejection from the nucleus \citep[see][for a comprehensive discussion]{Krause2004, Appleton2018}. These anomalous spiral arms are two large arm-like gaseous structures that extend through most of the visible disc of the galaxy. These arms do not correlate with the galaxy’s underlying stellar spiral structure and remain poorly understood. However, they are somehow related to the present or past jet activity of the AGN, which has triggered enhanced synchrotron radiation along the jet’s path over time, exciting shocks, and X-ray emission.

The action of a jet is thought to be responsible for the ``anomalous" radio-continuum spiral arms, which appear several kpc from the centre and extend through the outer disc. Our \halpha\ map demonstrates a clear spatial correlation with the radio and X-ray emission. In Figure \ref{RGB}, the circumnuclear accretion disc's location and the jets' observed geometry correspond to the black dashed lines. 

\subsubsection{The filaments and shocks arcs}

Using optical emission-line ratios of \nii\ and \halpha, \cite{Cecil1992} demonstrated evidence of potential braided structures in the central region, consistent with shocks forming filaments and arcs. These observations revealed that the inner jet is only a part of the overall picture in the central region of NGC~4258. The study also identified a bright, distorted loop of compact \halpha-emitting star clusters and optical ``bow-shock-like" structures. These structures are faintly seen in \halpha\ emission at varying projected angular distances from the nucleus. SITELLE observations show evidence of the presence of these features, excluding the N-shock. The \halpha\ loop, filaments, and shocks are represented by the blue lines, while X-ray and radio hotspots are indicated by blue circles.

\cite{JVicente2010}, making use of IFS, explored the central region of NGC~4258 with a FoV of \mbox{31.35\arcsec$\times$27.55\arcsec} at  \mbox{$R\sim1300$} and sampling of 0.95 \arcsec/pixel, covering a wavelength range between 5600–6850\AA. They found that emission line ratio maps show a thin ring-like region of high values indicative of shocks surrounding the galaxy nucleus and the border of the bipolar ionized gas component overlaps this front shock on the near side of the galaxy. They also report an interesting northern nuclear loop first detected by \cite{Cecil2000}. In our images, we highlighted this structure with a dashed-dotted blue line in the central region.

\subsubsection{Normal spiral arms}

NGC~4258 has been the subject of several observations by various authors using different instrumentation and telescopes \cite[e.g.][]{Cecil1992, Courtes1993, Dutil1995, Cecil2000, Krause2007, Bresolin2011}. While these observations have varying spatial resolutions, they consistently reveal a clear distribution of external and some internal spiral arms, a pattern also evident in observations made with SITELLE. In the \halpha\ map, distinct regions to the north and south show a clear spatial correlation of intensity with radio emission. As mentioned earlier, the spiral arm \halpha\ regions coincide with unpolarized radio emission – our BPT maps confirm these regions are photoionized by stars, supporting the interpretation by \cite{Krause2004} and \cite{Krause2007}, suggests that clumps of strong emission along the spiral arms are likely star-forming regions emitting unpolarized thermal radiation. In contrast, the radio emission along the jets is polarized and of non-thermal origin. The green labels SSP and NSP in our Figure \ref{RGB} indicate the locations of the normal spiral arm to the south and north of NGC~4258.

\section{Data analysis}\label{properties}

Once the data cubes are free of the sky and stellar continuum contribution, one can study the ionized gas distribution in NGC~4258 by fitting the emission lines in all three filters. For each spaxel, we used \texttt{ORCS}\footnote{\texttt{ORCS}, Outils de Réduction de Cubes Spectraux, which is an analysis engine for the SITELLE spectral cubes. The code allows extracting integrated and individual spectra, fitting a sinc gauss function to emission lines and much more.  For more information see \protect\url{https://github.com/thomasorb/orcs}. },  a \texttt{PYTHON}  module designed specially to fit the spectra cubes obtained with SITELLE \citep{Martin2015}.

For the fitting process, we assumed that all lines share the same velocity and velocity dispersion. The flux ratio between the \nii\lin6584\AA\ and \nii\lin6548\AA\ pair and \oiii\lin4949\AA\ and \oiii\lin5007\AA\ pair are fixed to 3, close to the theoretical ratio \citep{Storey2000}.

\begin{figure}
    \includegraphics[scale=0.48]{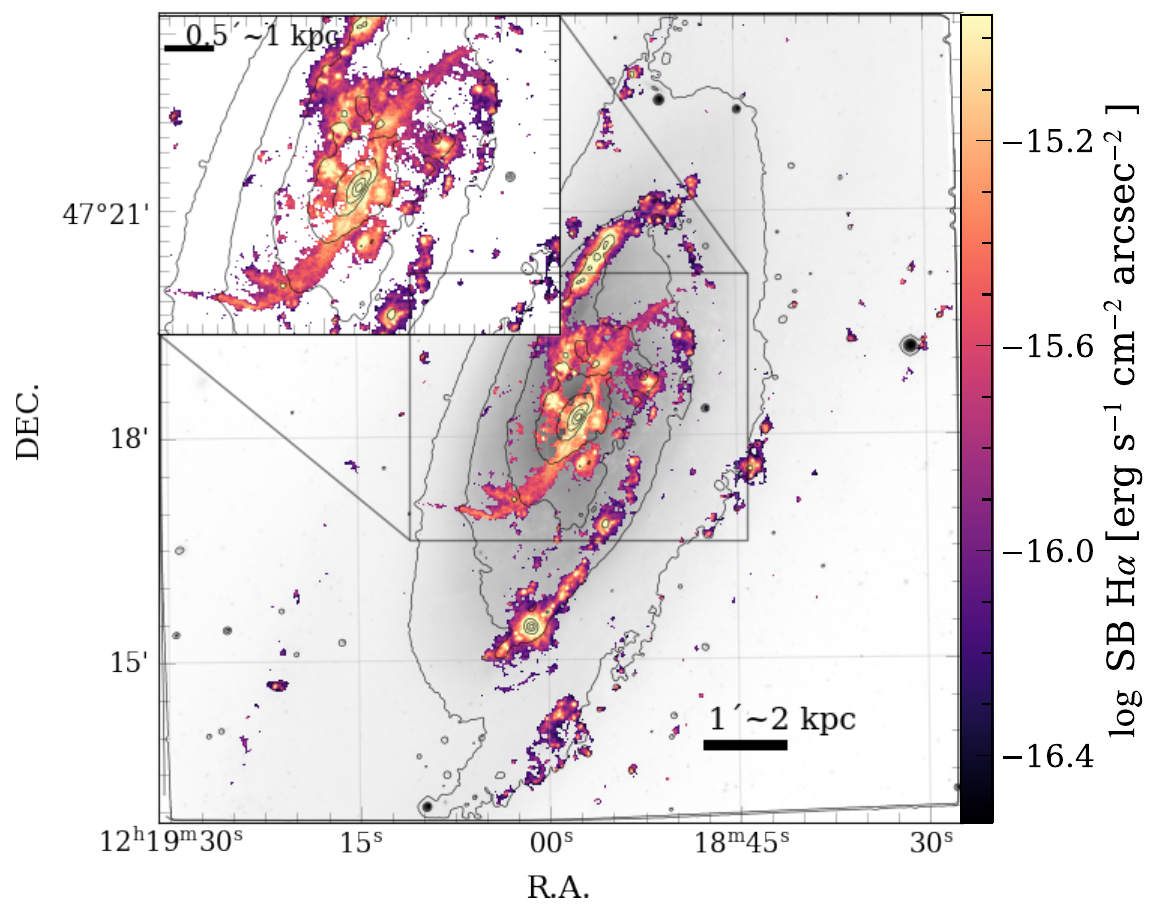}
    \includegraphics[scale=0.48]{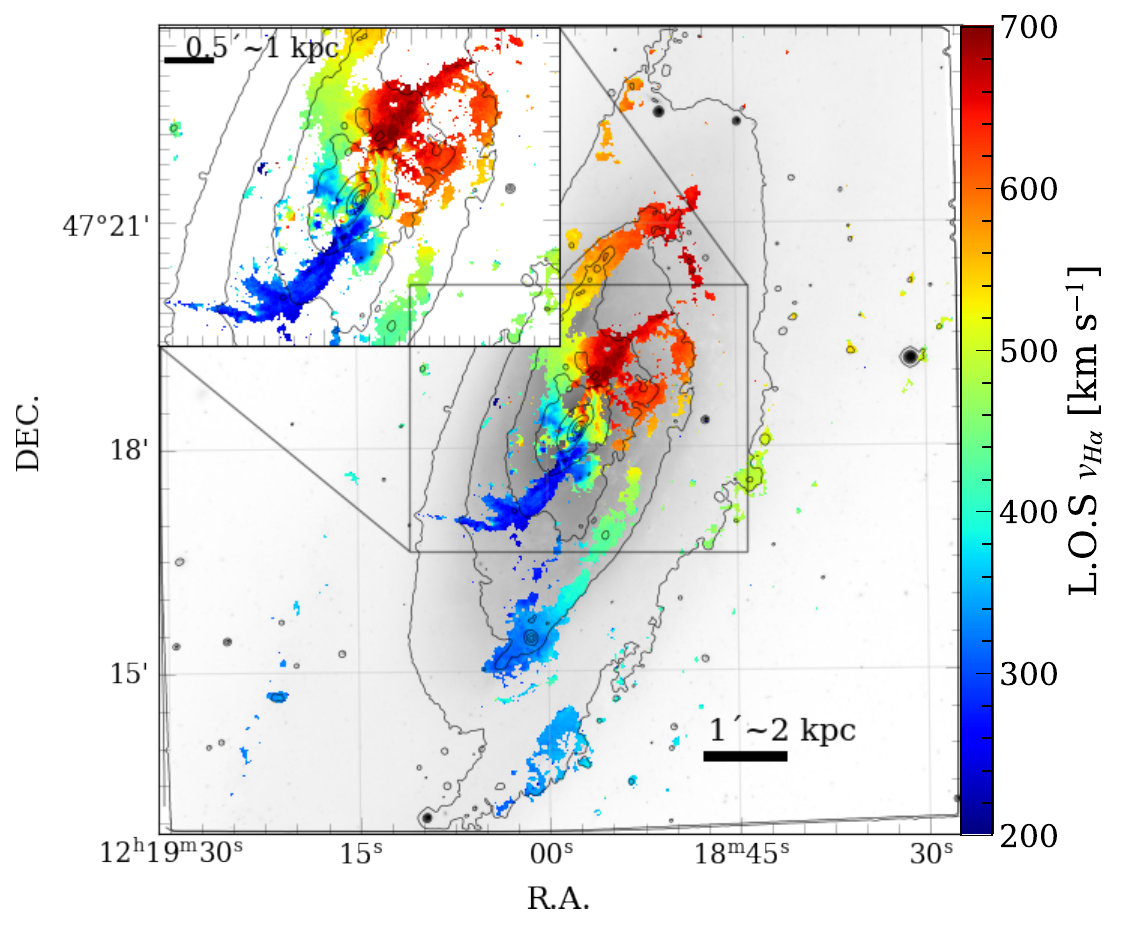}
    \includegraphics[scale=0.48]{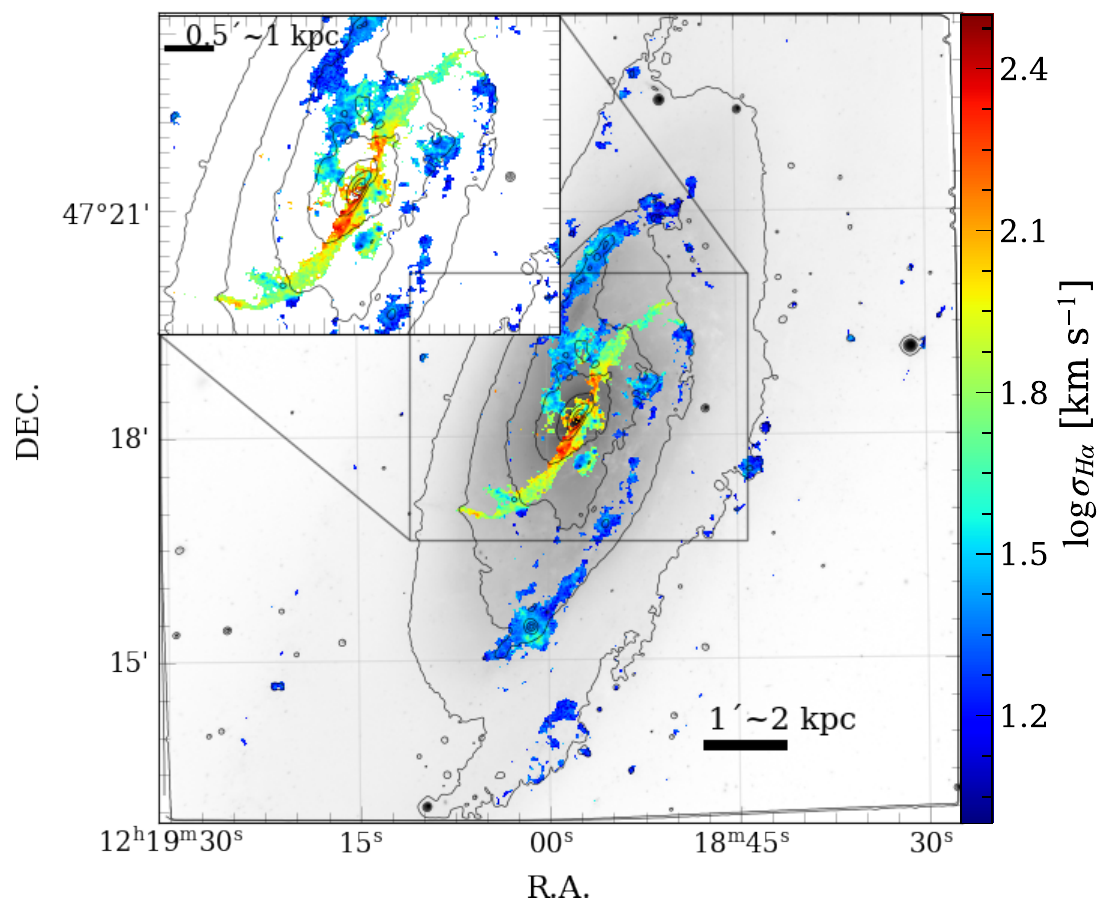}
    \caption[]{Top to bottom: Surface brightness \halpha\ map, line-of-sight velocity, and velocity dispersion for NGC~4258 derived from the fitting of the \halpha\ emission line. Spaxels with S/N<3 have been masked. The background image and the contours have been plotted as a reference and come from the SN3 deep image.}
    \label{kine_maps}
\end{figure}

The uncertainties on the returned parameters given by \texttt{ORCS} are consistent with traditional methods like Monte-Carlo or MCMC algorithm. Additional information returned by \texttt{ORCS} includes the line-of-sight velocity and the velocity dispersion. The maps for \halpha\ line are presented in Figure \ref{kine_maps}. Additional surface brightness maps for different emission lines are provided in Figure \ref{flux_maps}. The velocity provided by \texttt{ORCS} has an offset of $\sim70$ \uvel\ and a gradient of \mbox{$\sim10-20$ \uvel} across the field of view. We correct our velocity map using a modelled velocity of the sky in the field of view obtained by fitting the night-sky OH lines \citep{Martin2018}. 

\subsection{Reddening correction}

The dust interstellar extinction is determined by comparing ratios of hydrogen recombination lines with their theoretical values. In the optical domain, a reliable technique includes deriving the ratio of different Balmer lines, such as \halpha/\hbeta\ or \hbeta/\hdelta. The expected value of the Balmer decrement is determined by quantum physics and, assuming a fixed electron temperature and density, any deviation from this expected value may be attributed to dust extinction. In this study, we calculate the visual extinction (\Av) using \halpha/\hbeta, assuming a \cite{Cardelli1989} extinction curve with R$_v$ = 3.1.

\begin{figure}
\centering
\includegraphics[scale=0.49]{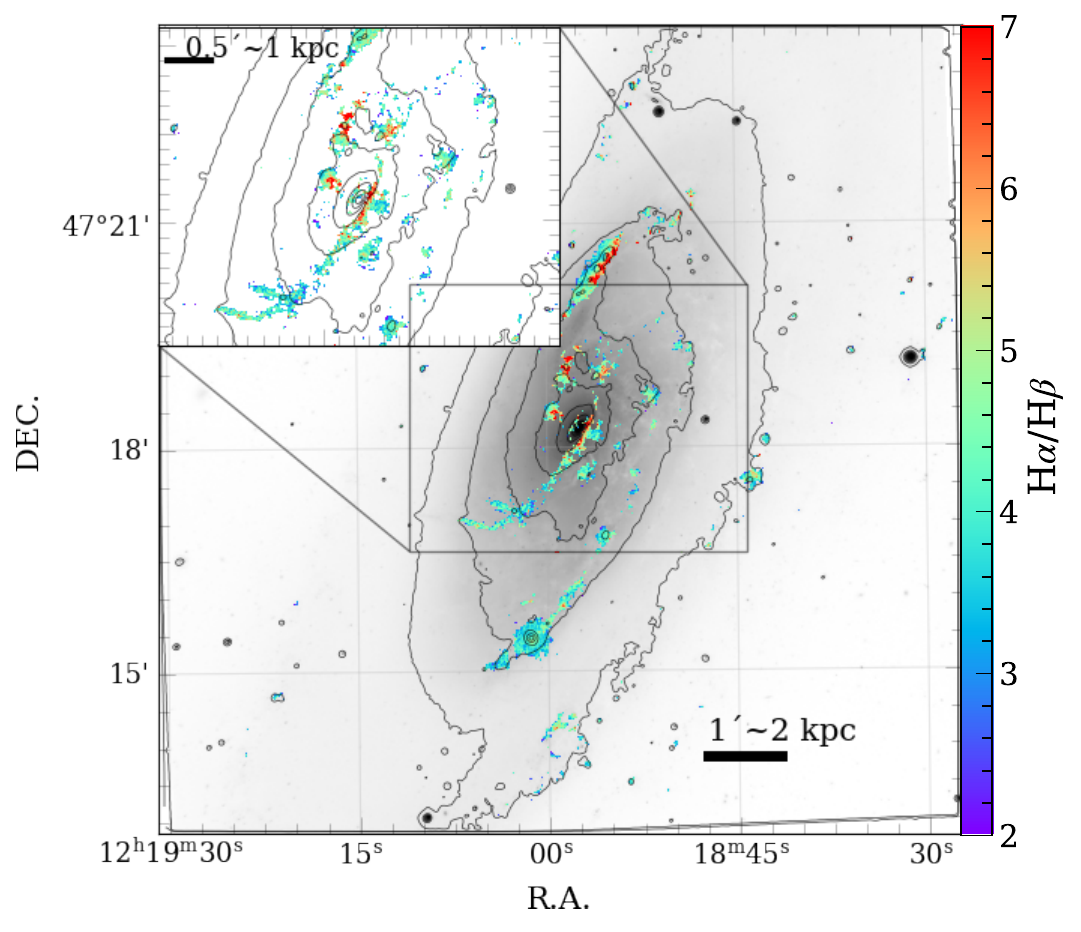}
 \includegraphics[scale=0.49]{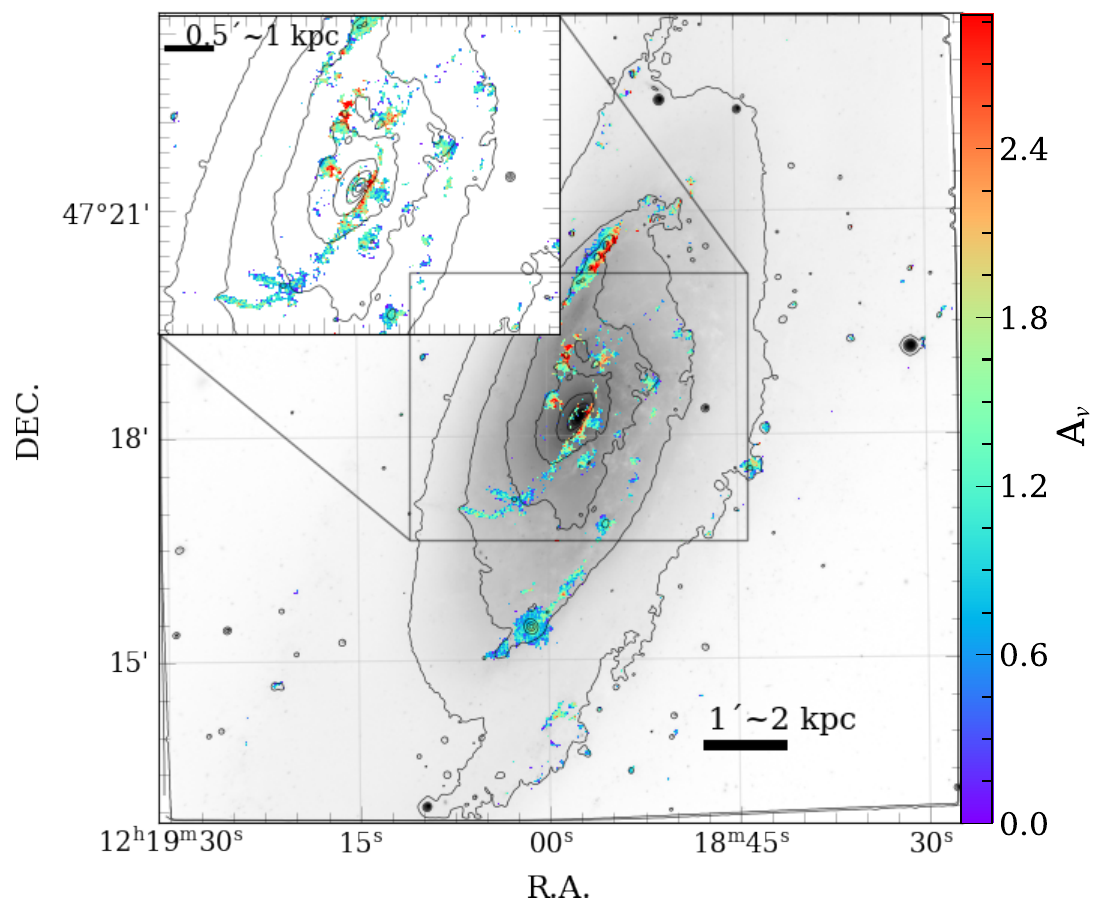} 
    \caption{Balmer decrement for \halpha\ and \hbeta\  emission lines (Top) and visual extinction ($A_{V}$) (Bottom) assuming the theoretical ratios for case B recombination \mbox{$F(\mathrm{H}\alpha) / F(\mathrm{H}\beta) = 2.86$} and the extinction curve of \protect\cite{Cardelli1989} with Rv=3.1.}
    \label{extinction_baldec}
\end{figure}

For \Te=10000 K and \ne=100 cm$^{-3}$, the theoretical \halpha/\hbeta\ ratio is close to 2.86 \citep{Osterbrock2006}. Because extinction is stronger at \hbeta\ than at \halpha\, its effect is to increase the observed ratio. In cases where the flux ratio is lower than the theoretical value, we assume a 0 value for the extinction. Maps of the  Balmer decrement for \hbeta\ and \halpha\ emission lines and visual extinction ($A_{V}$) are shown in Figure \ref{extinction_baldec}.

The expected values for zero extinction range between 2.7 and 3.0 for \hii\ regions (HIIRs) with temperatures between 5000 K and 20000 K, without strong density dependence \citep{Osterbrock2006}. Reported values for HIIRs and star-forming galaxies derived from the aperture of the Sloan Digital Sky Survey (SDSS) show lower and upper values around 2 and 7, respectively, with an average of 4 with the majority of data points clustered around this value \citep{Groves2012}. Upon closer examination of the extinction map in Figure \ref{extinction_baldec}, significant fluctuations in the extinction for individual spaxels become apparent. In some cases, the Balmer decrement ratio was below the theoretical value, while in others, we observed high values around $\sim7$.

Uncertainties associated with fluctuations in the observed Balmer decrements could be attributed to alignment issues between each cube, corrections for the underlying stellar population, or low signal-to-noise ratio (S/N) affecting the \hbeta\ line. To address these problems, several tests were conducted. Initially, data cube alignments were improved with SN3 as references, utilizing star and emission peak identifications. Following the realignment process, a significant improvement of the Balmer decrement was observed being more consistent with the range of values reported in the literature  \citep[see][]{Groves2012}.

Considering the contribution of the stellar absorption as the potential cause of large Balmer decrement ratios, we fitted the underlying stellar population using the youngest single stellar population (SSP) in the MILES library (60 Myr) and an older SSP at approximately 10 Gyr. Despite differences in the fitting process between these SSPs resulting in maximum variations of around $\sim$1, it cannot explain the large differences observed.

\begin{figure}
\centering
	\includegraphics[scale=0.48]{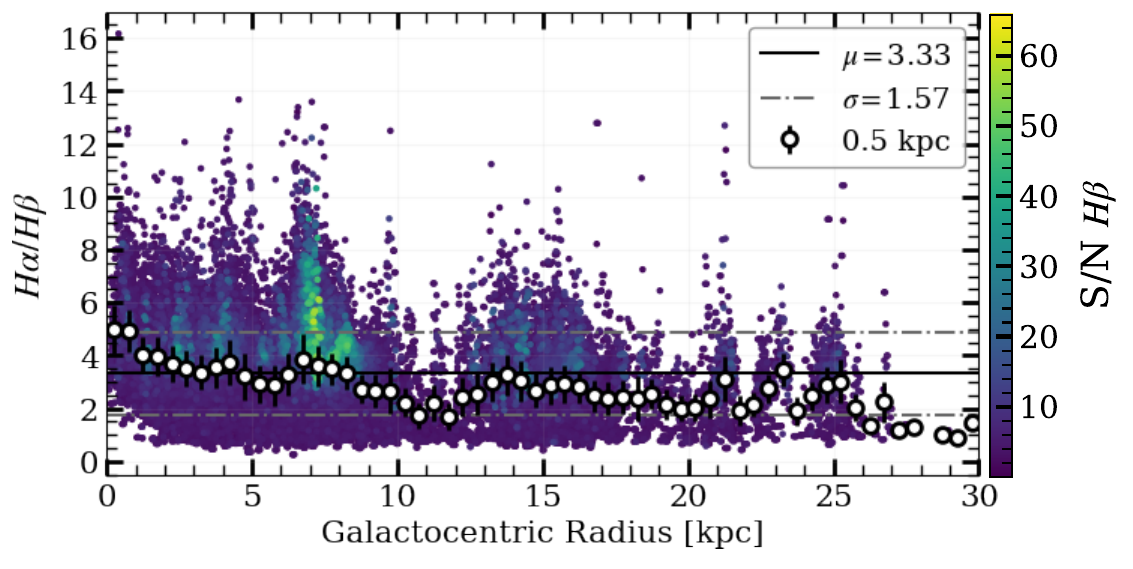}
	\includegraphics[scale=0.48]{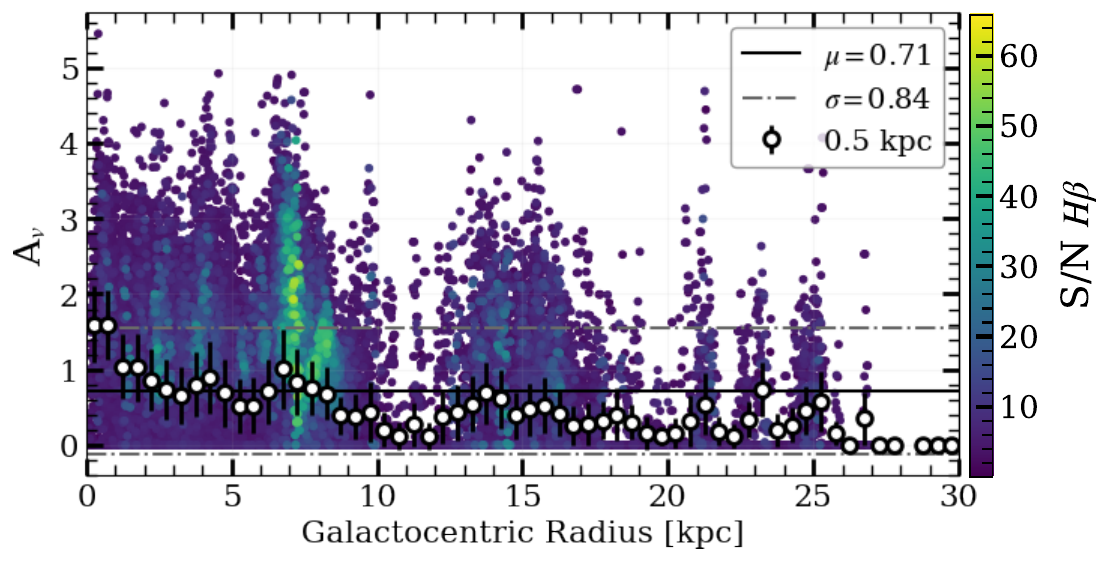}
    \caption{Balmer decrement for \hbeta\ and \halpha\ emission lines and visual extinction ($A_{V}$) assuming the theoretical ratios for case B recombination \mbox{$F(\mathrm{H}\alpha) / F(\mathrm{H}\beta) = 2.86$} and the extinction curve of \protect\cite{Cardelli1989} with Rv=3.1. Horizontal lines show the mean and standard deviation of the points whose values are given in the inset.}
    \label{extinction_baldec_RG}
\end{figure}

Furthermore, an investigation into the impact of S/N on the \hbeta\ line was conducted, as it may be responsible for high values in the ratio of \halpha\ to \hbeta. Figure \ref{extinction_baldec_RG} presents the variation of the Balmer decrement and extinction (\Av) with the deprojected galactocentric radius in the top and bottom panels, respectively. The colour bar indicates the S/N of the \hbeta\ line, with only points having S/N(\hbeta)>3 plotted. Notably, even for high S/N values (>30), large ratios persist, confirming that variations in the Balmer decrement are genuine.

As shown in Figure \ref{extinction_baldec_RG}, we observe that, for high S/N(\hbeta), \halpha/\hbeta\ can reach values up to 2.5 times the theoretical value. These elevated values are more pronounced in the circumnuclear region and toward the northern spiral arm, specifically between 7-8 kpc from the centre of the galaxy, corresponding to extinctions \Av\ between 2-3 magnitudes. This is 2.4$\sigma$ above the average extinction. The variation along the galactocentric radius becomes smoother in bins of sizes 500 pc, respectively. Notably, this smooth behaviour decreases within the first 5 kpc from the centre. In general, we found a median value in the Balmer decrement of $3.33\pm1.47$ and an extinction \Av\ of $0.71\pm0.84$.

The lack of information regarding the exact location, aperture, or emission lines in previous studies \cite[e.g.][]{Oey1993, Bresolin1999, Diaz2000, Bresolin2011} hinders direct comparisons. However, we examined the reported Balmer decrement value in the Sloan Digital Sky Survey (SDSS) for the HIIR located at R.A.=184.730564483 and DEC.=47.280001310, with an aperture of 3.0\arcsec. The SITELLE spectrum in the same region yields a ratio of $4.3\pm0.2$, while SDSS reports $4.1\pm0.3$. These values are consistent considering the errors.

The aforementioned test enables us to confirm the high values in the Balmer decrement and extinction, particularly in the circumnuclear region and toward the north spiral arm (Nsp). The significant variations in the Balmer decrement and high extinction may stem from fluctuations of the ISM in the circumnuclear region, where Seyfert activity is more prominent. These variations could be linked to the complexity of these star-forming regions and the presence of the active galactic nucleus (AGN). In earlier studies of the central region of this galaxy, \cite{Cecil1995} reported Balmer decrement values between 4 and 6 to the south-east and north-west, with the higher value located in the north-west. 

\subsection{Emission line diagnostic classifications}

Distinguishing emission regions according to the gas excitation mechanism is a fundamental concept in exploring star-forming galaxies. For this purpose, specific nebular line ratios are often used to construct the so-called BPT diagrams, originally suggested by \cite{Baldwin1981} and \cite{Veilleux1987}. The BPT diagrams are valuable tools to classify and understand the ionization sources of the gas in galaxies, not only for individual observations but also for spatially resolved galaxies. Standard BPT diagnostic diagrams rely on the line ratios \oiii\lin5007/\hbeta\ versus \nii\lin6583/\halpha\ (BPT-NII), \sii\lin\lin6717,6731/\halpha\ (BPT-SII), or \oi\lin6100/\halpha. High excitation values correspond to elevated \oiii\lin5007/\hbeta\ ratios and low values of \nii\lin6584/\halpha\ and \sii\lin\lin6717,6731/\halpha. These ratios are chosen because they are sensitive to the excitation mechanisms and physical conditions of the ionized gas, and they are close in wavelength, reducing the effect of reddening.

High values for \oiii\lin5007/\hbeta\ are expected when ionization is predominantly produced by UV photons, especially with a high ionization parameter. These diagrams have been extensively studied in the literature, from the local universe to high-z \cite[e.g.,][]{Kewley2013b}. Alongside empirical demarcation curves, \cite[e.g.,][]{Kauffmann2003,kewley2006, Stasinska2006, CidFernandes2010, CidFernandes2011, Kewley2013a}, these diagrams allow us to distinguish between different ionization sources in the interstellar medium. The BPT diagrams have not only been used to study surveys with single-aperture observations \cite[e.g.,][]{Teimoorinia2018, Wylezalek2018, Zewdie2020} but also to dissect properties of ionizing sources in galaxies using maps derived by IFS \cite[e.g.,][]{DAgostino2018, Mingozzi2019, Belfiore2022}.

\begin{figure}
	\includegraphics[scale=0.48]{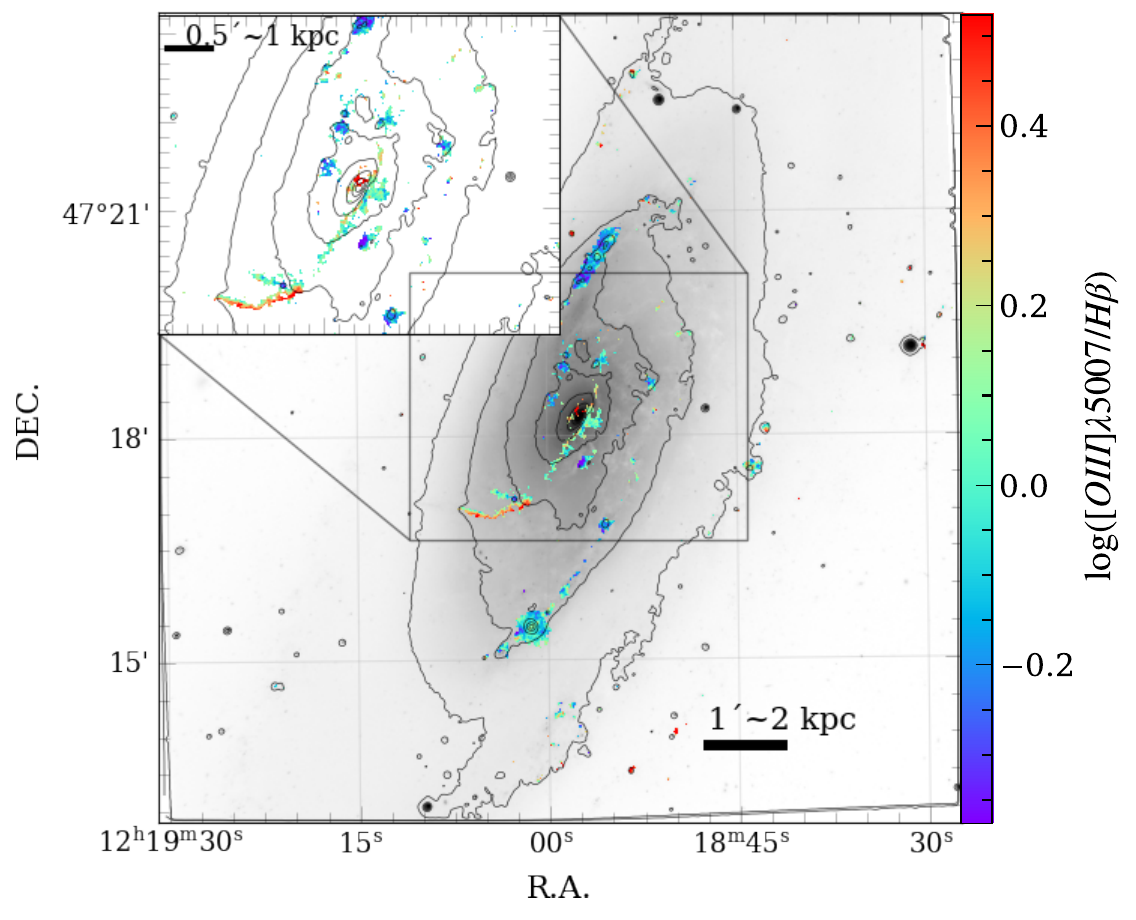}
	\includegraphics[scale=0.48]{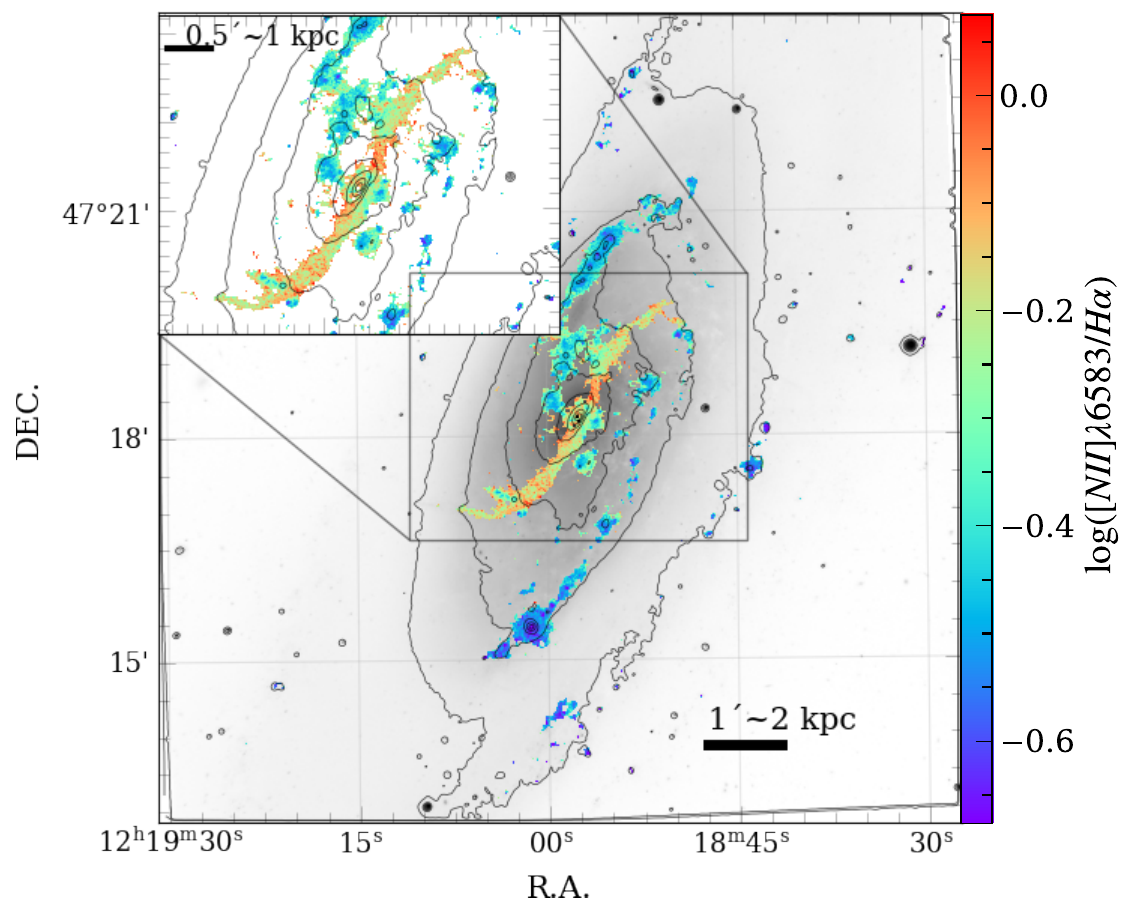}
        \includegraphics[scale=0.48]{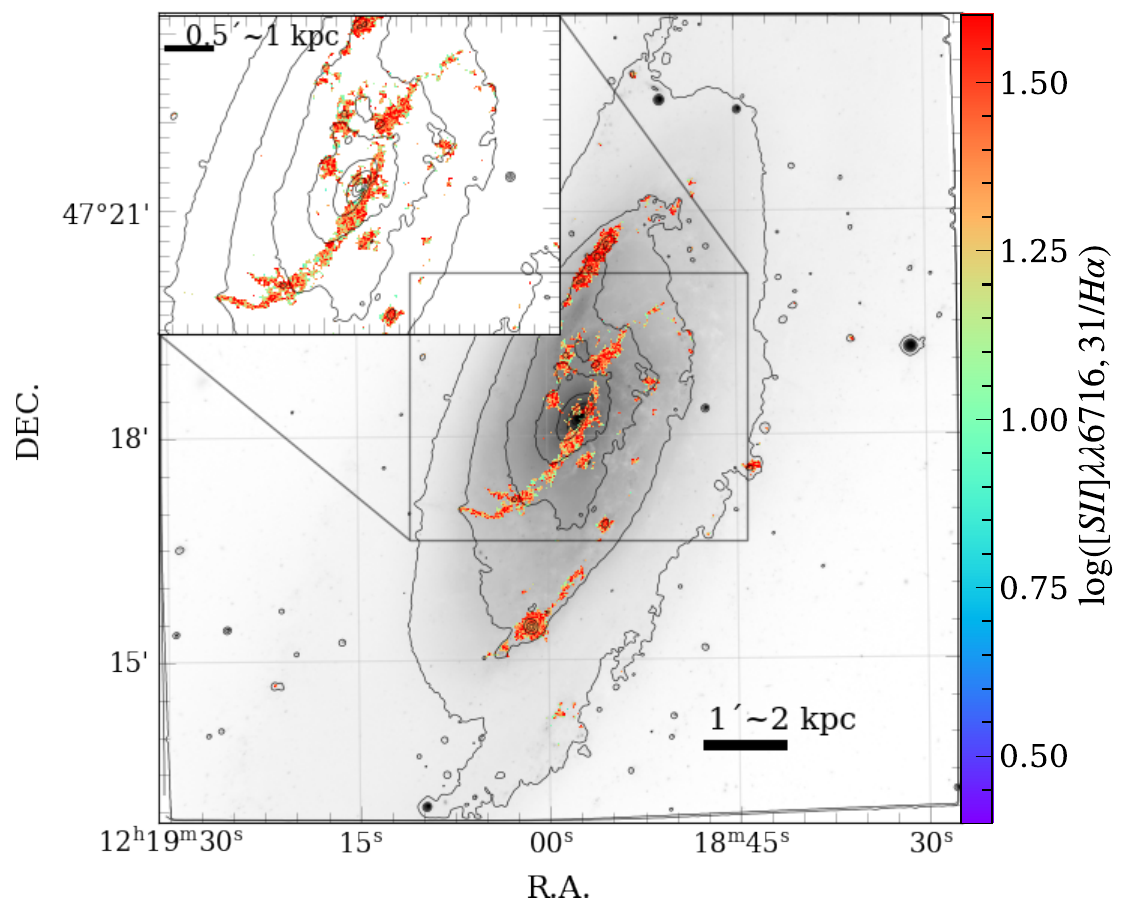}
	 \caption[]{Maps of the diagnostic line ratios in logarithmic scale. The label inset indicates the line ratios  \oiii\lin5007/\hbeta, \nii\lin6584/\halpha\ and \sii\lin\lin6717,31/\halpha. The contours have been plotted as a reference and come from the SN3 deep image.}
    \label{_diag_ratio_maps}
\end{figure}

To investigate the excitation mechanisms acting in NGC~4258, we derived the line ratio maps for \oiii\lin5007/\hbeta, \nii\lin6584/\halpha, \sii\lin\lin6717,6731/\halpha. For this work, we apply a consistent S/N cut of 3 on all emission lines in the BPT diagrams. The maps corresponding to the  \oiii\lin5007/\hbeta, \nii\lin6584/\halpha\ and \sii\lin\lin6717,31/\halpha\ emission line ratios are presented in Figure \ref{_diag_ratio_maps}.  According to these maps, the central region shows higher \nii\lin6584/\halpha\ and \sii\lin\lin6717,31/\halpha\  line ratios compared to the values distributed in the spiral arm. These high values follow the morphological structure of the collimated jet very well. 

\begin{figure*}
	\includegraphics[scale=0.55]{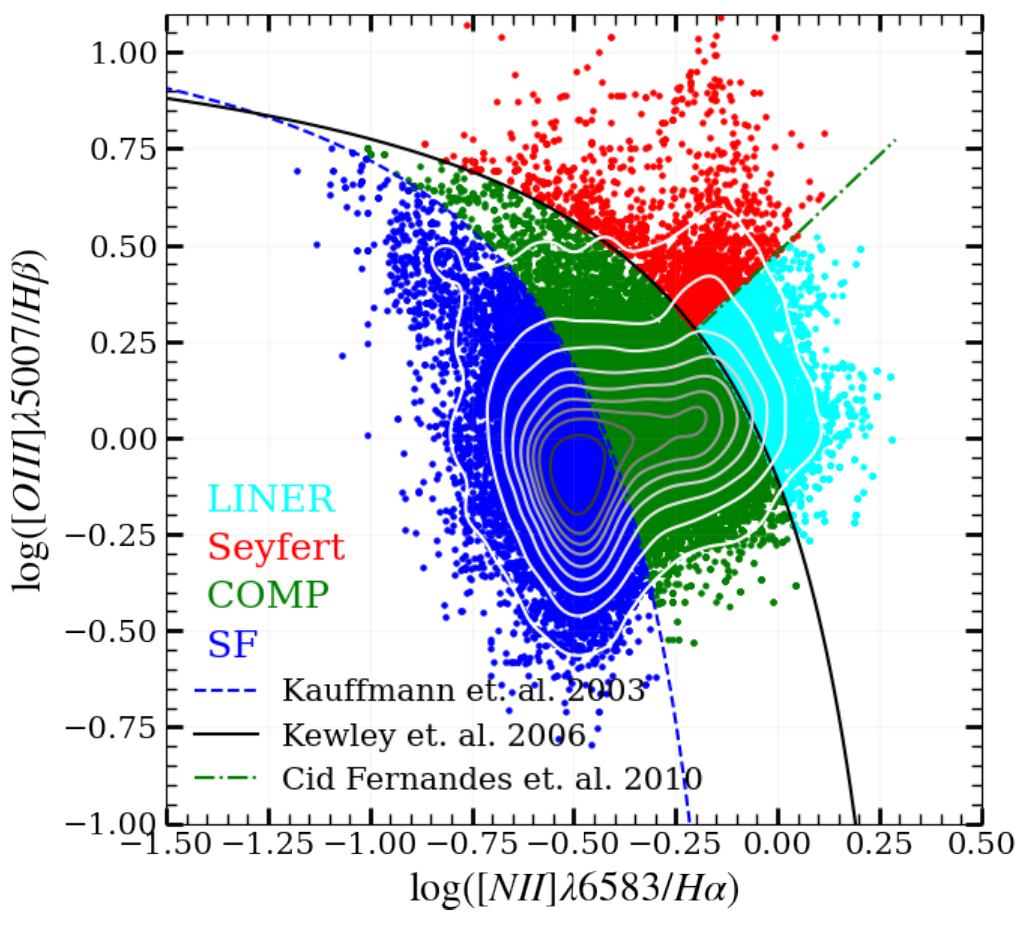}\includegraphics[scale=0.55]{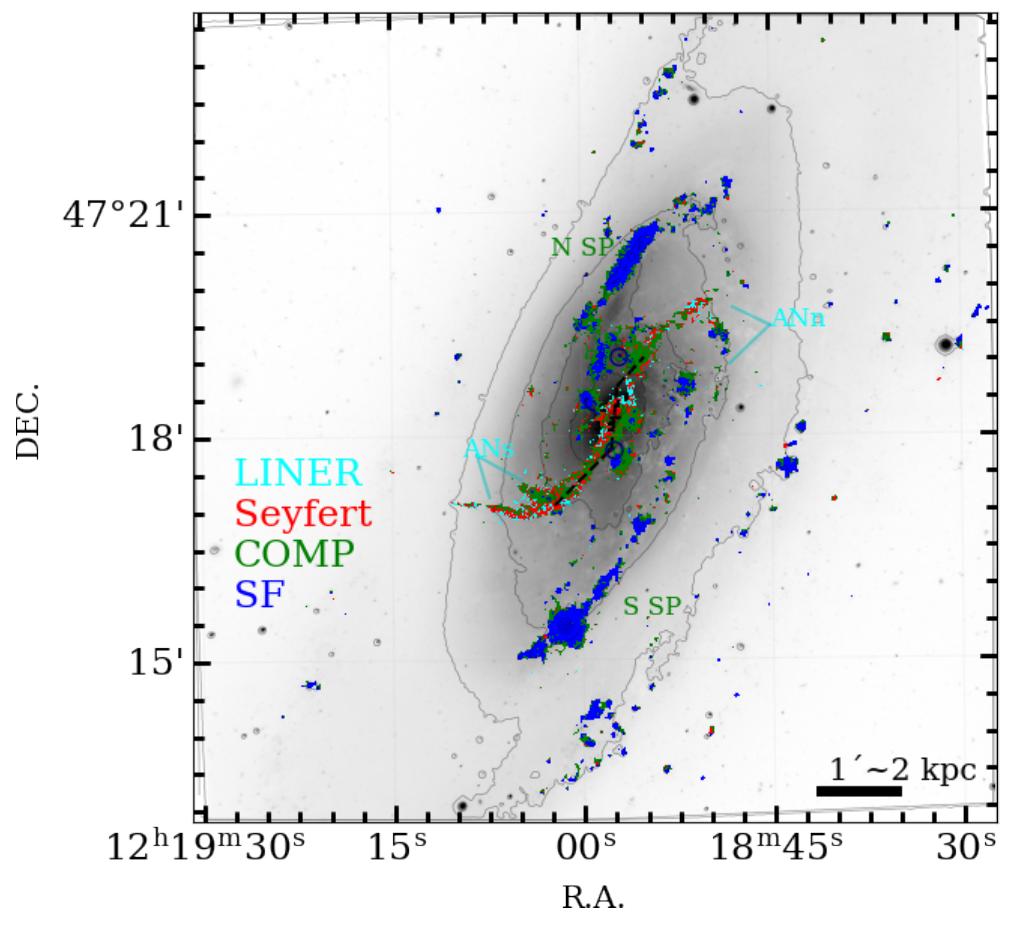}
	\includegraphics[scale=0.55]{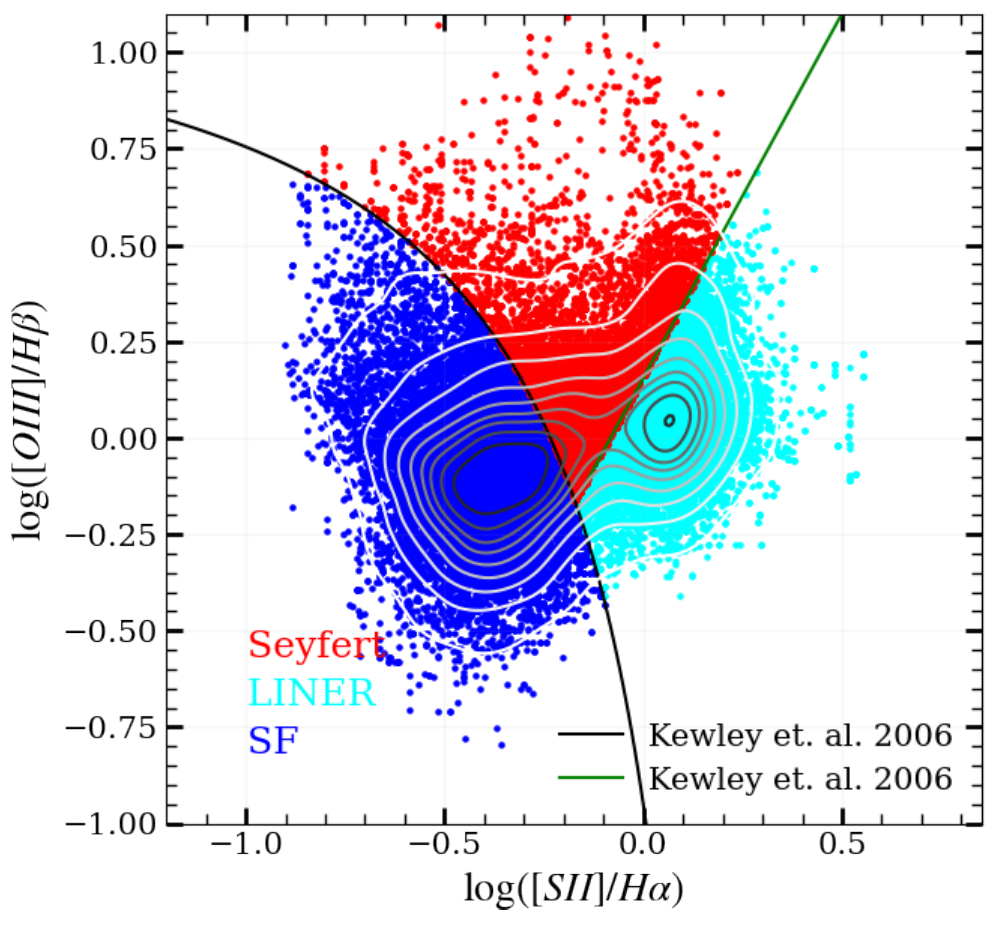}\includegraphics[scale=0.55]{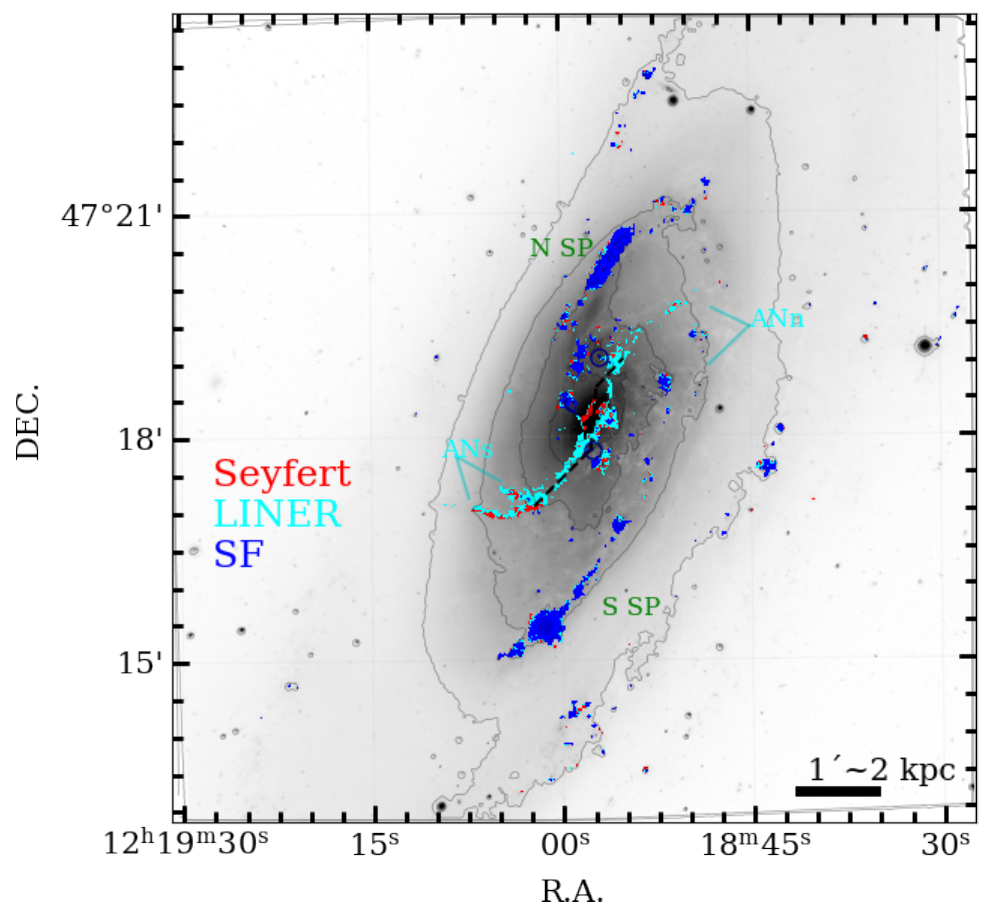}
    \caption{Left panels: NII- and SII-BPT diagrams for NGC~4258. The \protect\cite{kewley2006} classifications are shown as the solid lines. The black solid lines separate the star-forming determined by the upper limit of the theoretical pure stellar photoionization models from those emissions dominated by an AGN. The green line represents the Seyfert-LINER demarcation in the \oiii\lin5007/\hbeta\ vs. \sii\lin6717,31/\halpha\ diagram. In the \oiii\lin5007/\hbeta\ vs. \nii\lin6583/\halpha, \protect\cite{Kauffmann2003} classification line is shown as the blue dashed line, whereas, the dashed-dotted line corresponds to the separation between Seyfert–LINER given by \protect\cite{CidFernandes2010}. Iso-contours are point densities shown in grayscale. Right panels: The distribution of the ionizing source according to the BPT and its location in the galaxy. The colour code in the BPT and maps is the same, indicating the different locations of the spaxels in the FoV classified according to their ionizing source. The points correspond to fluxes with S/N$\geq$3 in all lines. \textbf{A more detailed zoom of the central region is presented in Figure \ref{bpt_centre}}}
    \label{bpt_diagrams} 
\end{figure*}

We created spatially resolved BPT-NII and BPT-SII diagrams by combining the maps of different line ratios. The diagrams and the corresponding spatial distribution in the galaxy are shown in Figure \ref{bpt_diagrams}. These BPT diagrams allow us to explore the distribution of the ionization mechanism not only in the central region of NGC~4258 but also in the outskirts of the galaxy. 

As previously noted by \cite{Appleton2018}, the strong segregation between the minor-axis filament and the southern arm of the jet is evidence that two different heating mechanisms are present in the centre of NGC~4258, namely young massive stellar clusters and nuclear activity.
 
\subsubsection{\hii\ regions}

Classical \hii\ regions are gas clouds ionized by short-lived hot OB stars, typically associated with the latest burst of star formation. They are often selected based on demarcation lines defined in NII- and SII-BPT diagrams. Star-forming regions located below this demarcation have been shown to form a tight sequence in metallicity \citep{Dopita2000,kewley2006}, known as the star-forming or \hii\ ``abundance sequence" or ``star-forming sequence",  from low metallicities (high \oiii/\hbeta\, low \nii/\halpha) to high metallicities (low \oiii/\hbeta\, high \nii/\halpha). It has been used as a tool for investigating the metallicity, ISM, and ionizing radiation field in star-forming galaxies in the local Universe using IFS and also as a function of cosmic time, independent of the large systematic errors \cite[e.g][]{Kewley2013b}.

We found that most spaxels fall within the star-forming region and they are located mainly along the spiral arms, with a minor contribution in the nuclear region, particularly in the area of the \halpha\ filaments, as previously noted by \cite{Appleton2018}. We found some spaxels in the shorter bifurcation of the south-east jet falling in the region classified as star-forming in the BPT diagrams; their velocity dispersion is around 30 \uvel, which has not been previously reported.

\subsubsection{Composite}

The region on the BPT diagram that falls between the \cite{Kauffmann2003} and \cite{kewley2006} demarcation lines is considered to result from a mixture of ionization, involving both star formation and an additional hard component, mainly here from AGN activity. For the spaxels in the region of composite objects according to the NII-BPT diagram, we find the spatial location coincident with the jet aligning with the emission in the radio and X-ray in the central region. The N-Shock arc is partially detected, making it difficult to distinguish its origin. On the other hand, the arc in the southern part of the galaxy, referred to as ``S-Shock," is detected and located very close to a star-forming region. To the south of the spiral arms, we observe a complex of compact star formation regions surrounded by emission, probably related to shocks. Potentially, these could have originated from supernova remnants surrounded by structures resembling arcs in the \halpha\ image.

\subsubsection{AGN}

Contemporary works on the mixing of star formation and AGN consider emission lines in the region below the \cite{Kauffmann2003} line on the BPT diagram as purely originating from star formation. In contrast, emission lines above the \cite{kewley2006} line are considered to be dominated by harder sources, such as AGN or shocks. Additionally, \cite{kewley2006} proposed low ionization line criteria for separating Seyferts and LINERs. Other authors have suggested different classifications  \citep{Stasinska2006,CidFernandes2010,CidFernandes2011}, especially in the NII-BPT diagram. Here, we used the classical \cite{kewley2006} line to separate LINERs from Seyfert emission in the SII-BPT diagram, whereas, in the NII-BPT diagram, we applied the demarcation given by \cite{CidFernandes2011}.

From our analysis of the BPT diagrams, we find a region located in the centre of NGC~4258 as an expected product of the accretion of material towards the interior of the black hole and also an interesting region in the ``anomalous spiral arm"  towards the south which is dominated by emission coming from AGN with the high values of \oiii/\hbeta. In the SII-BPT diagram, we find an equal distribution of the emission of AGN in red and LINERs in Cyan. Nuclear activity could be producing an ionization cone in NGC4258.

In the centre of the galaxy, we can distinguish the northern loop, detected for the first time by \cite{Cecil2000} and described by \cite{JVicente2010}. The morphology of the loop in our images better follows the appearance of data presented by \cite{JVicente2010}, although the nature of this loop remains uncertain. \cite{JVicente2010} argues that it can be produced by gas bubbles originating from the interaction of the jet with the dense interstellar medium, as proposed by \cite{Cecil1992}. We observed, in the images of \halpha, that the contrast between the eastern and western sides of the loop seems weaker than in \cite{Cecil2000} as in \cite{JVicente2010}, and the BPT diagrams support that this loop comes entirely from ionization by AGN. The trailing strands of the jet at the south-east at the end of the bifurcation have high excitation, as does the region near the nucleus.

\begin{figure*}	\includegraphics[width=\columnwidth]{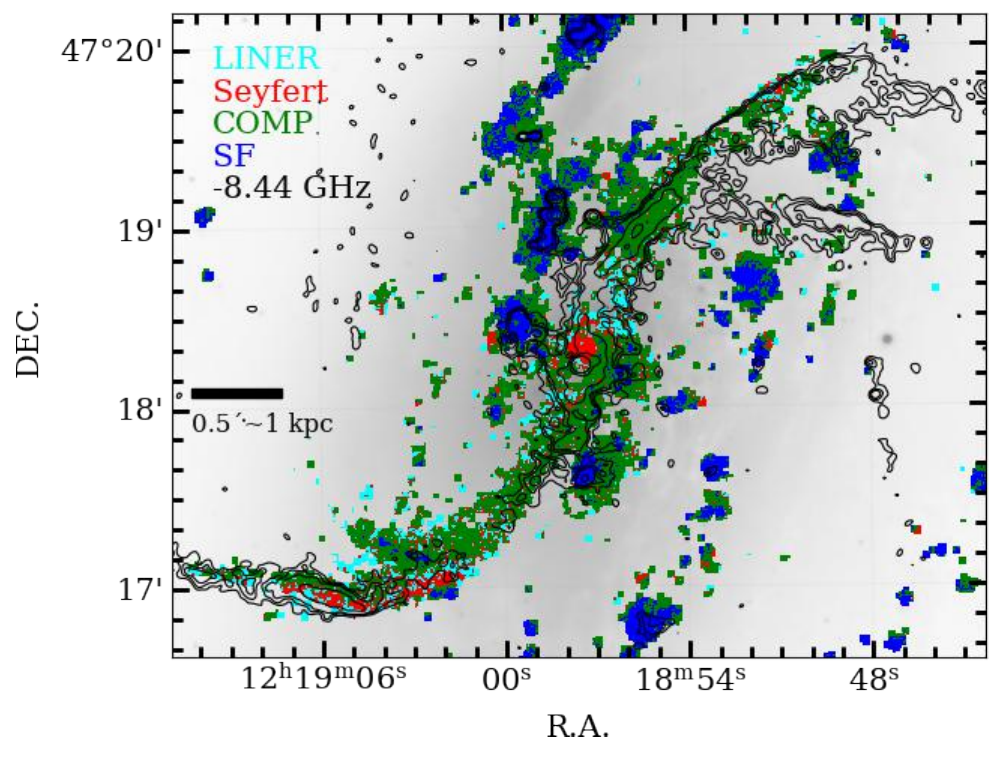}\includegraphics[width=\columnwidth]{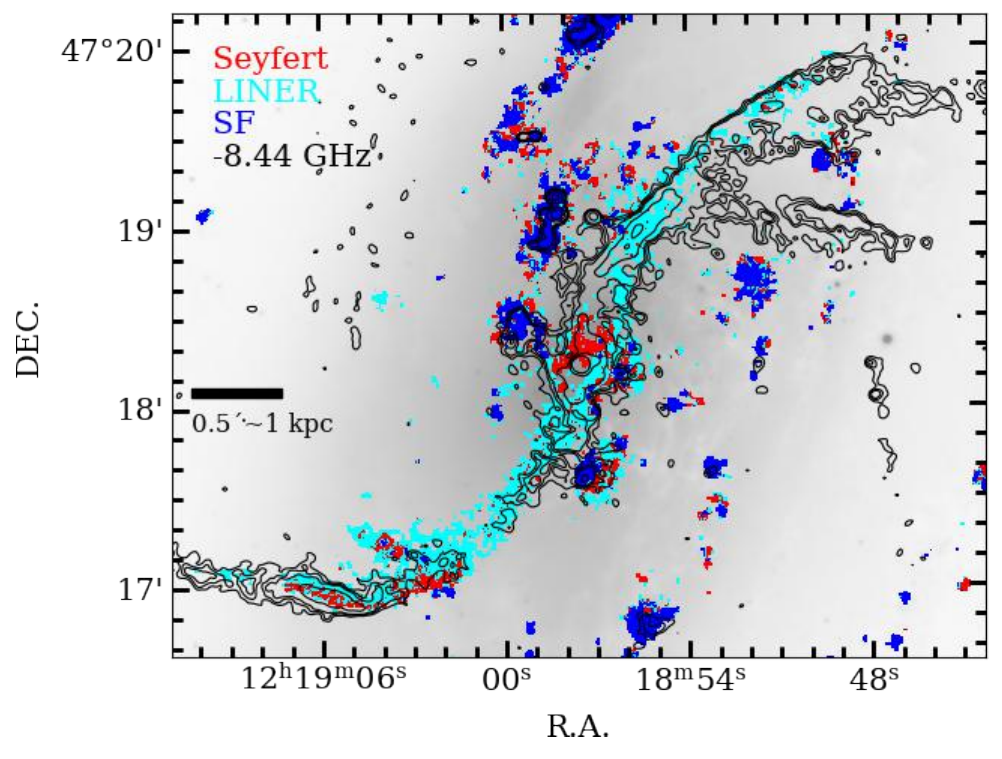}
    \caption{Maps showing the centre of the galaxy and the spaxels classified according to their ionization source based on NII- and SII-BPT diagrams. In black, the contours correspond to observations using the VLA at 8.44 GHz, as reported by \protect\cite{Krause2004}.}
    \label{bpt_centre}
\end{figure*}

The central region of NGC~4258 is more interesting compared to the typical spiral arms showing star-forming regions. In Figure \ref{bpt_centre}, we present the central region and its classification based on the ionization source obtained from the BPT diagrams. On this figure, we superimposed the 8.44~GHz from the VLA contours as reported by \cite{Krause2004}. The superposition assigns clearly which part of the radio emission belongs to the spiral arms and which belongs to the jets. The clumps of strong emission along the spiral arms are well correlated with the regions of highest intensity in \halpha. However, \cite{Krause2004} has suggested that this radio emission may come from two processes: unpolarized thermal radiation and polarized and nonthermal. In the case of the first process, they suggest an emission that comes from star-forming regions, while the second comes from the Jet. The highest polarized emission occurs at the centre and along the northern jet (35\% and 65\% of the total intensity), mainly to the northwest of the bifurcation, whereas small regions of low percentage polarization (20\%) in the southern jet are detected. 

Comparing our BPT spatially resolved and figures 3 and 4 presented by \cite{Krause2004};  where they show the location and length of the vectors proportional to the linearly polarized intensity. We observed the main non-polarized emission comes from star-forming regions, where some of the intensity peaks in radio emission and blue points in the map overlap. On the other hand red, green, or cyan regions overlap very well with the maximum radiation corresponding to the galactic centre (red pixels) and the jet, matching very well the polarized intensity radio emission intensity. Here, using spatially resolved BPT diagrams, we confirm the suggestion of the origin of polarized radio emission proposed by \cite{Krause2004}.

\subsubsection{Mixing sequence and Star Formation Rate.}

NGC~4258 exhibits active star formation primarily concentrated in its spiral arms. However, most of the research has predominantly focused on the central region. Some estimates of the star formation rate (SFR) have been conducted using HIIRs detected in radio emission, resulting in a total SFR of \mbox{0.078 \usfr} for the uncontaminated spiral arms \citep[9 regions;][]{Hyman2001}.

In contrast, a different approach involving modelling the spectral energy distribution (SED) within the central 3.4 kpc$^2$ by \cite{Ogle2014} yielded an SFR of \mbox{$\sim$ 0.084 \usfr}. Additionally, from the luminosity of the 7.7$\micron$ PAH feature, they estimated a rate of \mbox{0.069 \Msol y$^{-1}$}. However, they argue that using a CO-based molecular gas mass density of \mbox{$\Sigma($H$_2)$=300 \Msol pc$^{-2}$}, the Kennicutt-Schmidt (KS) law \cite{Kennicutt1998} would predict a value 25 times greater for the central region, which is not observed \cite[see][for a deeper discussion]{Ogle2014}. So, the nuclear SFR in NGC~4258 appears to be quite modest, as suggested by previous studies.

We estimated the galaxy-wide SFR from the \halpha\ emission line. The simplest approach to calculate the SFR using the \halpha\ emission line involves excluding regions exhibiting AGN ionization and integrating only areas associated with HIIRs. This is because not all \halpha\ emission originates from star-forming regions, and direct transformation to SFR may not be accurate. While this approach has limitations, particularly in traditional narrow-band imaging or long-slit spectroscopy, in the last few years, innovative techniques have been introduced in the literature. These techniques are accessible only with IFS and allow us to correct the \halpha\ flux by accounting for the AGN contribution \citep{Davies2014, Davies2017, Thomas2018, DAgostino2018, Neumann2019, Smirnova2022}. In the following, we briefly discuss the method and the application to NGC 42458 and SITELLE data.

In nearby active galaxies, elevated AGN-ionization is typically observed near the nucleus, with star-forming regions predominantly located in the outskirts of the galaxy. Moreover, the proportion of emission attributed to AGN activity gradually diminishes with increasing distance from the galaxy's centre. The intricate interplay between star formation and AGN activity is reflected in the BPT diagrams, resulting in a curve known as the ``Mixing Sequence". This sequence exhibits a strong dependence on the distance to the galaxy's centre \citep{Davies2014, Davies2016}.

The ``mixing sequence" serves as a compelling proxy for estimating the SFR using \halpha\ emission corrected by AGN contamination. Recent studies on nearby galaxies hosting AGNs have demonstrated the efficiency of this approach \citep{Neumann2019, Agostino2021, Smirnova2022, Molina2023}. In this study, we implemented the ``mixing sequence" and applied extinction corrections to derive both spatial and galaxy-wide SFR values.

The primary assumption of this method is that the measured emission-line fluxes along the line of sight ($\mathbf{F_{obs}}$) are a combination of underlying emissions attributed to pure star formation ($\mathbf{F_{SF}}$) and those originating from the AGN ($\mathbf{F_{AGN}}$). This is achieved by utilizing resolved spatial BPT diagrams. Thus, \mbox{$\mathbf{F_{obs}}=f_{SF}\cdot\mathbf{F_{SF}}+f_{AGN}\cdot\mathbf{F_{AGN}}$}, where $f_{AGN}$ and $f_{SF}$ are non-negative linear coefficients representing the AGN and star-forming ionization fractions, respectively. An important additional basic constraint of \mbox{$f_{AGN}+f_{SF}=1$} ensures that flux is preserved. 

We employ a Monte Carlo Markov Chain (MCMC) exploration, facilitated by the Python package \texttt{emcee}, to minimize $\mathbf{F_{obs}}$ in terms of $f_{SF}$ and different basis points that best represent the emission-line flux dataset among all possible combinations. Different methodologies for obtaining $f_{SF}$ or choosing base points may introduce errors of up to 15\%. Therefore, the MCMC exploration enables us to obtain more accurate estimates of uncertainties \citep[e.g.][]{Smirnova2022}. One can pick basis vectors of emission line ratios to characterize pure AGN and star-forming ionization and treat the composite data points of the ``mixing sequence" as a linear combination of the basis vectors:

\begin{equation}
\begin{pmatrix} \text{\nii/\halpha} \\ \text{\oiii/\hbeta} \end{pmatrix}_{\text{obs}} =
f_{SF}\begin{pmatrix} \text{\nii/\halpha} \\ \text{\oiii/\hbeta} \end{pmatrix}_{\text{SF}}+
(1-f_{SF})\begin{pmatrix} \text{\nii/\halpha} \\ \text{\oiii/\hbeta} \end{pmatrix}_{\text{AGN}}
\end{equation}

The emission line columns are normalized to either \halpha\ or
\hbeta\, which minimizes the impact of internal extinction within the
host on the results. In summary, the idea behind the method is to pick emission-line basis vectors at the extreme ends of the ``mixing sequence", which should best reflect the assumption of two completely independent excitation mechanisms.

Finally, the SFR for each spaxel is obtained from the \halpha\ luminosity corrected by extinction and multiplied for the star-formation fraction and using the relation from \cite{Kennicutt1998a}: 

\begin{equation}
\frac{SFR}{\text{\usfr}}=7.9\times10^{-42}\frac{\text{L(\halpha)}}{\text{\ulum}}
\end{equation}

\begin{figure*}
\includegraphics[scale=0.45]{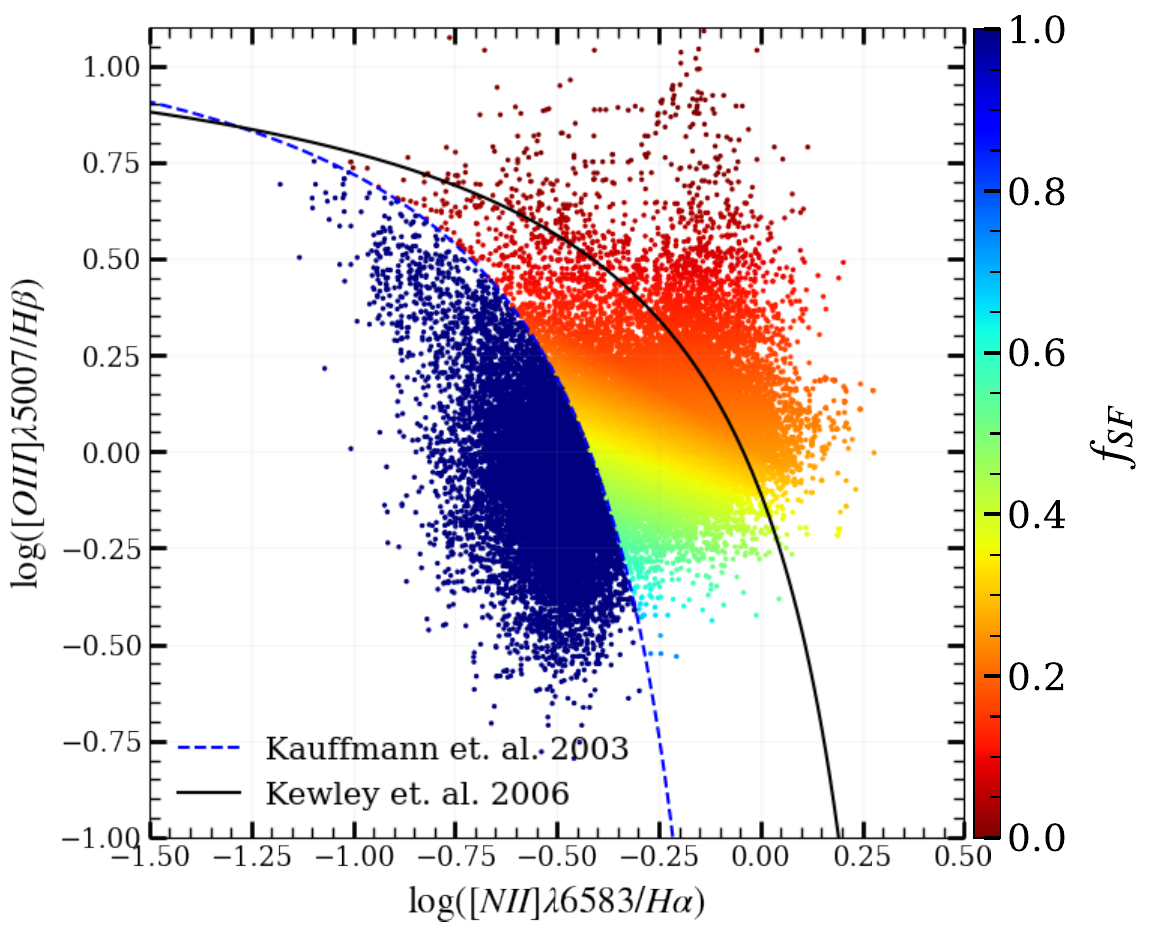}
\includegraphics[scale=0.49]{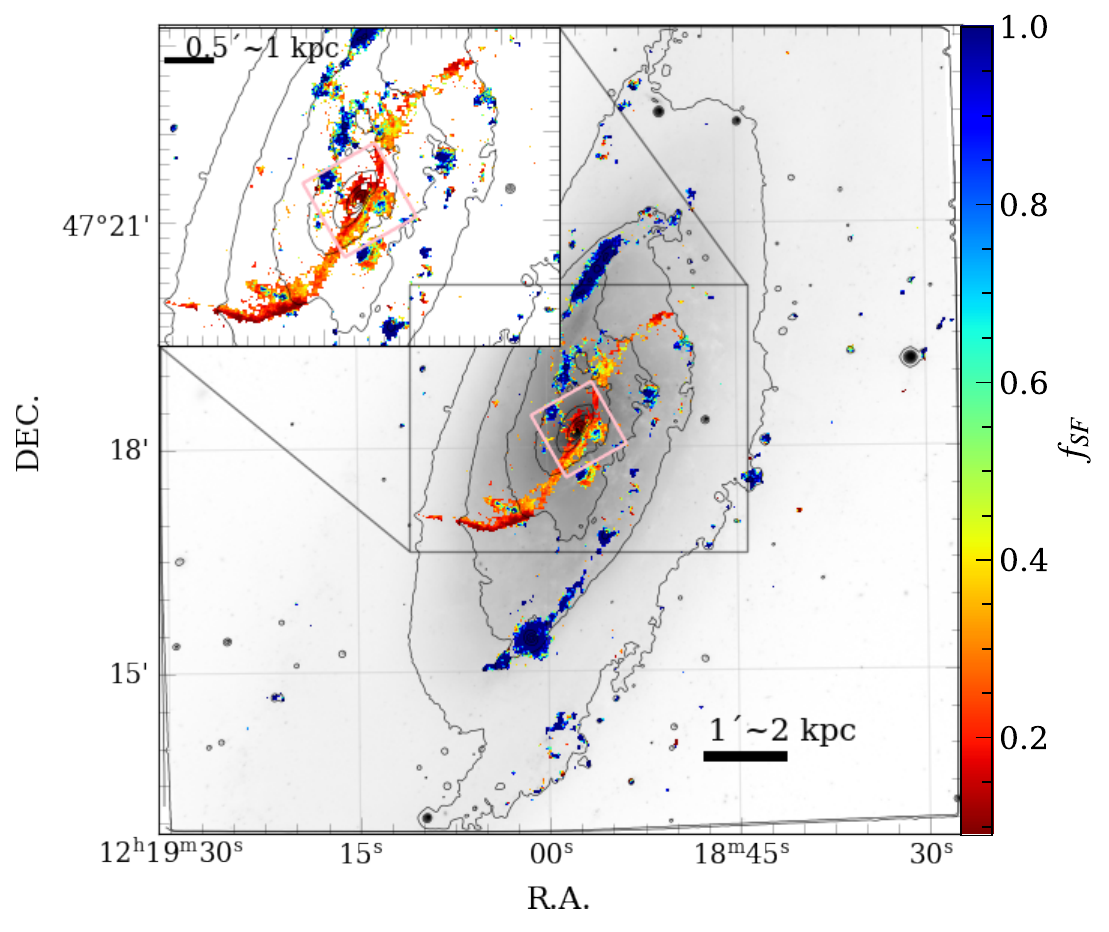}\includegraphics[scale=0.49]{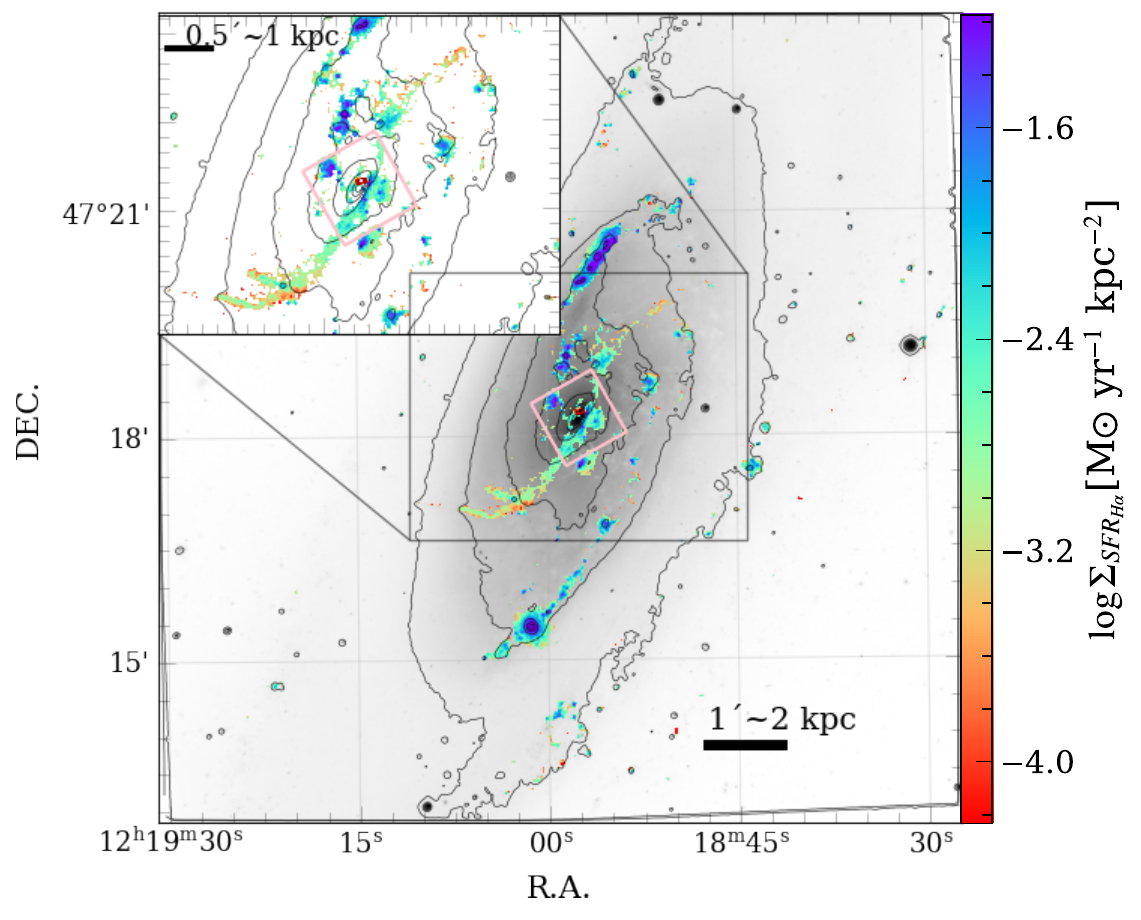}
\includegraphics[width=\columnwidth]{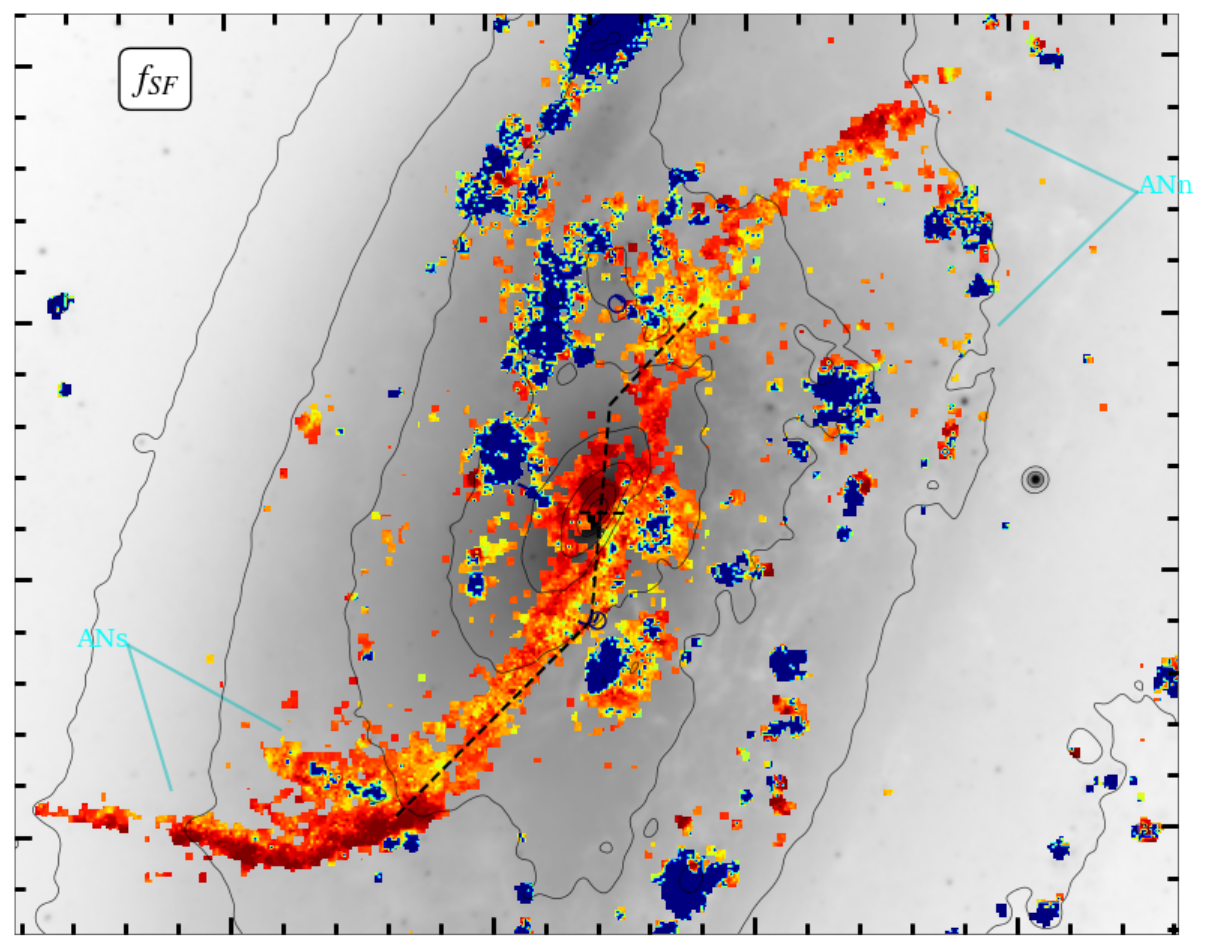}\includegraphics[width=\columnwidth]{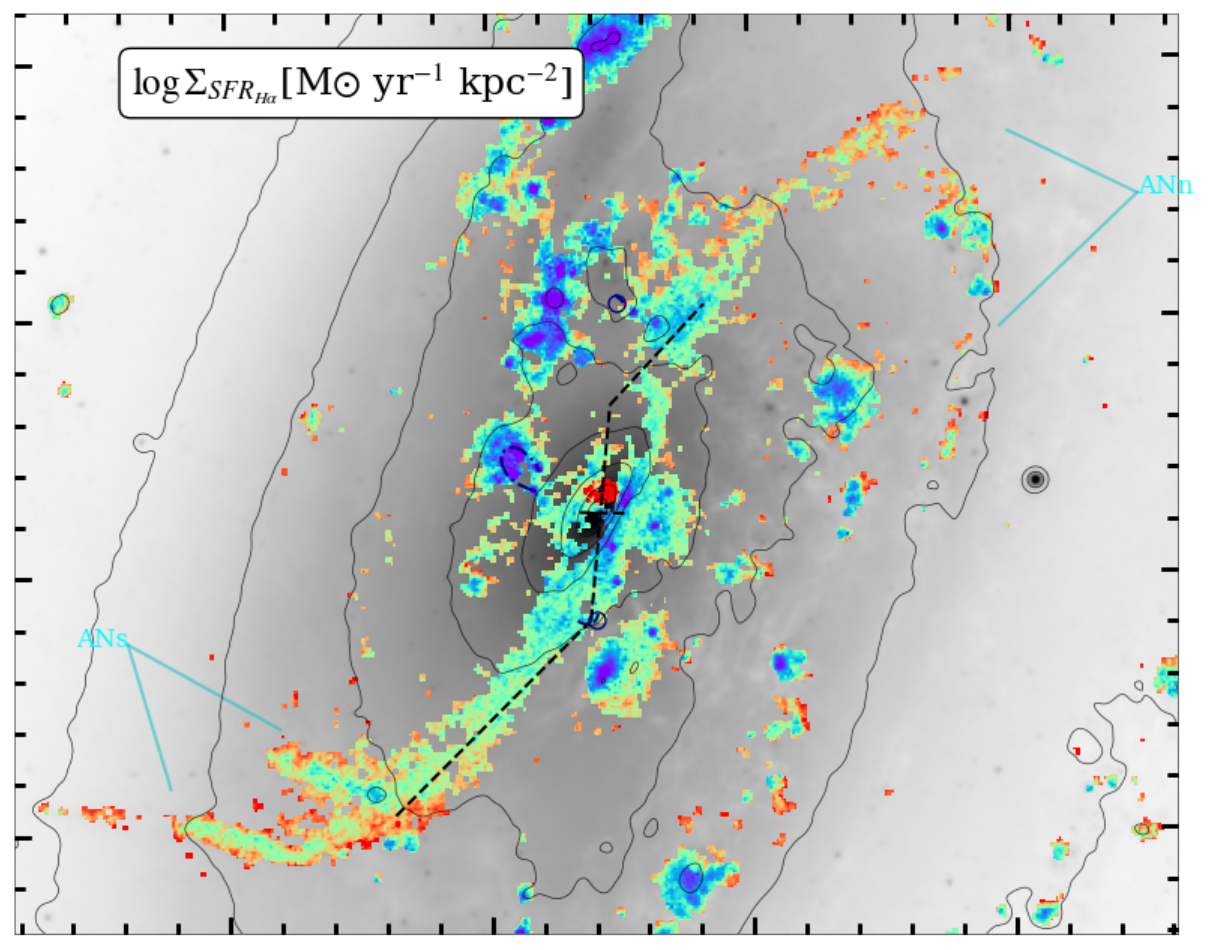}
\caption[]{Top Panel: BPT diagram of NGC~4258 illustrating a mixing sequence between ionization by \hii\ regions and AGN. Middle Panels left and right, respectively: Distribution of $f_{SF}$ across the galaxy, with blue points representing ionization by star-forming regions and red points indicating AGN ionization and Distribution of Star Formation Rate (SFR) per pixel across the galaxy, respectively. The inset pink region corresponds to approximately 3.4 kpc$^{2}$, as demonstrated by \protect\cite{Ogle2014}. \textbf{In the bottom panels, we plot the zoom of the central region.  Only spaxels in the BPT diagram with all lines involved in the ratios have been considered.}}
    \label{sfr_frac_ratio}
\end{figure*}

In Figure \ref{sfr_frac_ratio}, we observe the distribution of the BPT diagram as a function of $f_{SF}$ and its spatial distribution across the galaxy. The ``mixing sequence" aligns with the overall spatial pattern of the galaxy, similar to other nearby active galaxies in the local universe \citep[e.g][]{Davies2014,Davies2016,Richardson2014,Neumann2019}, with high AGN-ionization predominantly situated near the nucleus. The fraction's distribution correlates with the distance to the galaxy's centre, nearly diminishing at the nucleus and gradually increasing towards regions along the spiral arms.

Along the jet, the fraction remains below 30\%, indicating a minimal contribution of ionization due to star formation. However, at the end of the jet and in the region of the bifurcation in the south-east, there is a noticeable contribution from star-forming regions, with values consistent with 1 in the region of the shortest bifurcation. In contrast, the longest bifurcation is dominated by AGN ionization. This observation might be linked to the illumination of the biconical-shaped jet and its interaction with the interstellar medium outside the plane of the disc. Simultaneously, the star-forming regions visible at the bifurcation are within the disc, and their distribution could be associated with a projection effect.

The distribution of the star formation rate in the galaxy shows higher values towards the northern spiral arm compared to the southern one, the extinction is also higher in that direction, while in the central regions, the values of SFR are lower. Making extinction and ``mixing sequence" corrections to the \halpha\ luminosity, we derive a galaxy-wide SFR of 3 \usfr and, to compare to other studies, for the central 3.4 kpc$^{2}$ region we derive a SFR of 0.3 \usfr. These values place NGC~4258 within the typical values found for star formation in grand-design, spiral galaxies without AGN ranging between 1.0 and 13 \usfr\ \citep{Grosbol2012}. The propagated errors in flux, extinction, and star formation fraction used to estimate the SFR lead to a final error of approximately 20\%. The major contribution to the error comes from the extinction correction. 

The comparison between the SFR derived exclusively from spaxels within the star formation zone on the BPT diagrams and the SFR from all spaxels, after being corrected by $f_{SF}$, reveals that 85\% of the galaxy-wide SFR originates from the former. This suggests that the majority of star formation activity is concentrated within the spiral arms. In the central 3.4 kpc$^{2}$, spaxels classified within the star formation region contribute 54\% to the total SFR in this zone. It is essential to emphasize that a significant portion of these spaxels in the star formation region in the central zone is associated with the \halpha\ filament.

Given the elevated extinction values observed in the northern part of the galaxy, we also estimated the SFR by considering an average extinction value (\Av=0.7), consistent with most of the spaxels classified in the star-forming region. In this scenario, the galaxy-wide SFR in units of \usfr\ is 1.5 and 1.3 for all spaxels and only those in the star-forming region, respectively. For the central region, the corresponding values are 0.1 and 0.07, which closely align with the value reported by \cite{Ogle2014}. These differences in the SFR could be indicative of a reservoir of material available for forming new stars, particularly in the northern part of the galaxy. 

Finally, since this method may underestimate the SRF, especially in the outermost zones where weak lines (e.g. \hbeta) may not be detected with sufficient S/N, assuming all Halfha emission detected with S/N>3, an upper limit in the extinction, (\Av=1.52) and excluding the jet region, we derive an upper limit on the SFR of \mbox{$\sim7$ \usfr}.

Utilizing observations in redder wavelengths could be beneficial in enhancing our understanding of these high extinction values, especially in regions far from the nucleus, within the deprojected distance range of 5 to 10 kpc which looks to be dominated by star-forming clusters and minimal interaction with the jet and radio structures.

\subsubsection{Radial and Azimuthal variations in line ratios}

\begin{figure}
\centering
	\includegraphics[width=\columnwidth]{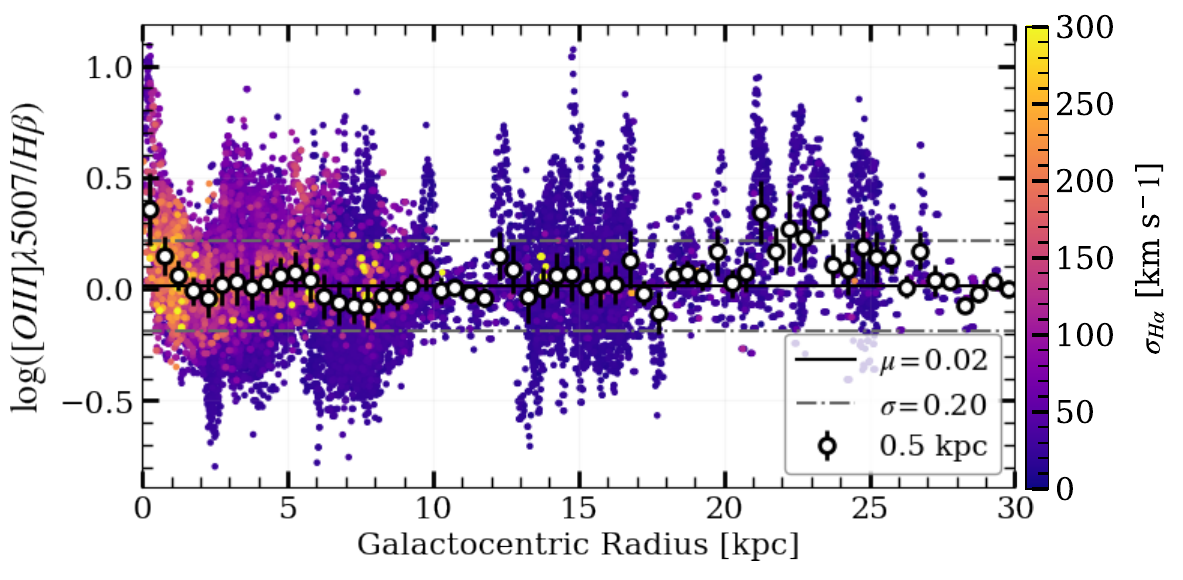}
	\includegraphics[width=\columnwidth]{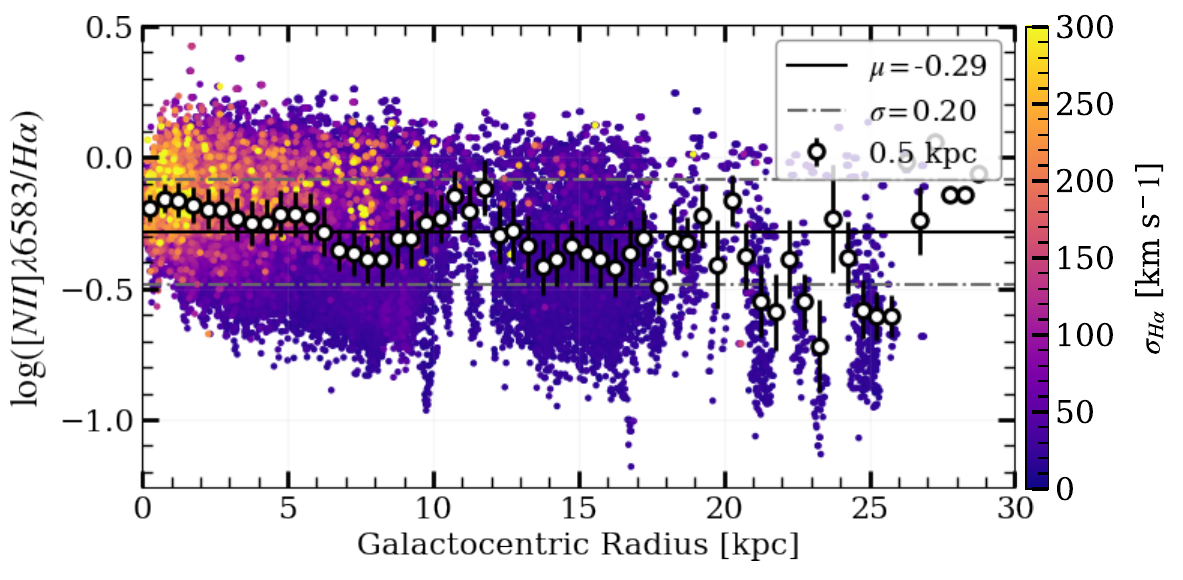}
	\includegraphics[width=\columnwidth]{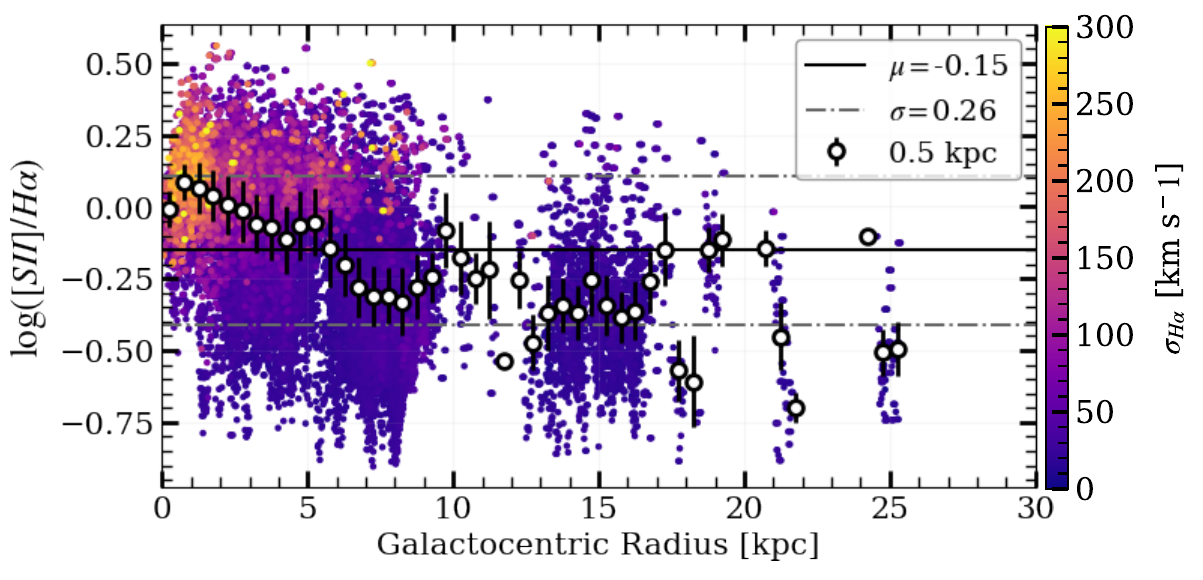}
	 \caption[]{Diagnostic line ratios in logarithmic scale as a function of deprojected galactocentric radius. The colour bar corresponds to the velocity dispersion measured in the \halpha\ line. The horizontal lines correspond to the average and standard deviation, the values are given in the inset label of the plot. Horizontal lines show the mean and standard deviation of the points whose values are given in the inset.}
    \label{_diag_ratio}
\end{figure}

Figure \ref{_diag_ratio} illustrates the variation of the line ratios used in the BPT diagrams relative to the deprojected galactocentric radius and the velocity dispersion. The line ratios \nii/\halpha\ and \sii/\halpha\ exhibit a smooth decrease in the first 5 kpc, while \oiii/\hbeta\ drops sharply in the first 3 kpc and then maintains an approximately constant level. This trend is validated using bins of sizes 1 and 0.5 kpc, represented by the green and cyan circles, respectively, as shown in the figure. The most substantial fluctuations occur at scales of a few parsecs. Similar findings were reported by \cite{DAgostino2018}, demonstrating how variations in spaxel binning alone can induce changes in properties such as the fraction of spaxels contributing to the luminosity from ionization by star-forming regions, particularly in the presence of an AGN, at different scales.

On the other hand, the velocity dispersion in the central zone \mbox{($<3$ kpc)} shows values around 200 \uvel. In this zone, the velocity dispersions follow a trend similar to that found by \cite{Cecil1992}, especially in the part of the south-east jet and the centre of the galaxy. The kinematics along the jet exhibit a large difference in velocity dispersions compared to the classical spiral arms of NGC~4258, mainly composed of HIIRs. In the external zones ($>5$ kpc), the velocity dispersion decreases and the kinematics are dominated by the HIIRs, with values around 30-50~\uvel.

In addition to the flux ratio to characterize the ionizing sources, we investigated the flux ratio between \sii\lin6717 and \sii\lin6731, which is sensitive to the electron density (\ne). We convert the ratio of the \sii doublet to \ne\ adopting an electron temperature of \mbox{10000 K} following Eq.6 of \cite{Sanders2016}. This prescription provides estimate limits of \ne\ within the \mbox{$1-10^5$ cm$^{-3}$} range corresponding to 1.4484 and 0.4375, respectively.  

\begin{figure}
\centering
	\includegraphics[width=\columnwidth]{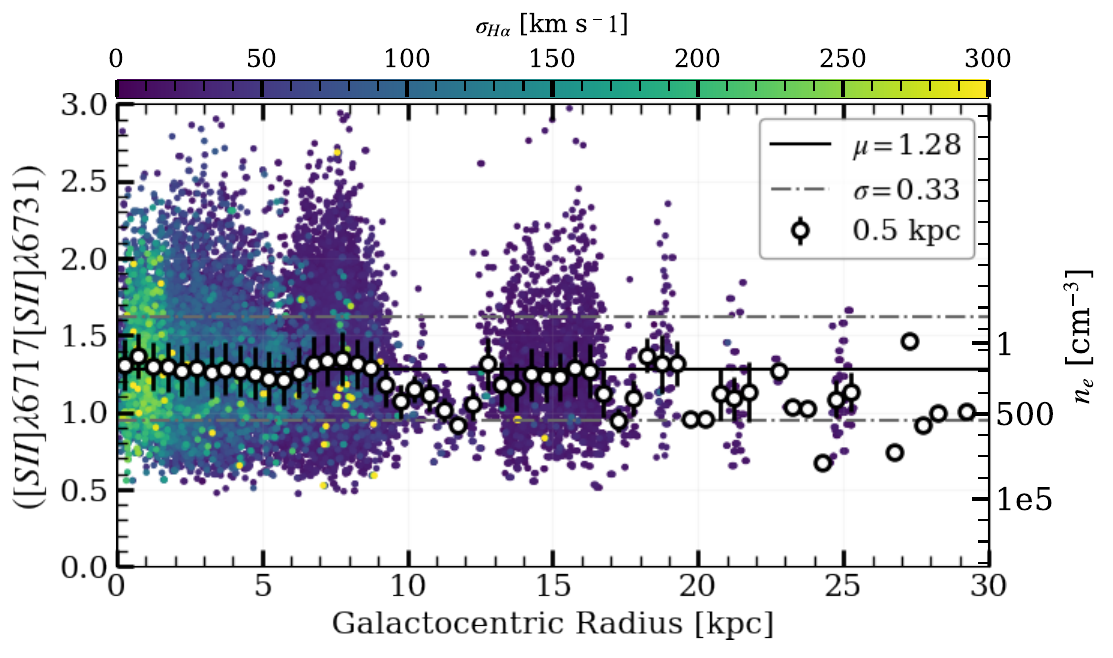}
	 \caption[]{\sii\lin6717/\sii\lin6731 ratio with the corresponding \ne\ scale given in the right axis, following the conversion of \protect\cite{Sanders2016}. \textbf{The colour bar corresponds to the velocity dispersion in the \halpha\ line.}}
    \label{_ne_ratio}
\end{figure}

Figure \ref{_ne_ratio} displays the radial variation of \sii\lin6717/\sii\lin6731 ratio and the corresponding \ne\ scale in the right axis the colour bar corresponds to the velocity dispersion. For individual spaxels, we found a larger variation with densities ranging between the limits of \cite{Sanders2016} prescription. It is necessary to take caution because although we exclude spaxels with S/N<3, a larger S/N may be required when measuring semistrong line ratios due to the propagation of the uncertainties of both lines in the ratio.

On the other hand, with integrated values at scales of 0.5 kpc we estimated an average \ne\ corresponding to \mbox{$\sim200$ cm$^{-3}$} and almost constant in the first 10 kpc.  If we only use spaxel classified as photoionized by stars this value decreases to \mbox{$\sim60$ cm$^{-3}$}. This result is consistent with spatially resolved studies of 9 nearby AGNs as reported by \cite{Mingozzi2019}. 

\begin{figure}
	\includegraphics[width=\columnwidth]{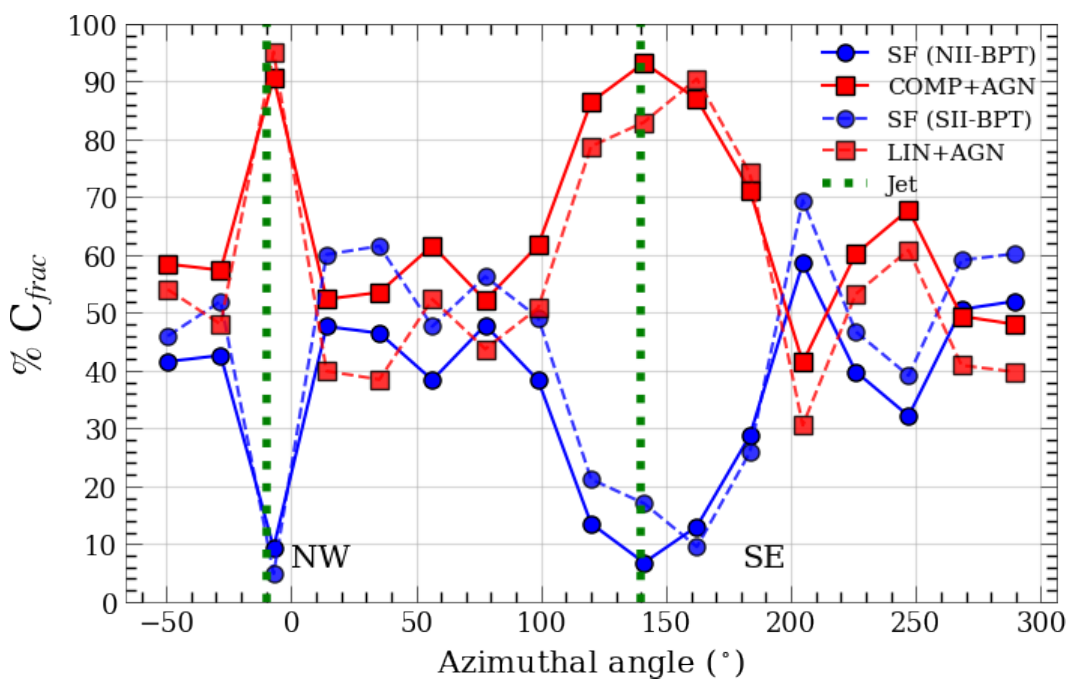}
	 \caption[]{
  Covering fraction of the pixels classified as AGN+composite (red) and star-forming (blue) according to BPT diagrams (BPT-NII with continuous lines, BPT-SII with dashed lines) as a function of the projected azimuthal angle.
  The green dotted line represents the peak of the AGN+composite points corresponding to the location of the Jet. We use \mbox{azimuth=0\degree} (North-West) and 180\degree (South-West) along the major axis.}
    \label{diag_ratio_az}
\end{figure}

We also explore the distribution properties of azimuthally projected pixels in bins of 15\degree. For this, we use the diagnostic diagrams and compare the fraction of spaxels classified according to their ionization source over the total spaxels detected in that bin \mbox{($C_{frac}=Npix_{BPT(class)}/Npix_{tot}$)}. This gives us a complementary idea about the geometry of the ionized gas and its properties for interpreting observations at high redshift. Figure \ref{diag_ratio_az} displays the azimuthal distribution of the percent covering fraction. 

This plot shows its variation as a function of angle in polar coordinates. In the figure, we can see two dominant regions in the spaxel fraction corresponding to Seyferts and composite objects are evident. The two peaks of the Syfers+composite fraction align with the Position angle (P.A.) of the galaxy (in pink) and are associated with the jet to the north-west and south-east, exhibiting a minimum of star formation. This geometry has been interpreted in terms of outflows produced in a biconical path by the nucleus. It's worth noting that the first emission peak, situated around \mbox{-10\degree}, is narrower and corresponds to the north-west region. The second peak, located around \mbox{140\degree}, corresponds to the south-east, and this peak shows an aperture angle larger than that of the north-west. This asymmetric could be attributed to the non-uniform illumination of the interstellar medium at the bifurcation of the jet in the south-east region by radiation from the ANG or projection effects.

\cite{Cox1996} suggested that ionization in the ``anomalous spiral arms" of NGC~4258 is produced via the bar shock and is unrelated to a jet from the black hole. On the contrary, \cite{Cecil2000} suggest that they are linked dynamically by the precession of the central engine and they are probably a fossil record of changing jet activity in NGC~4258, which has recently moved out of the plane of the galaxy. Given the presence of spaxels ionized by Syfers in the part of the jet, our results are in agreement with the latter study.

The fraction of spaxels classified as star formation decreases rapidly in the zone of the jets, reaching minimum values, while the maximums are situated between two ranges: \mbox{-30\degree\ to 20\degree} and \mbox{100\degree\ to 190\degree}, corresponding to the direction of the spiral arms. Some intermediate peaks occur adjacent to the jet in the north and south directions and for the \halpha-filament. This behaviour is also observed for the velocity dispersion measured in \halpha, with the largest values located along the jet and the narrowest components originating from spaxels in star-forming regions. The jet seems to be deflected at a ridge of highly excited gas, suggesting an opening angle to the nuclear radiation of up to \mbox{90\degree} in the south-east and 50\degree\ in the north-west. These values are consistent with those deduced for ionization cones in other active galaxies \citep[e.g.][]{Wilson1994,Falcke1996}.

\subsubsection{Spatially and kinematically resolved BPT}

Spatially and kinematically resolved BPT diagrams allow us to explore the dominant contribution to ionization in each spaxel in the galaxy and their connection with the velocity dispersion. A similar approach has been used by other authors \cite[e.g.,][]{Westmoquette2012, McElroy2015, Karouzos2016} to constrain the properties of galaxies. In \cite{Dettmar1990}, large velocity anomalies and very large velocity dispersion of the emission lines were reported on a small scale (<10\arcsec, in NGC~4258) with an asymmetric brightness distribution. They argued that the change in ionization conditions with distance from the nucleus suggests a physical relation to the jet emanating from the active nucleus. \cite{Cecil1992}, using Fabry-Perot observations in \halpha, showed that the velocity dispersion along the south-east jet averages $\sim80$ \uvel with a ratio of \nii/\halpha>0.5 being considerably above that of the HIIRs ($\sim40$ \uvel). Here, we combine the spectral resolution in \halpha\ and the spatially resolved BPT diagrams to have a general view of its connection. We found similar results with the nuclear emission lines being asymmetric in the north-west south-east direction, broad velocity dispersion and the ionization within this structure changes to lower excitation with increasing distance from the nucleus.

\begin{figure*}
\centering
	\includegraphics[scale=0.45]{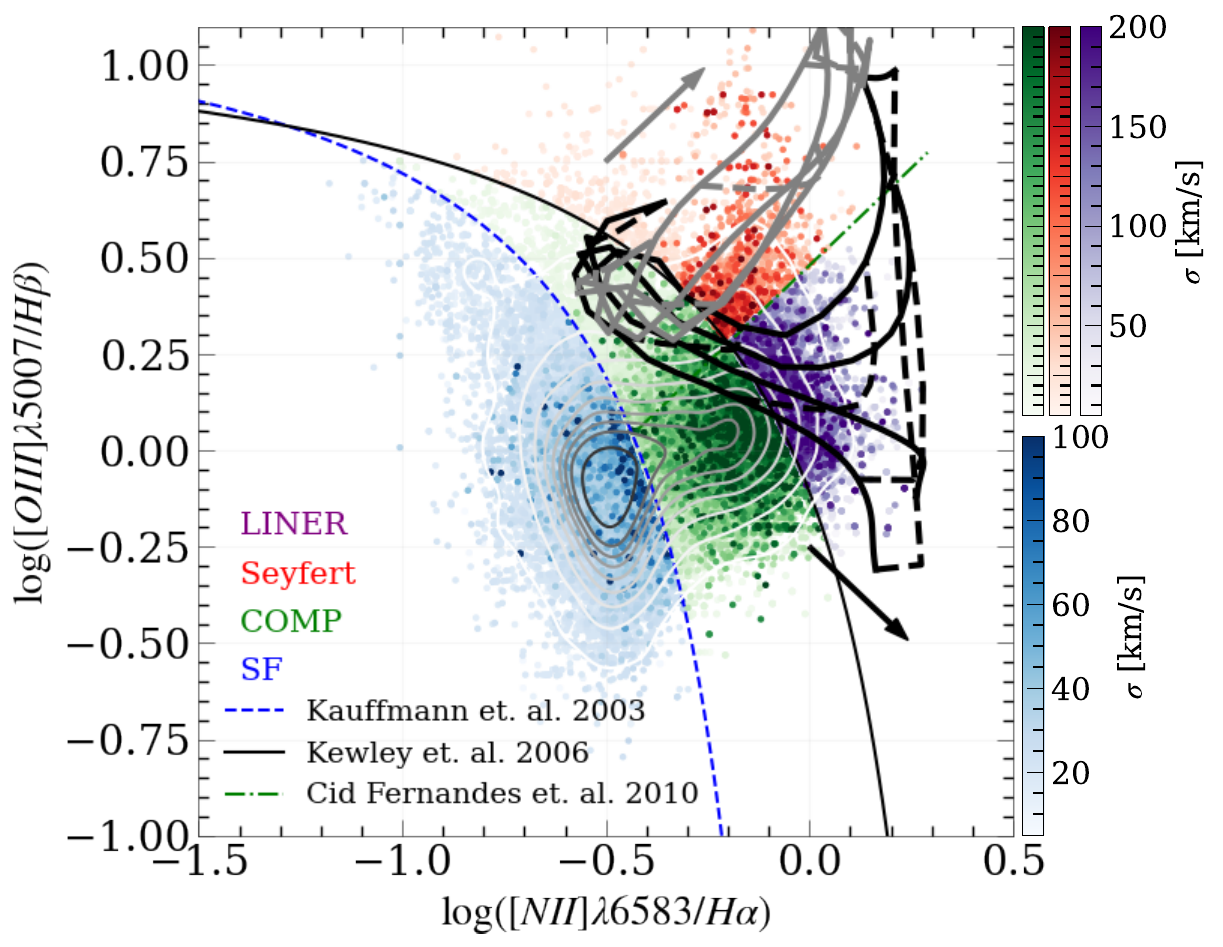}\includegraphics[scale=0.45]{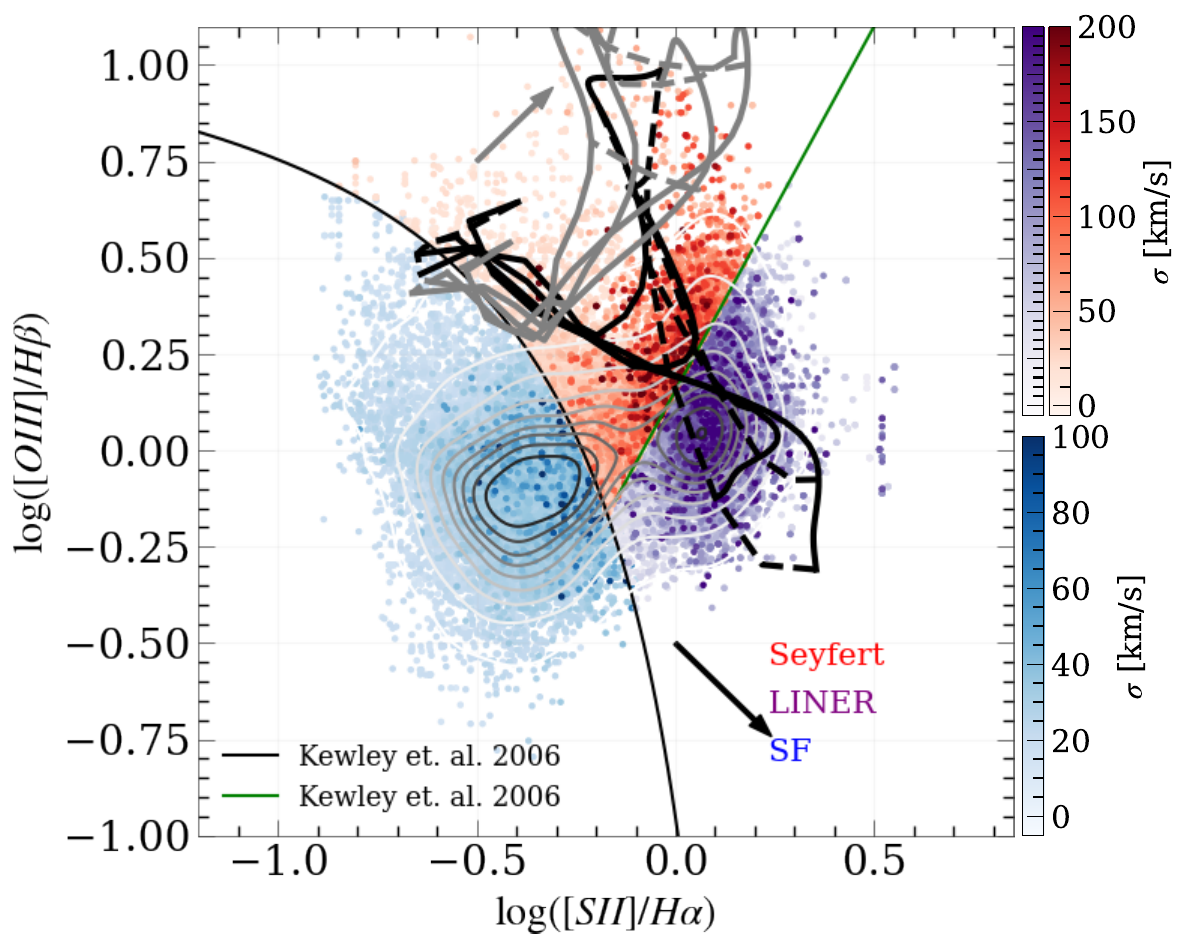}
	\includegraphics[width=\columnwidth]{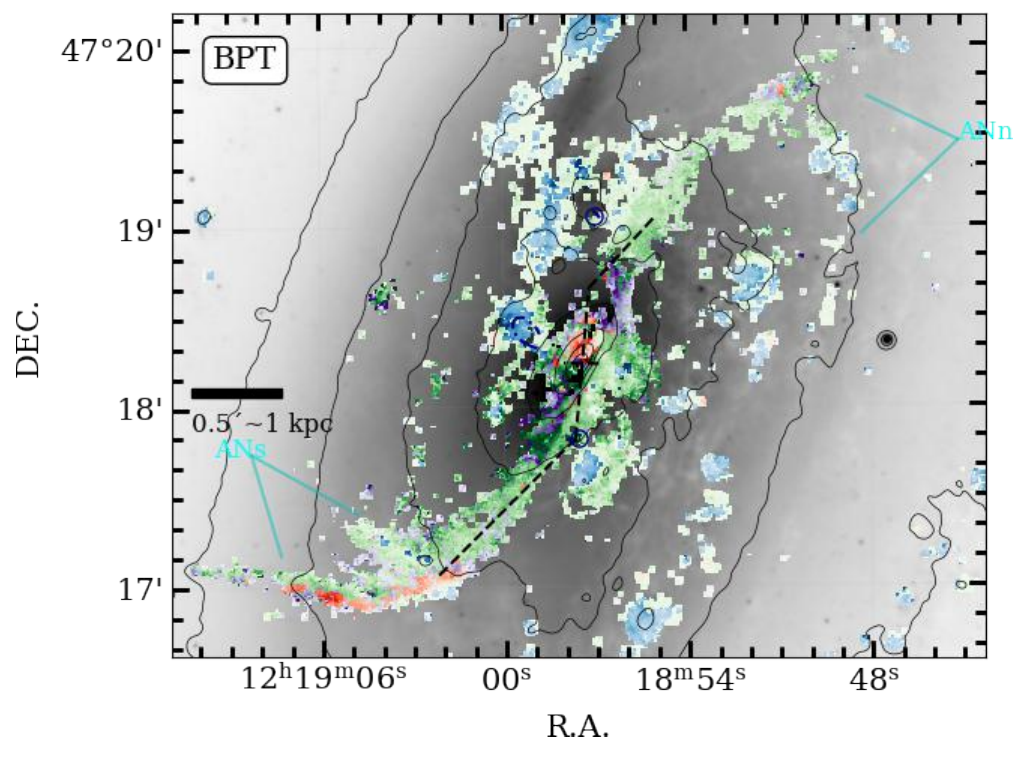}\includegraphics[width=\columnwidth]{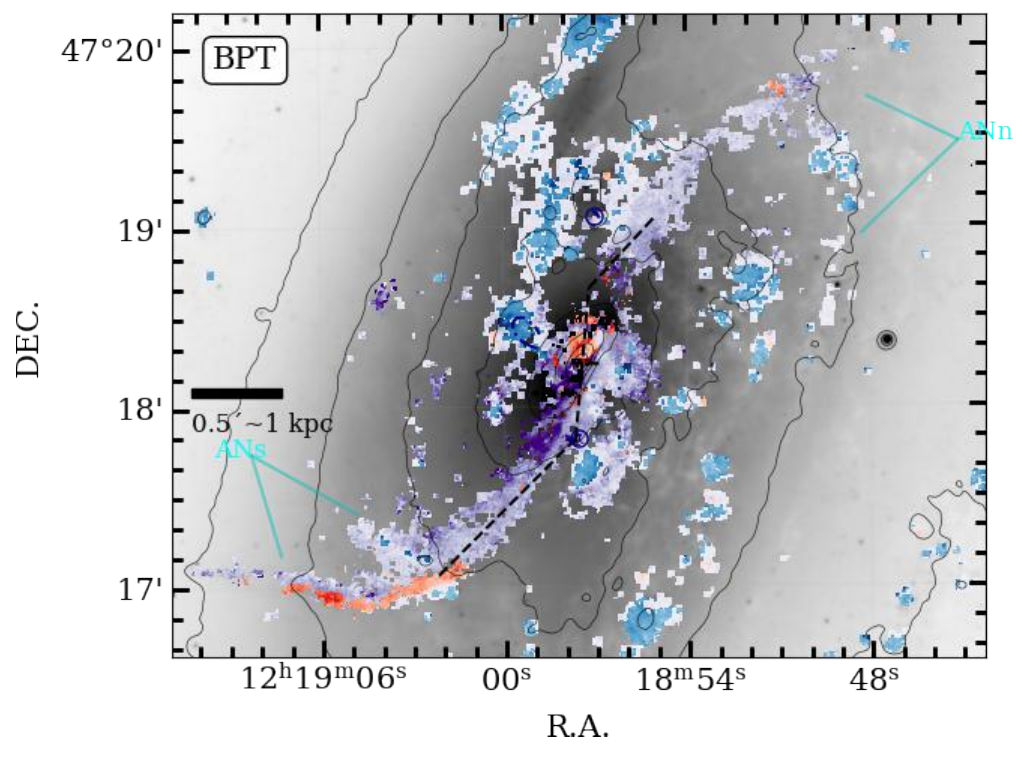}
    \caption{The top panels display the NII- and SII-BPT diagrams, while the bottom panels present maps showcasing spaxel locations in the centre of the galaxy, classified according to their ionizing source and colour-coded based on the \halpha\ velocity dispersion in \uvel, the colours are the same as in the BPT diagrams, with darker shades indicating higher $\sigma$(\halpha). The grey and black lines in the diagrams represent models of photoionization by fast shocks from \protect\cite{Allen2008}. The black lines depict photoionization with only front shocks, and the grey lines consider pre-ionization by a precursor. Solid lines represent shock winds of 200, 400, 500, and 1000 \uvel, while dashed model curves represent magnetic field intensities of 0.0001, 1.0, 5.0, and 10. Grey and black arrows indicate the direction of increasing wind velocity in each model.}
    \label{bpt_diagrams_velocity} 
\end{figure*}

Figure \ref{bpt_diagrams_velocity} displays diagnostic diagrams (NII-BPT, SII-BPT) with the velocity dispersion in \halpha\, along with its distribution in the galaxy presented in the last row of the figure. The central region stands out as being dominated by ionization from Seyferts and LINERS, characterized by a larger velocity dispersion compared to the HIIRs. In the outer part of the south-east jet, we observe the presence of ionization by Seyferts, with velocities ranging between 80 and \mbox{160 \uvel}. Similar values are identified in the central region, previously recognized as the trailing strands in \cite{Cecil1995}. According to their findings, this region should be ionized by shocks with velocities around 90 \uvel. Our observations align with this interpretation and present a complementary image—the central region, also ionized by nuclear emission, exhibits a velocity dispersion around \mbox{150 \uvel}. This value corresponds closely to the velocity dispersion of stars derived from the Ca triplet in the nuclear region, as presented by \cite{Cecil1995} and \cite{Terlevich1990} with values of 171 and \mbox{141 \uvel}, respectively.

In Figure \ref{bpt_diagrams_velocity}, we have overlaid model predictions for photoionization by fast shocks from \cite{Allen2008}. The black lines depict photoionization with only front shocks, while the grey lines consider pre-ionization by a precursor. Solid lines represent shock velocities of 200, 300, 500, and 1000 \uvel, and dashed model curves represent magnetic field intensities of 0.0001, 1.0, 5.0, and 10, assuming solar metallicity and a pre-shock density of 1 cm$^{-3}$. Upon comparing the model grids to our data in the BPT diagrams, we observe that the models accurately reproduce galaxy LINER points without the need for precursors.

After a visual inspection of the integrated profiles of the different regions, we noticed potential broad components, so we fitted multiple Gaussians to the integrated spectra and compared them with the fits obtained using a single Gaussian function using the Bayesian information criterion (BIC). We use a $\Delta$BIC>10 in cases where the fit improves significantly, and two components prevail over a single one. These fits were made using \texttt{ORCS} and its \texttt{Bayesian fitting implementation}.

The integrated spectra of the \halpha-filaments and the shock structure in the north exhibit the best fit with a single Gaussian component, displaying negative $\Delta$BIC values and velocity dispersions of $\sim$35 and $\sim$70 \uvel, respectively. These regions align with the diagnostic diagrams in the star formation and composite objects zone. The N-shock lacks a clear detection, making its velocity dispersion only a lower limit. The S-shock is visible in the \halpha\ contour at a lower spatial resolution of the HST. We measured a velocity dispersion for a broad component of $\sim$120 \uvel\ and one narrow component of $\sim$50 \uvel, and we found that this region falls into the zone of composite objects in the BPT diagram. 

Notably, the loop structure is more favourably fitted with a two-component model over a single Gaussian component, as indicated by a $\Delta$BIC$\sim$50. The fit reveals a narrow component with a velocity dispersion of $\sim$42 \uvel\ and a velocity difference of $\sim$75 \uvel\ towards the red for the wide component of the model, possessing a velocity dispersion of $\sim$140 \uvel.

\begin{figure*}
	\includegraphics[scale=0.45]{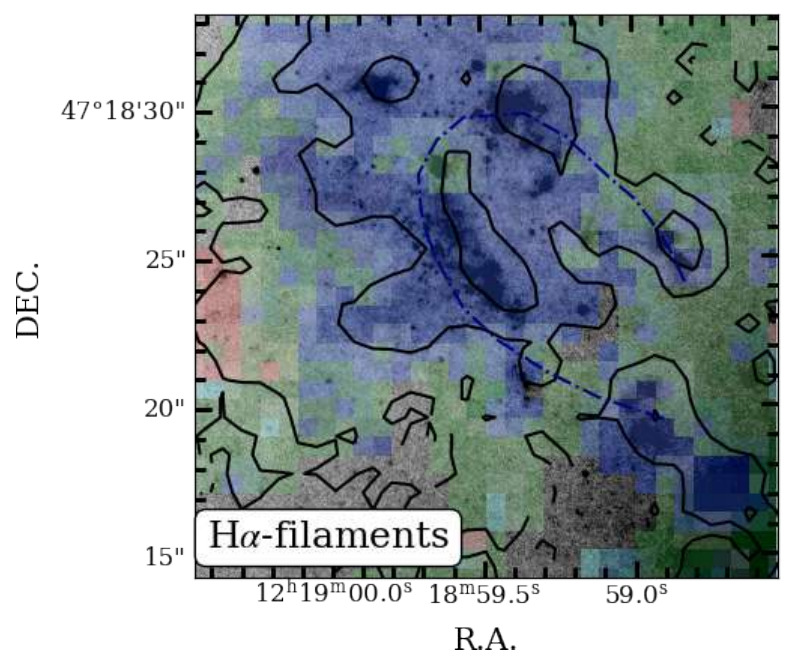}\includegraphics[scale=0.45]{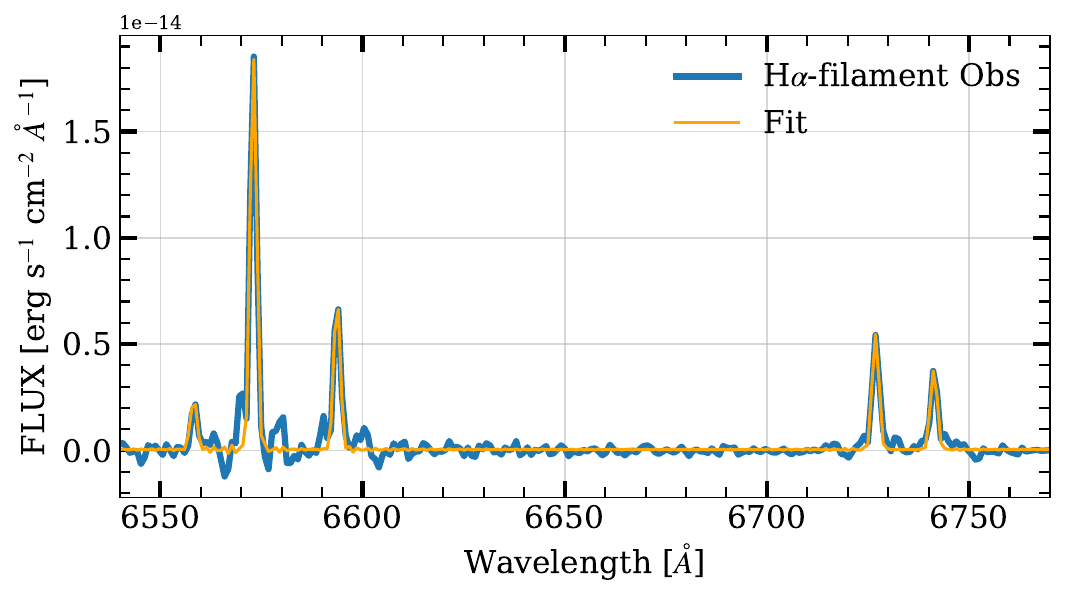}
	\includegraphics[scale=0.45]{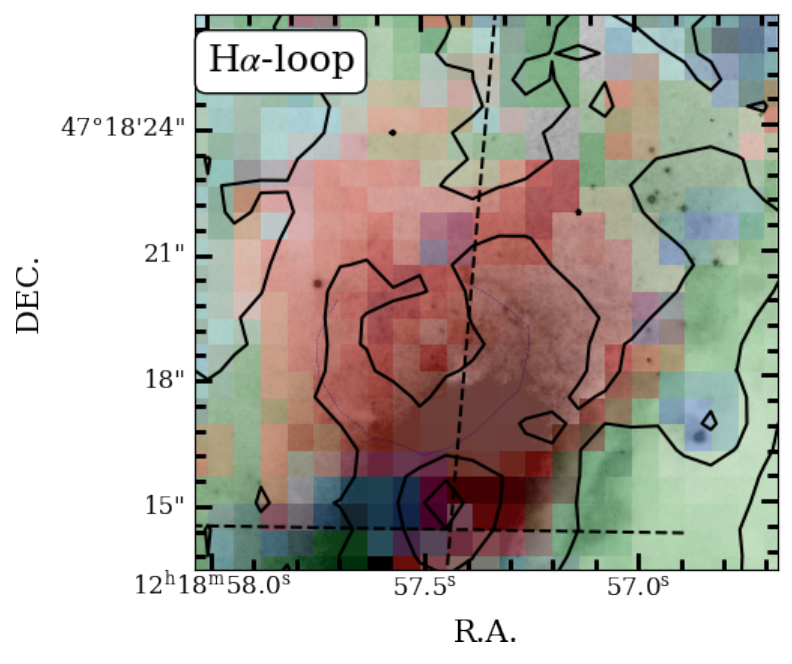}\includegraphics[scale=0.45]{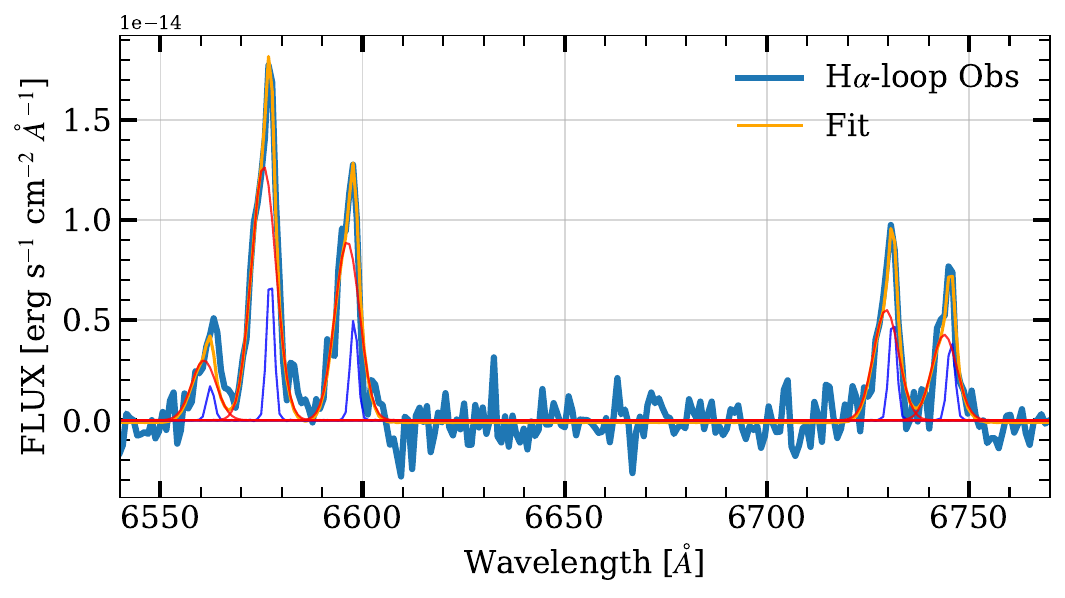}
	\includegraphics[scale=0.45]{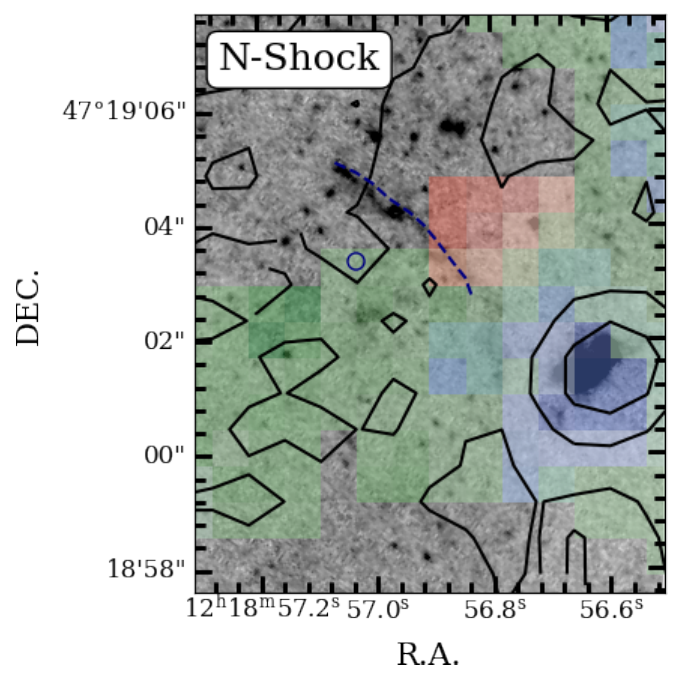}\includegraphics[scale=0.45]{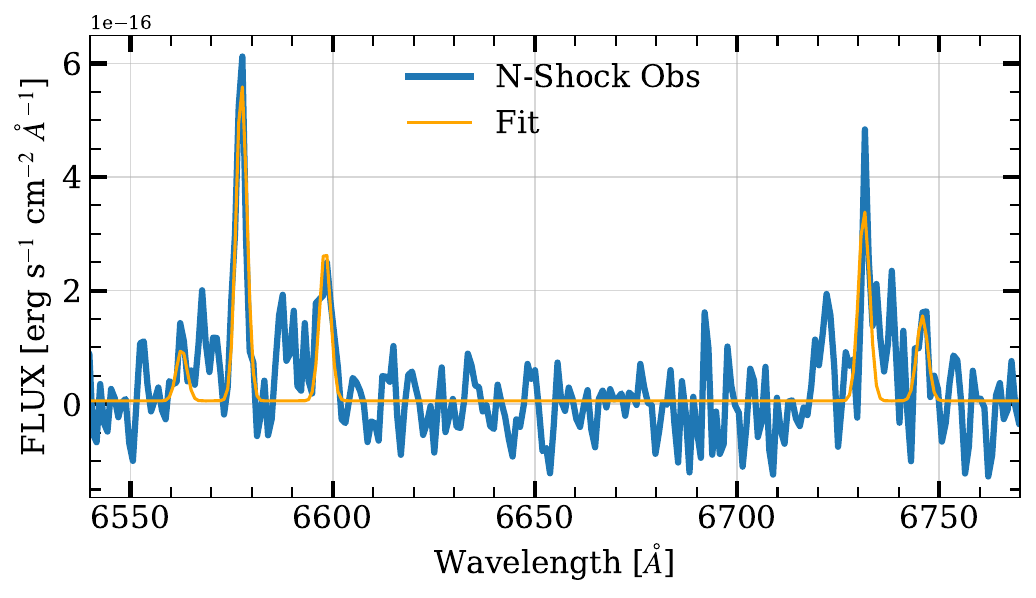}
\includegraphics[scale=0.45]{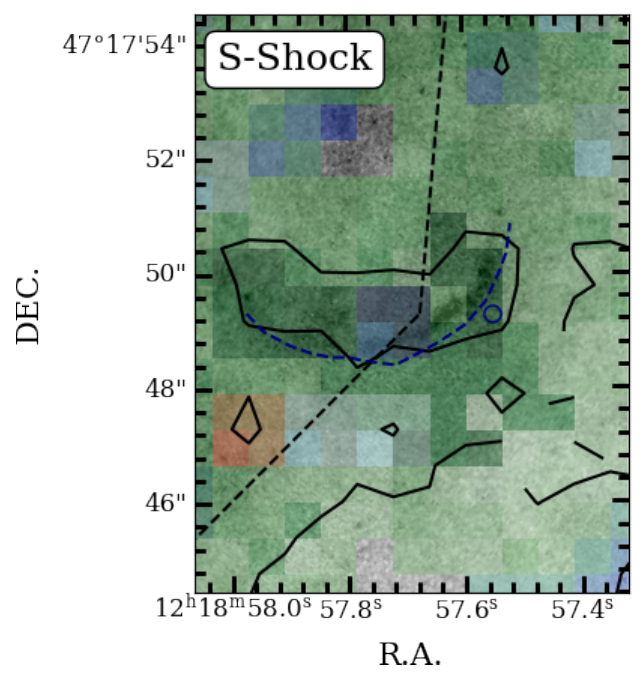}\includegraphics[scale=0.45]{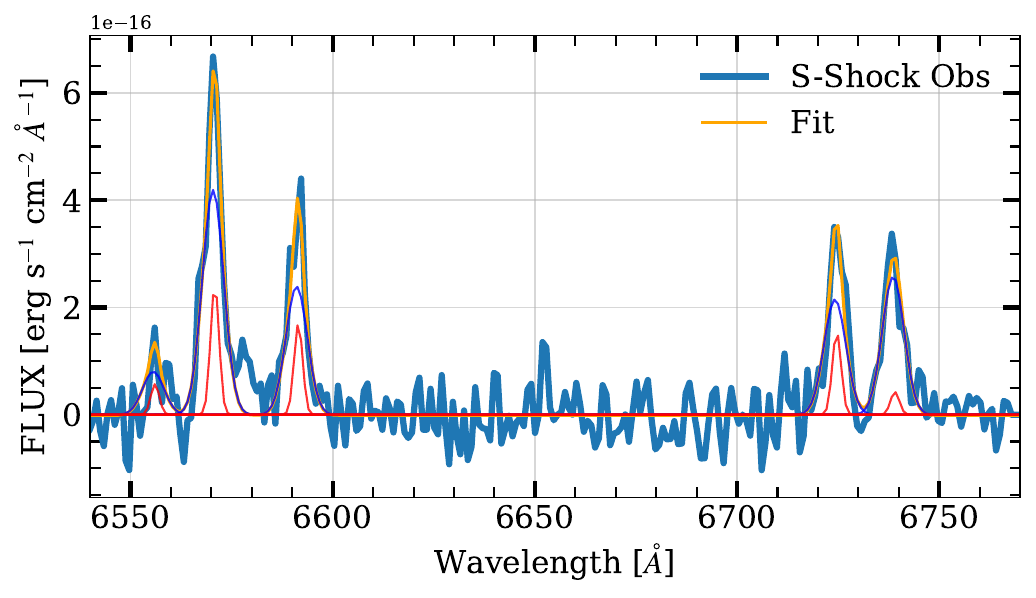}
    \caption{Zoom-in on the main features previously reported in the literature, as described in the main text and observed with SITELLE from top to bottom: \halpha-filaments, \halpha-loop, north and south arc shocks. The colour represents the NII-BPT ionizing source, red (AGN), blue (star-forming) and green (composite objects), the dashed lines depict features visible in HST observations, and the contours correspond to the \halpha\ emission line map derived from our measurements. The colours are the same as in Figure \ref{bpt_diagrams_velocity}, but the intensity has been decreased to visualize the background image and the corresponding large-scale structure. The right panels display the integrated spectrum along spaxels intercepted by the structures (dashed lines) with their corresponding fit. The \halpha-filament and N-shock are fitted with a single Gaussian component, whereas the \halpha-loop and the S-shock are best represented with two Gaussian components shown in red and blue, respectively. The total fit is plotted in yellow. }
    \label{bpt_diagrams_velocity_regions} 
\end{figure*}

In Figure \ref{bpt_diagrams_velocity_regions}, we focus on specific regions previously highlighted in the literature, such as the \halpha-loop, \halpha-filaments, and arc shocks \citep[referred to as bow-shocks by][]{Appleton2018} to the north and south of the centre. The \halpha-filaments are consistent with the photoionization by massive stars, exhibiting narrow components in the emission lines. The shock structures fall into the zone of composite objects in the BPT diagrams. 

The kinematic information reveals a notable increase in velocity dispersion, reaching approximately $\sim$100 \uvel\ for the \halpha-loop, suggesting a potential double component, as evidenced by the profile. This loop is also close to the centre therefore, the double peak could be related to the rotation integrated across the loop. The loop shows ionization by AGN activity, as indicated by the NII-BPT and SII-BPT diagrams and is located near two blue clumps corresponding to star-forming regions. The observation of the \halpha\ loop near the nuclear region coincides with the most intense radio emission. \cite{Falcke1996} proposed that this structure represents a double helix. This implies that the strands of emission lines are generated at the surface of the radio jet, possibly through an interaction between the jet and the ISM.

\subsubsection{Emission fluxes and Line Profiles along the P.A.}

\begin{figure}
\centering
	\includegraphics[width=0.5\textwidth]{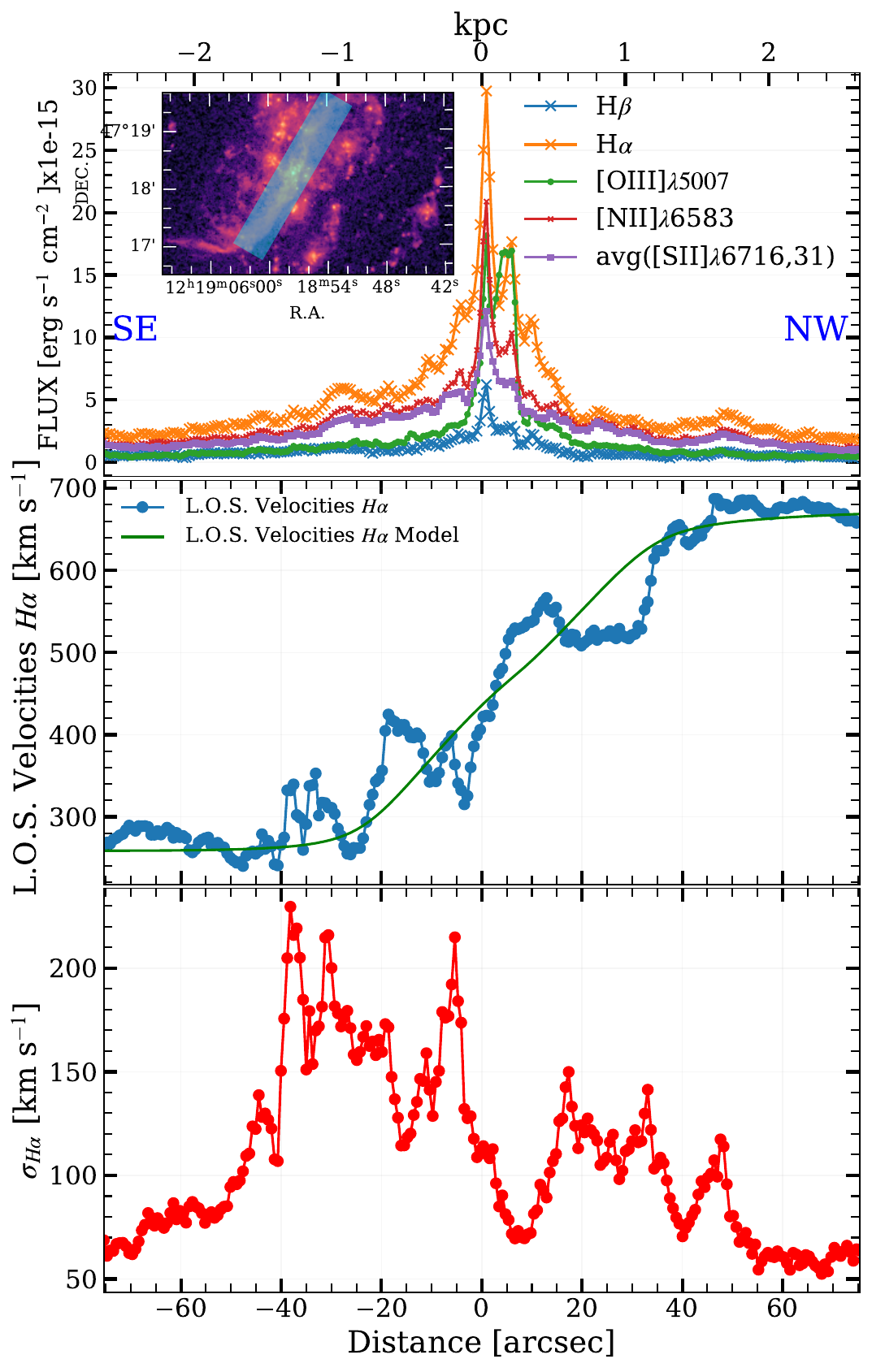}
    \caption{Emission line fluxes and velocities derived along the central region using a P.A.=150\degree\ and width of 12.5\arcsec $\sim$ 0.45 kpc. For the doublet of \sii\, we plotted the average value. The inset plot shows the central region in \halpha\ with the area used to extract the profiles.}
    \label{jet_profiles_variation_lines} 
\end{figure}

We explore the variation of fluxes and kinematics along the P.A. of the galaxy, not only to the south-east jet but also to the north-west, extending up to 80\arcsec (corresponding to 3 kpc). We present the emission-line fluxes and velocities derived from fits with a single sinc-Gaussian function in Figure \ref{jet_profiles_variation_lines}. The top panel displays the summed line fluxes of \hbeta, \oiii\lin5007, \halpha, \nii\lin6583, and \sii\lin\lin6717,31. The inset shows the region across which the 1D profiles have been extracted, using the P.A. of the galaxy. The variation of the fluxes is similar for \hbeta, \halpha, and \sii\lin\lin6717,31, with some peaks in the \halpha\ emission to the south-east part of the jet, corresponding to star-forming regions. The maximum fluxes coincide in the centre, while \oiii\lin5007 and \nii\lin6583 are enhanced to the north-west in the region occupied by the loop in the central 100 pc. The velocity dispersion is high to the south-east with more variation and decreases in the north-west direction, these results are consistent with those found by \cite{Cecil1995}.

The line-of-sight velocity varies between 370 and 760 \uvel\ from the southeast to the northwest. We obtain a velocity of \mbox{$441\pm7$ \uvel} using the location of the continuum peak in the centre, which is in agreement with the 450 \uvel\ reported by \cite{Drehmer2015}, determined from the stellar kinematics using the peak of the continuum in the central region, and the $445.6\pm0.2$ and \mbox{443 \uvel}\ reported by \cite{Gonzalez-Lopezlira2019} using HI data and globular clusters, respectively. Making a visual inspection of the HI velocity map \cite[][HALOGAS Survey]{Heald2011}, we note that the centre has a larger systemic velocity than the outskirts. But this will be discussed in a future paper. 

On the other hand, when we include the entire FoV in the determination of the systemic velocity after masking spaxels with \mbox{S/N$<$5}, we obtain \mbox{$473\pm4$ \uvel}. This value is in agreement with the values reported by \cite{Herrnstein2005, Argon2007}, and \cite{Humphreys2013} using Very Long Baseline Interferometry observations, corresponding to approximately 470, 472, and \mbox{474 \uvel}, respectively. \cite{Cecil1992} reported a value of \mbox{$472\pm4$ \uvel} for \halpha\ Fabry–Perot  observations. This suggests that the jet may have some effect on the movements and rotation of the gas near the centre, indicating that the gas and stars are not entirely coupled in the velocity field. 

Modelling the velocity field, we found a systemic velocity of \mbox{$472\pm1$ \uvel} and we distinguished that the rotation of the jet-like arm is fairly consistent with that of the other arm, which is surprising if the arm is indeed a jet. The curvature at the end of the 'jet' is compatible with a trailing spiral arm. However, this arm appears as a bar of ionized gas, and the ionization parameters suggest ionization by an AGN, indicating it could indeed be a jet. This implies a radial component that may not be observed due to the orientation along the major axis. The kinematic model shows significant rotational deviations in two regions along the minor axis, resembling filaments and bow shocks. The gas could be either accreting towards the centre or outflowing out of the plane of the disk. The middle panel in Figure \ref{jet_profiles_variation_lines} shows the L.O.S velocities along the P.A. and the model in green.  Outside 10 kpc the galaxy seems to behave like a normal spiral galaxy, while closer to the nucleus is disturbed. 

\begin{figure*}
\centering
	\includegraphics[width=\textwidth]{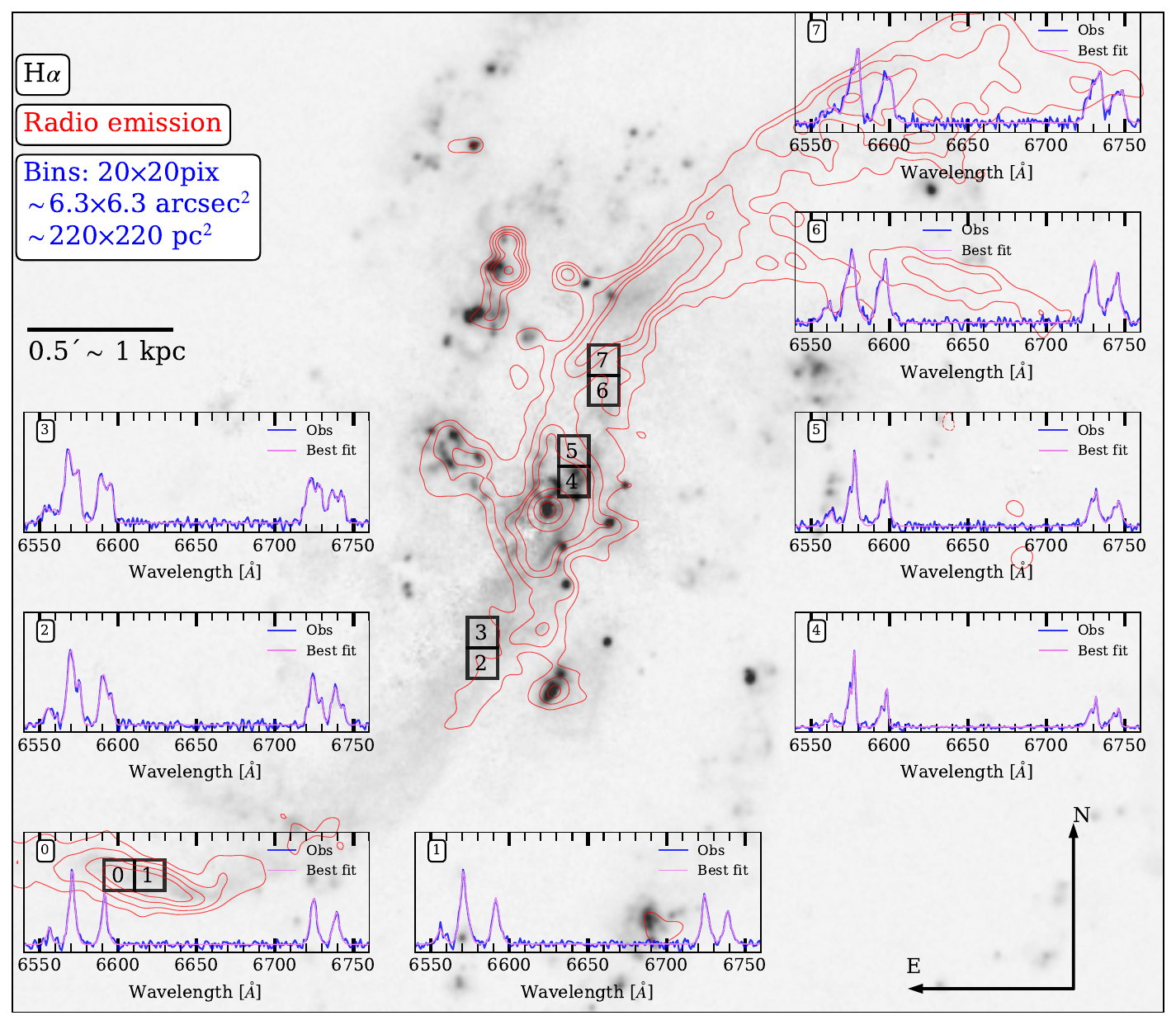}
    \caption{Both single and double profiles have been observed in \halpha\ emission within the central region of NGC~4258. Individual spectra are extracted from boxes measuring $20\times20$ pixels$^{2}$, corresponding to $220\times220$ pc$^2$. The plotted wavelength range shows the \halpha\, \nii\lin\lin6548,83 and \sii\lin\lin6717,31 lines. The criteria for selecting a better fit with two components for each emission line is the same as the previously described method, relying on the $\Delta BIC$ value. The more intricate profiles are concentrated in the centre, where distinctly two resolved components are observable. The black squares show the location of some bins as examples where the profiles show a double component, especially along the jet. The number of the integrated spectrum corresponds to the number shown in each square. We present the \halpha\ image from SITELLE in the background and radio contours in red.}
    \label{jet_profiles} 
\end{figure*}

In addition to integrated regions, we explore the kinematics and profiles along the jet in the central region of NGC~4258 using boxes in bins of $20\times20$ spaxels. Figure \ref{jet_profiles} presents \halpha\ emission profiles along the jet, each covering an area of $6.3\times6.3$ arcsec$^{2}$ per bin, equivalent to a physical scale of $220\times220$ pc$^2$. The fits in the \halpha-filament structure are generally consistent with a single component, with only some bins exhibiting two components having a velocity dispersion between 40-50 \uvel and separations below 50 \uvel, indicative of multiple star-forming regions. Particularly towards the south-east of the jet, profiles featuring two Gaussian components become more prevalent, emphasizing the complexity of the kinematics in this region.

Most profiles are fitted with two components separated by around 200 \uvel\ within the central 40 \arcsec. However, as we move farther than 50\arcsec\ from the nucleus, the separation of components decreases to 90 \uvel\ on average, displaying a smooth transition from the centre to the external part of the galaxy in the south-east direction of the jet. In contrast, the components to the north-west exhibit an average separation velocity of around 100 \uvel. The higher velocity dispersion values and the presence of multiple components in this direction suggest a potential outflowing component from the AGN, with the separation possibly attributed to braided jets produced by nuclear activity. These findings align with previous observations, including a long-slit spectrum along the minor axis by \cite{Rubin1990}, narrow-band Fabry-Perot observations by \cite{Cecil1992}, and images from \cite{Falcke1996} and \cite{Appleton2018}.

 A deeper detailed kinematic analysis is outside the scope of the current work, more details will be presented in a future study to explain the curvature at the end of the jet by testing the hypotheses of the jet being in or out of the disk.

\section{Discussion}\label{Discu}

The primary objective of this study is to investigate the spatially resolved distribution of ionized gas at multiple scales and radial distances, covering a broader area than previous spectroscopic campaigns. We employed new integral field spectroscopy (IFS) from the SITELLE instrument at the Canada–France–Hawaii Telescope (CFHT). This work is part of the ongoing SIGNALS survey, in which NGC~4258 stands out as the only galaxy featuring its distinctive ``anomalous" structure. Our findings provide significant insights into the dynamics of AGN feedback mechanisms and their impact on the host galaxy, NGC 4258. These results directly address our initial questions about how AGN-driven outflows interact with the interstellar medium (ISM) and help elucidate the nature of the ``anomalous arms." 

The spatially and kinematically resolved maps of ionized gas have revealed how the low-luminosity AGN (LLAGN) in NGC 4258 drives energy into its surroundings. These findings address our initial questions by showing that AGN-driven outflows extend beyond the nuclear region, influencing large parts of the galaxy up to several kiloparsecs. The outflows interact dynamically with the ISM, as evidenced by changes in ionization states, gas velocity dispersions, and the structure of outflow channels. This interaction is particularly evident in the ionization maps where outflows from the central AGN energize the gas far into the spiral arms, modifying the conditions for star formation and leading to the observed variations in star-forming activity.

Our detailed analysis of the "anomalous spiral arms" using emission-line diagnostics has provided crucial insights into their composition and origins. These structures are not merely extensions of the traditional spiral arms but are significantly influenced by the central AGN activity. The arms show elevated levels of ionization and exhibit kinematic properties consistent with being shaped by shocks and AGN-driven outflows. The data suggest that these arms are sites where the kinetic energy from the AGN is converted into thermal and radiative energy, impacting the star formation rates and gas dynamics in these regions.

By mapping the influence of the AGN across different scales within NGC 4258. This study underscores the importance of AGN activity in shaping galactic structures, even in galaxies classified as LLAGNs. The interplay between the energy output from the AGN and the ISM’s response provides a dynamic laboratory for studying the feedback mechanisms that regulate galaxy evolution. These processes are crucial for understanding the lifecycle of galaxies and the balance between star formation and the growth of supermassive black holes.

Comparing our observations with current theoretical models of AGN feedback, our results support hypotheses that suggest AGN outflows can regulate star formation over large distances within a galaxy. This study not only aligns with theoretical predictions but also provides empirical data that can refine these models, especially in the context of LLAGNs where direct observations of feedback mechanisms were previously limited.

Low-power jet feedback in LLAGN—High-velocity \mbox{H$\alpha$} dispersion in the \halpha{} pattern and the presence of "anomalous arms" place NGC~4258 among the class of spiral discs dominated by weak jets like NGC~4579 and 3C~293—where the mechanical energy injected by the jets heats \mbox{H$_2$} and drives kiloparsec-scale turbulence, thereby lowering the global star-formation efficiency. This similarity indicates that jets with powers of $\lesssim10^{23}\,\text{W\,Hz}^{-1}$ can regulate stellar growth even in intermediate-mass systems, strengthening the view that LLAGN provide continuous, rather than merely episodic, self-regulation for their host galaxies. \citep[e.g.,][]{Ogle2019, Bagchi2025, Drevet2023}

Additional mechanisms beyond jet-driven turbulence may also regulate star formation and sculpt the spiral structure in NGC~4258.  Recent simulations and IFU surveys for different galaxies show that bar-induced shear and shocks suppress gravitational collapse along bar dust lanes, producing velocity jumps of up to $\sim170$ \uvel\ and characteristic LINER/composite signatures in BPT diagrams \citep{Fragkoudi2025,SilvaLima2025}. NGC~4258 hosts an oval bar that appears to funnel gas inwards while simultaneously inhibiting star formation inside $\lesssim3$ kpc through oblique shocks against the differentially rotating disc. The spatial overlap between these bar-driven shock fronts and the jet axis suggests that gravitational shear and mechanical feedback operate in concert. 

On the other hand, magnetohydrodynamic simulations that include cosmic rays  suggest they can supply up to $\sim50$\,per cent of the total mid-plane pressure, launching cold–warm galactic winds and further lowering the star-formation rate by removing low-angular-momentum gas \citep{Butsky2018,Hopkins2021}.  A cosmic ray-driven pressure gradient can accelerate ionized gas to altitudes of several kiloparsecs, consistent with the diffuse halo observed above the "anomalous arms".  Cosmic rays therefore offer an energetic channel that complements both the jet’s kinetic power and the bar-driven shocks.

Finally  the influence of jet, bar and cosmic rays processes place NGC~4258 in a multi-phased feedback framework and serves as a benchmark for studying feedback LLAGNs linking sub-parsec (SMBH accretion) to disc-scale regulators (bars, cosmic rays) and, ultimately, to the M*-M$_{\mathrm{BH}}$ relation. Such mechanisms resemble those implemented in cosmological simulations of low luminous galaxies to reproduce the decline of the cosmic SFR density from $z \sim 2$ to the present day \cite{Pillepich2018,Nelson2019}.  Thanks to its proximity and rich multi-wavelength coverage NGC~4258 enable us to examine features that would be unresolved in more distant galaxies, providing insights into how SMBHs influence their environments across different  range of luminosities \citep{humphreys2013maser}.  Although optical observations of NGC~4258 reveal intricate spiral arms and dust lanes that seem typical of a spiral galaxy, integrating optical, X-ray, infrared, and radio data uncovers extensive details about its stellar populations, encompassing both old and young stars. Therefore, NGC 4258 provides an empirical calibration point for these models and shows that ``continuous-mode" feedback can coexist with moderate star-formation rates (3\usfr) without catastrophic quenching.

In conclusion, mechanical feedback from LLAGN produce radio jets and winds that inject energy into the interstellar and circumgalactic medium, driving outflows and heating gas. This process can remove or heat cold gas, making it less available for star formation and thus lowering the star-formation efficiency. LLAGN can provide ongoing, moderate feedback that maintains a balance between gas cooling and heating, preventing runaway star formation or excessive gas accumulation \cite{Murthy2022,Almeida2023}. LLAGN-driven winds and jets can quench a significant fraction of star formation (over 10\% in some models), especially if the feedback persists over millions of years \cite{Bruggen2009}.  Therefore, LLAGN play a crucial role in the self-regulation of their host galaxies by continuously injecting energy that heats and removes gas, as is demonstrated in NGC~4258, thereby suppressing star formation and controlling galaxy growth. This feedback is effective even in intermediate-mass systems and operates over long timescales, supporting the view that LLAGN are key to the ongoing regulation of galaxy evolution. 

NGC~4258 serves as a pivotal case for studying the interconnected roles of star formation, shocks, and AGN activity. While SITELLE operates in the optical range, JWST focuses on the infrared spectrum. The integration of these datasets enables a multi-wavelength approach to studying NGC~4258, resolving its stellar populations and the ISM. SITELLE maps the distribution of ionized gas, while JWST will trace molecular gas, together offering a comprehensive view of the ISM phases. The synergy between SITELLE and JWST data in the future will facilitate enhanced modeling of AGN feedback, ISM dynamics, and star formation processes. SITELLE’s broad spectral coverage supports the development of models that will be validated by JWST’s detailed data, ensuring more robust conclusions. Additionally, SITELLE provides spatially resolved velocity and dispersion maps of the ionized gas. These kinematic maps and resolved properties are indispensable for interpreting JWST’s high-resolution data within the context of galaxy-wide flows, such as AGN-driven outflows, inflows, or rotation patterns, thus offering a richer understanding of NGC~4258’s dynamic processes.

In the upcoming release of JWST’s NIRCam observations, targeting key emission lines will significantly advance our understanding of galactic processes.  Lines such as [FeII]  are crucial for tracing high-velocity shocks, while Pa$\alpha$ and Br$\alpha$ will confirm recent star formation, and molecular hydrogen lines will map the ISM complemented by SITELLE’s observations. Additionally, the identification of polycyclic aromatic hydrocarbons will help pinpoint both young stellar objects and older populations. Instruments like SITELLE at the CFHT, when combined with high-resolution imaging, enable us to investigate feedback processes at unprecedented scales. The combination of SITELLE's detailed optical data and forthcoming infrared observations from JWST will further clarify the interdependencies between AGN activity and star formation. 

Looking forward, while JWST observations will focus on parsec-scale details, SITELLE provides a "big picture" view of the ionized gas across different scales. This approach illustrates how localized processes are integrated into the overall kinematics and morphology of the galaxy, linking them to global properties such as ionization sources, the influence of shocks, and star-forming regions. Studying NGC~4258 with such comprehensive tools makes it an invaluable template for understanding AGN-ISM interactions. This detailed examination provides critical insights that will inform future IFS surveys with Fourier transform spectrographs with high-resolution to map emission lines in a large FoV coverage, including SIGNALS and others, enriching our understanding of galactic dynamics and evolution of galaxies.

\section{Conclusions and Summary}\label{sum_conclu}

Our main findings regarding NGC~4258, based on the SITELLE data, can be summarized as follows:

\begin{itemize}

\item We analysed the line ratios (\nii/\halpha, \sii/\halpha, \oiii/\hbeta) in BPT diagrams with respect to galactocentric radius and velocity dispersion. Both \nii/\halpha\ and \sii/\halpha\ show a gradual decrease within the first 5~kpc, whereas \oiii/\hbeta\ decreases sharply over the initial 3~kpc and then stabilizes. Significant fluctuations at parsec scales are consistent with previous observations in other Seyfert galaxies \cite[e.g.][]{DAgostino2018}. 

\item We estimated the star formation rate (SFR) by applying a correction factor related to the mixing-sequence observed in the BPT diagrams and by correcting the  \halpha\ emission for extinction. The total galaxy-wide SFR is approximately \mbox{3 \usfr},  while the central 3.4 kpc$^2$ exhibits an SFR of \mbox{0.3 \usfr}. These values are consistent with previous SED-fitting studies that covered the same region.

\item Spatially and kinematically resolved BPT diagrams reveal distinct ionization sources in NGC~4258. The central region is dominated by Seyfert and LINER-like ionization, with notably higher velocity dispersion compared to areas dominated by \hii\ regions. A correlation emerges between velocity dispersion and ionization source, with decreasing ionization observed at larger distances from the nucleus. This kinematic information also traces the jet’s velocity and the complexity of the line profiles; broad or double-peaked profiles reflect jet–ISM interactions, which may have driven the formation of the ``anomalous spiral arms." Combining the spectral resolution in the \halpha\ line with spatially resolved BPT diagrams provides a comprehensive view of the ionization environment, highlighting asymmetric nuclear emission lines and their evolution with distance. The data reveal an arc-like structure offset from the galactic plane, indicative of AGN-driven outflows interacting with the ISM at projected distances of up to 5~kpc.

\item In the lower "anomalous arc", we find  evidence of star-forming sources, possibly due to shocks quenching star formation or due to heavy extinction. While small-scale shocks can compress gas and trigger further star formation, large-scale shocks can suppress it. Here, we infer that shocks likely quench star formation, as we do not observe significant extinction in this region.

\item Using spatially resolved BPT diagrams, we confirm the origin of polarized radio emission proposed by \citet{Krause2004}. Strong emission clumps along the spiral arms correlate with regions of intense \halpha\, indicative of star-forming regions emitting unpolarized thermal radiation. Meanwhile, radio emission along the jet is polarized and non-thermal, originating from the AGN. Observed localized regions of shock-induced star formation suppression contrast with areas where shocks have triggered star formation. The ``anomalous spiral arms" likely result from AGN-driven jets, offering a clear example of such feedback in action.

\item The central zone (<3~kpc) shows velocity dispersions around 200 \uvel, consistent with earlier findings, particularly in the southeast jet and the galaxy center. Along the jet and ``anomalous spiral arms", velocity dispersion differs significantly from that in the classical spiral arms (>5~kpc), where \hii\ regions exhibit values of 30-50\uvel. Emission from the ``anomalous spiral arm" aligns with AGN activity and shocks, reaching velocity dispersions of up to 250 \uvel\ as far as 6~kpc from the nucleus toward the southeast bifurcations. Both the inner and outer spiral arms host star-forming regions with evidence of shocks, which are well explained by fast shock photoionization models.

\item NGC~4258 prominently displays ionized gas along the jet, prompting an azimuthal analysis of pixel distributions in15\degree\ bins. The covering fraction $\%C_{frac}$ reveals two primary regions aligned with the galaxy’s position angle. Peaks in the fraction of spaxels classified as Seyfert or composite—coincident with the jet—suggest outflows emanating along a biconical path from the nucleus. Meanwhile, spaxels predominantly ionized by star formation drop rapidly in these jet zones, reaching minimum fractions. Maximum values occur between \mbox{-30\degree\ to 20\degree} and between \mbox{100\degree\ and 190\degree}, corresponding to the direction of the spiral arms. These angles match ionization cones observed in other active galaxies \citep{Wilson1994}, supporting the interpretation by \citet{Falcke1996} for NGC~4258. Beyond 10~kpc, NGC~4258 resembles a typical spiral galaxy; closer to the nucleus, disturbances and the ``anomalous spiral arms" arise from the interaction of conventional spiral arms with the AGN jet, quenching star formation along its trajectory. The ionization source in the ``anomalous spiral arms" of NGC~4258 remains debated, with possibilities ranging from bar shocks \citep{Cox1996} to dynamic coupling with the central engine’s precession \citep{Cecil2000}. The presence of small (<100~pc) Seyfert-ionized regions in the anomalous arm supports a dynamic connection to evolving jet activity.

\item Our data suggest that the low-power jets have a direct impact on the disk gas out to ~5–6 kpc, driving shocks (as seen by high \sii/\halpha\  and broad lines) that both trigger and quench star formation in different regions. The anomalous spiral arms appear to be jet-driven features, consisting of shocked gas and sparse star formation, distinct from the normal spiral density-wave arms.

\item Investigations into the kinematics and ionization state of the gas reveal how AGN-driven energy regulates star formation and galactic evolution. This is particularly important for understanding how low-luminosity AGNs (LLAGNs) contribute to maintaining the balance between inflow and outflow processes \citep{wilson2001xray}. The "anomalous spiral arms" of NGC~4258 align with regions where radio jets or outflows interact with the galaxy’s disk. Observations of ionized gas allow us to map these regions and determine the strength and structure of the outflows. Such studies provide valuable insights into the efficiency of feedback processes in LLAGNs and their influence on both nuclear and large-scale galactic structures \citep{ceccarelli2001jets}. Comparing the observed ionization characteristics of NGC~4258 with theoretical models refines our understanding of LLAGN environments, which are shaped by AGN photoionization, shock heating from jet-ISM interactions, and turbulent mixing layers. These processes are revealed through kinematic data and the morphology of emission line profiles, in agreement with \citet{wilson2001xray}.

\item  We interpret the observed phenomena as resulting from jet-induced perturbations in the galactic disk. The intense and extended velocity spreads perpendicular to AGN jets and ionization cones are currently observed only in galaxies hosting low-power jets with an inclination sufficiently low relative to the galactic disk to impact and significantly affect its material. In line with cosmological simulations, our results demonstrate that low-power jets can influence their host galaxy on parsec scales and extend their impact to kiloparsec scales.

\end{itemize}

\subsection{Summary}\label{sum_conclu2}

NGC~4258 shows a complex interplay among its spiral arms, bars, shocks, and feedback from the jet, which collectively regulate star formation and the structure of the galaxy. Whereas the spiral arms mainly gather gas and star-forming regions, the bars and shocks can suppress it locally, and feedback and turbulence further modulate star formation, resulting in diverse patterns across this galaxy.  These characteristics make NGC~4258 a potential benchmark for studying low-luminosity AGN feedback in spiral galaxies, providing valuable insights for cosmological simulations of low-luminosity galaxies. 

In summary, the ionized gas in NGC~4258 provides critical insights into the interactions between the LLAGN and its host galaxy, illuminating the nature of feedback processes and their influence on galactic structure and evolution. This study marks the first comprehensive mapping of the main nebular lines \oii\lin3727, \hbeta, \oiii\lin4959, \oiii\lin5007, \nii\lin6548, \halpha, \nii\lin6583, \sii\lin6716 and \sii\lin6731 using Integral Field Spectroscopy (IFS), covering nearly the entire galaxy. The data obtained from SITELLE significantly enhances our understanding by providing crucial spectral and spatial insights that complement and expand the interpretation of forthcoming James Webb Space Telescope (JWST) observations.

\section*{Acknowledgements}

We thank the referee for his helpful comments, which significantly improved this draft. D.F.A  acknowledges the support from the National Science Foundation under grant 2109124 for  SIGNALS: Unveiling Star-Forming Regions in Nearby Galaxies. This work is based on observations obtained at the Canada-France-Hawaii Telescope (CFHT) which is operated from the summit of Mauna Kea by the National Research Council of Canada, the Institut National des Sciences de l'Univers of the Centre National de la Recherche Scientifique of France, and the University of Hawaii. The observations at the Canada-France-Hawaii Telescope were performed with care and respect from the summit of Mauna Kea which is a significant cultural and historic site. Based on observations obtained with SITELLE, a joint project between Université Laval, ABB-Bomem, Université de Montréal and the CFHT with funding support from the Canada Foundation for Innovation (CFI), the National Sciences and Engineering Research Council of Canada (NSERC), Fond de Recherche du Québec - Nature et Technologies (FRQNT) and CFHT.
LRN is grateful to the National Science Foundation NSF - 2109124, the Dunlap Institute, and the Natural Sciences and Engineering Research Council of Canada NSERC - RGPIN-2023-03487 for their support. The Dunlap Institute is funded through an endowment established by the David Dunlap family and the University of Toronto.
S.D.P. acknowledges financial support from Juan de la Cierva Formaci\'on fellowship (FJC2021-047523-I) financed by MCIN/AEI/10.13039/501100011033 and by the European Union ``NextGenerationEU"/PRTR, Ministerio de Econom\'ia y Competitividad under grant PID2019-107408GB-C44, PID2020-113689GB-I00, and PID2020-114414GB-I00, and PID2022-136598NB-C32, and from Junta de Andaluc\'ia FQM108, and also from the Fonds de Recherche du Qu\'ebec - Nature et Technologies. \\

\section*{Data Availability}
Data cubes from the SIGNALS survey are publicly available at the Canadian Astronomy Data Centre \href{https://www.cadc-ccda.hia-iha.nrc-cnrc.gc.ca/en/search/?Plane.position.bounds@Shape1Resolver.value=ALL&Plane.position.bounds=M%20106&Observation.collection=CFHT&Observation.instrument.name=SITELLE&Plane.dataProductType=cube&Observation.type=SCIENCE#sortCol=caom2%3APlane.time.bounds.lower&sortDir=dsc&col_1=_checkbox_selector;;;&col_2=caom2%3AObservation.uri;;;&col_3=caom2%3AObservation.collection;;;&col_4=caom2%3AObservation.sequenceNumber;;;&col_5=caom2%3APlane.productID;;;&col_6=caom2%3APlane.position.bounds.cval1;;;&col_7=caom2%3APlane.position.bounds.cval2;;;&col_8=caom2%3AObservation.target.name;;;&col_9=caom2%3APlane.time.bounds.lower;;;&col_10=caom2%3APlane.time.exposure;;;&col_11=caom2%3AObservation.instrument.name;;;&col_12=caom2%3APlane.energy.bandpassName;;;&col_13=caom2%3APlane.calibrationLevel;;;&col_14=caom2%3AObservation.type;;;&col_15=caom2%3AObservation.proposal.id;;;&col_16=caom2%3AObservation.proposal.pi;;;&col_17=caom2%3APlane.dataRelease;;;
}{Bookmark URL, cadc-ccda} or visit the SIGNALS survey available in \url{https://signal-survey.org/} for more information. Maps and tools for analysis are available upon request to the corresponding author.


\bibliographystyle{mnras}
\bibliography{references} 



\appendix

\section{Spectra regions and Emission lines maps}
Spectra for the different regions as presented according to Figure \ref{_regions} and Emission line flux maps of NGC~4258 in \oii\lin3727 for SN1, \hbeta, \oiii\lin5007 for SN2 and \halpha, \nii\lin6583, \sii\lin6716,6731 for SN3. The fits have been obtained using \texttt{ORCS} routine after sky background subtraction and absorption correction of the stellar component.

\begin{figure*}
\centering
	\includegraphics[width=\columnwidth]{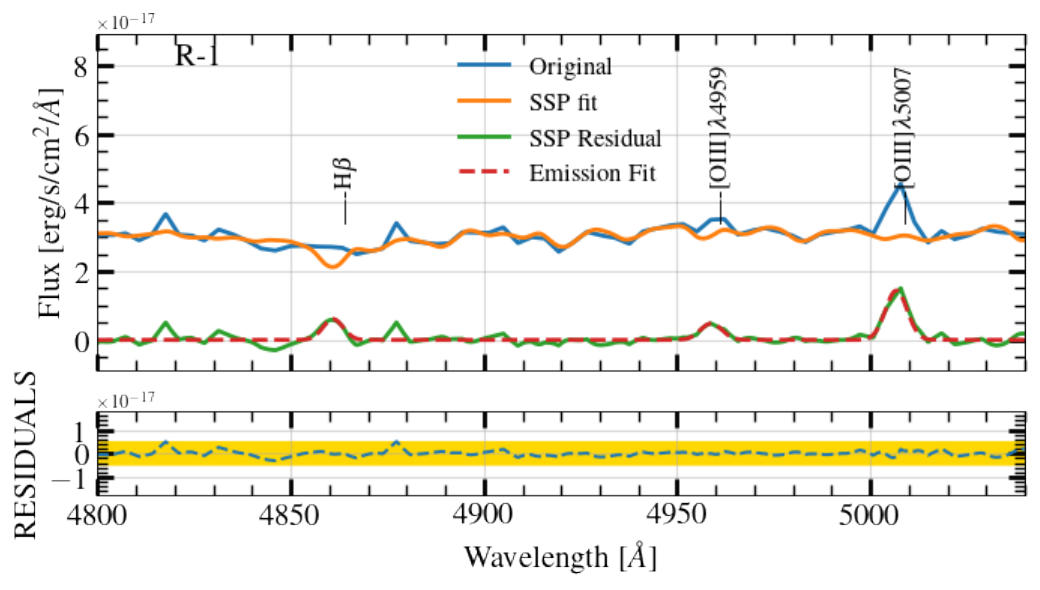}\includegraphics[width=\columnwidth]{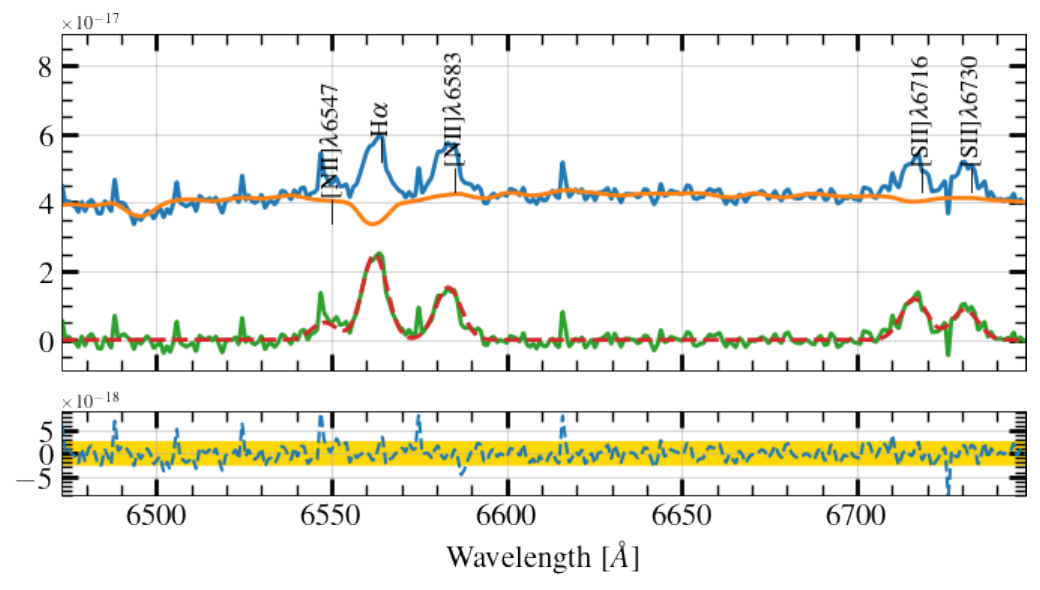}	
	\includegraphics[width=\columnwidth]{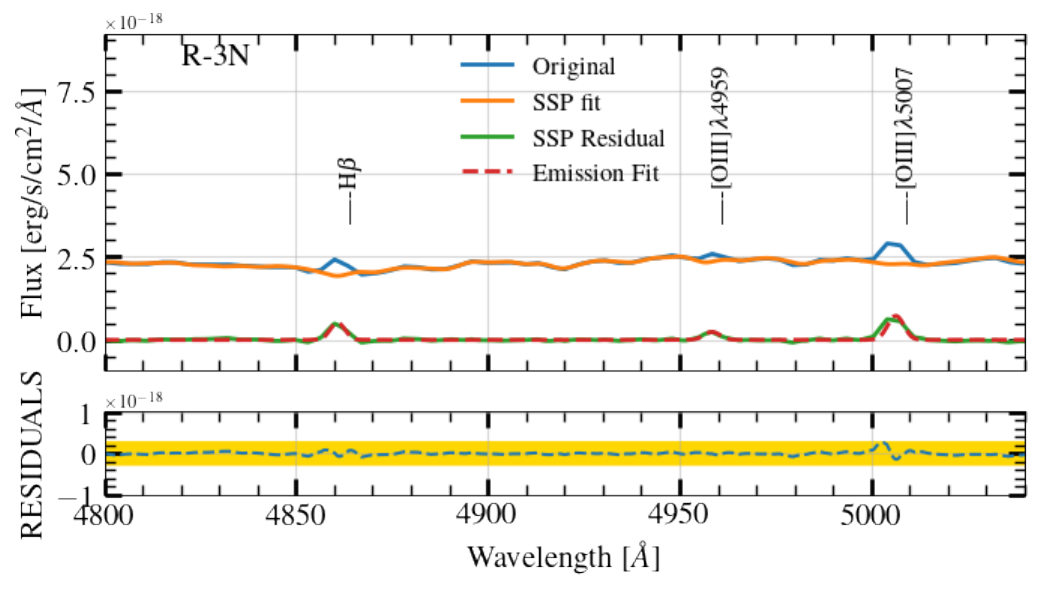}\includegraphics[width=\columnwidth]{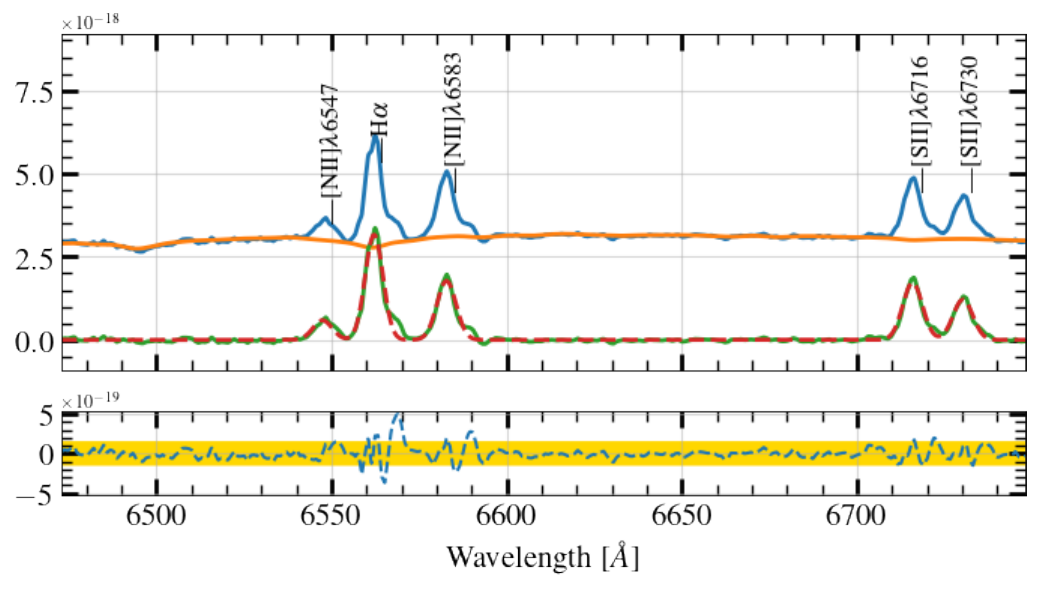}	
 	\includegraphics[width=\columnwidth]{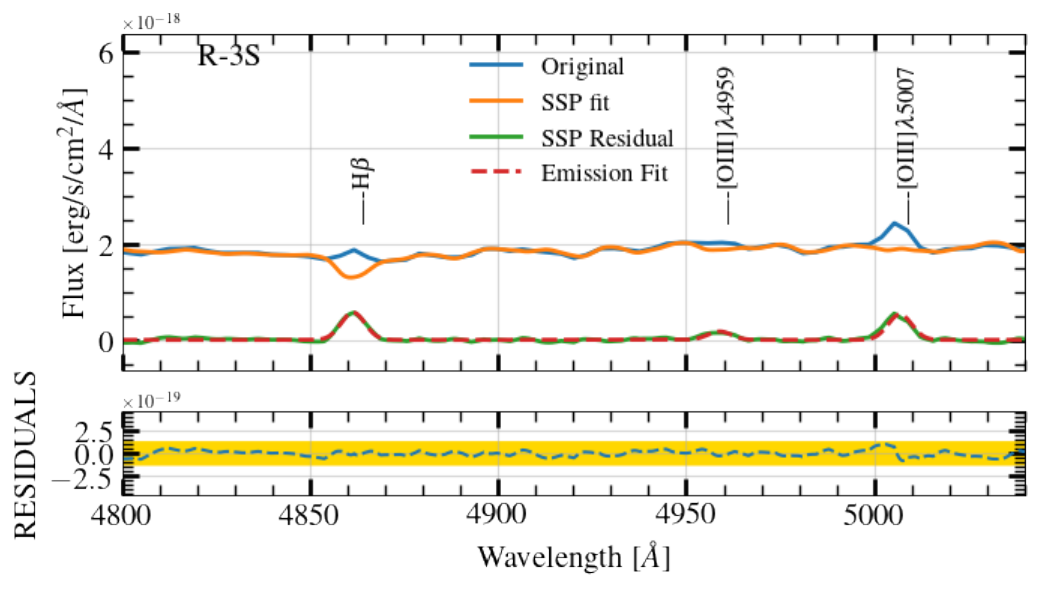}\includegraphics[width=\columnwidth]{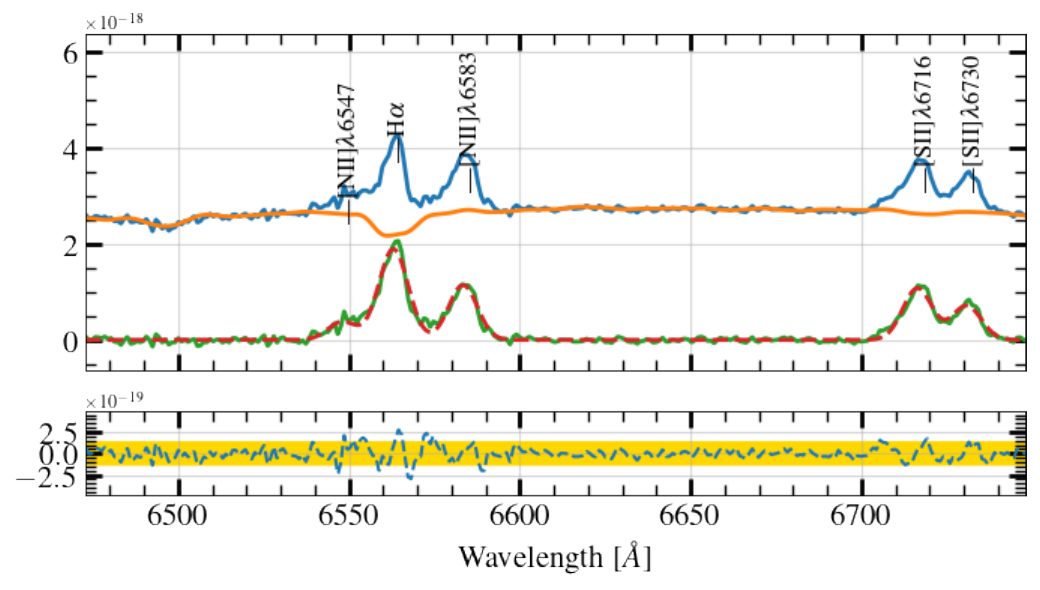}
  \includegraphics[width=\columnwidth]{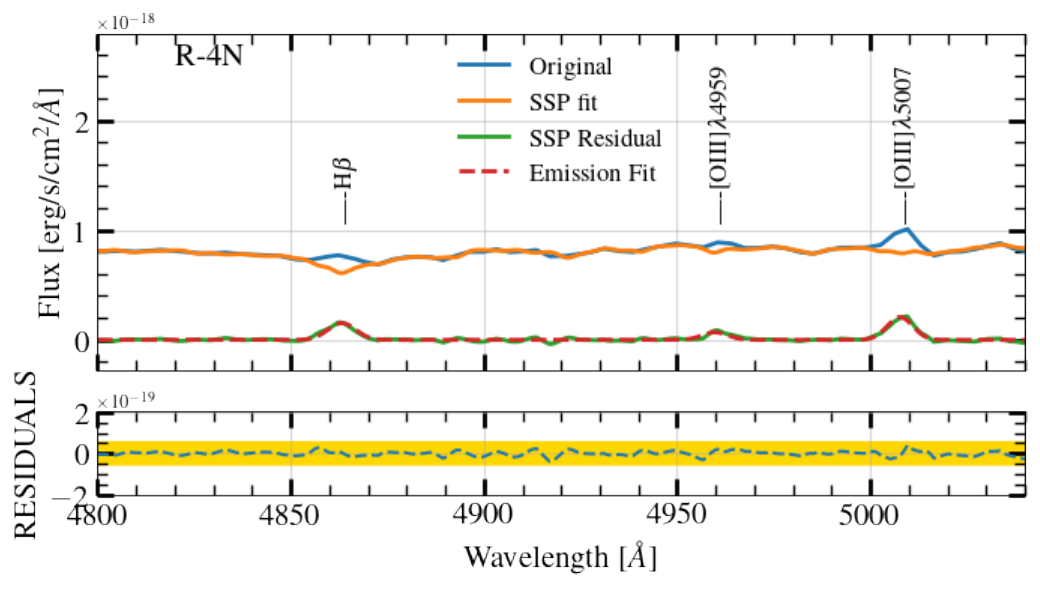}\includegraphics[width=\columnwidth]{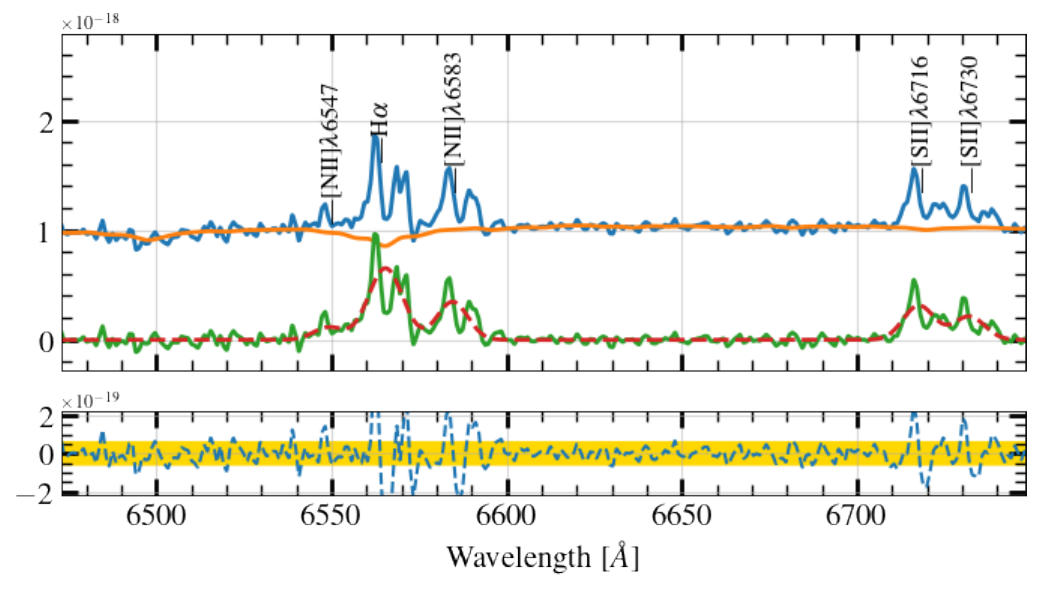}	
    \caption{Spectra for the different regions as presented according to Figure \ref{_regions}. The description is the same as in Figure \ref{_stellar_spec_fit}. We display the regions R1 to R-5N. For R-6 regions, we found that do not give significative information about the continuum correction given the lack of S/N in their detection, even after considering more spaxels compared with other regions.}
    \label{_stellar_spec_fit_t}
\end{figure*}

\begin{figure*}
\centering
 	\includegraphics[width=\columnwidth]{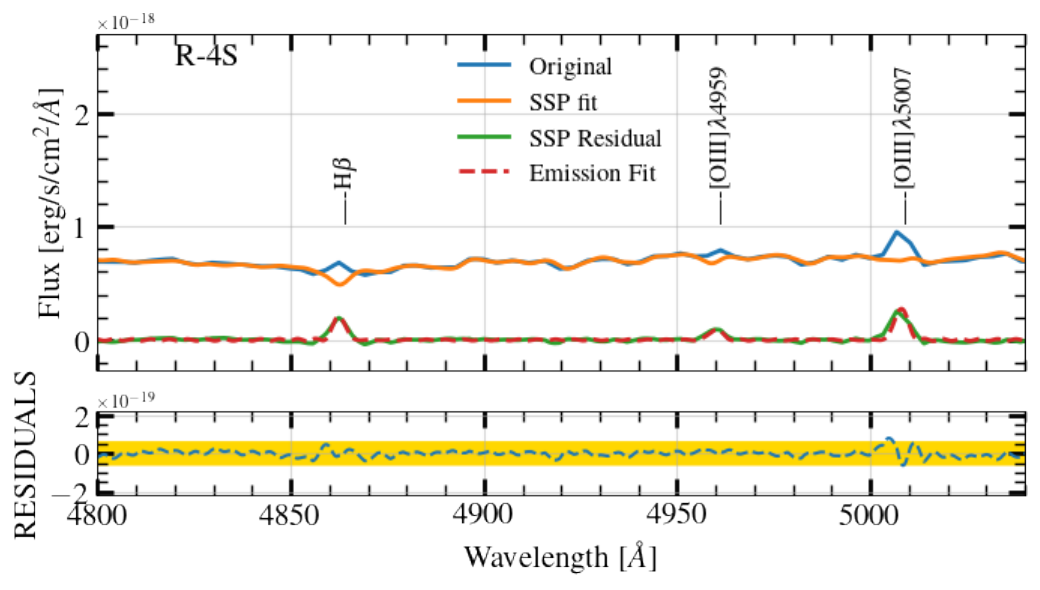}\includegraphics[width=\columnwidth]{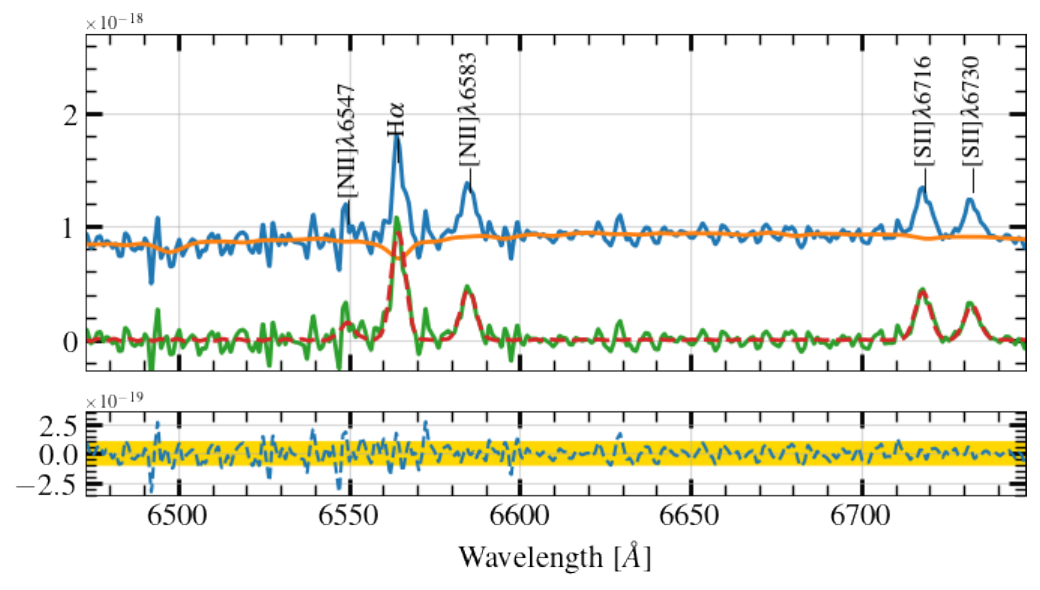}
    \includegraphics[width=\columnwidth]{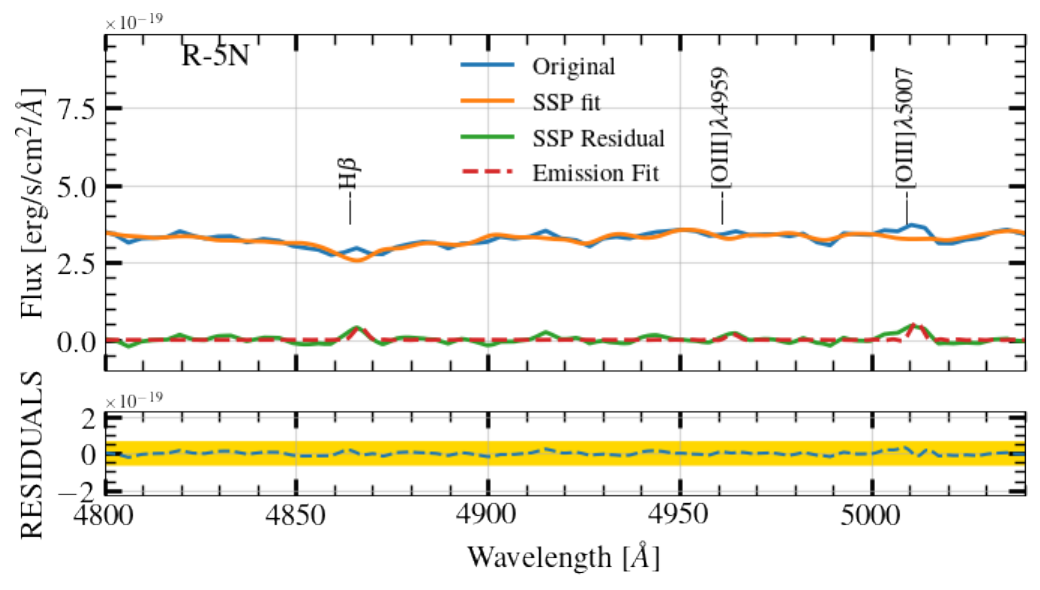}\includegraphics[width=\columnwidth]{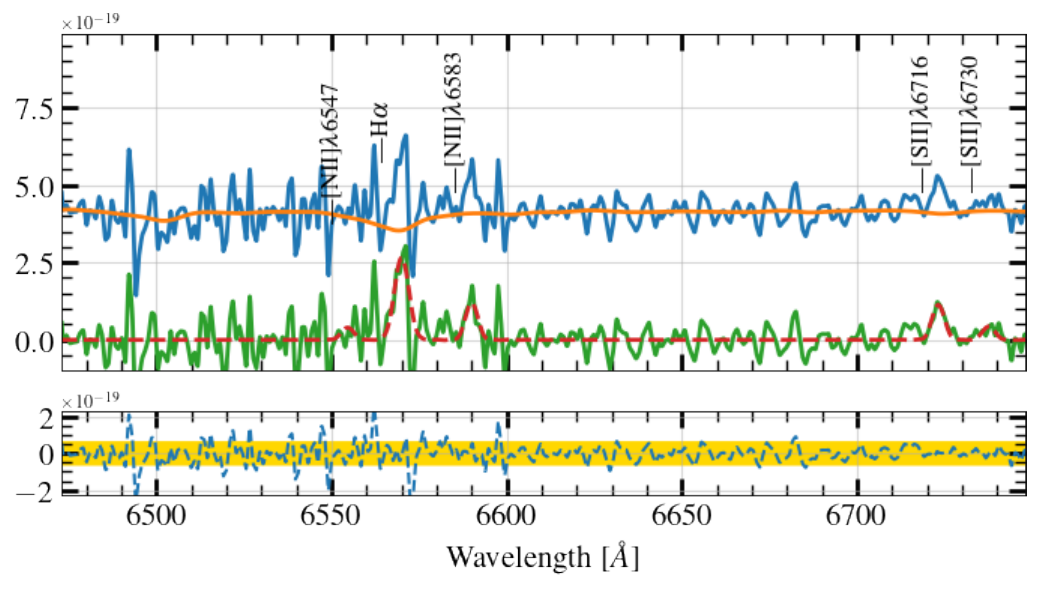}	
 	\includegraphics[width=\columnwidth]{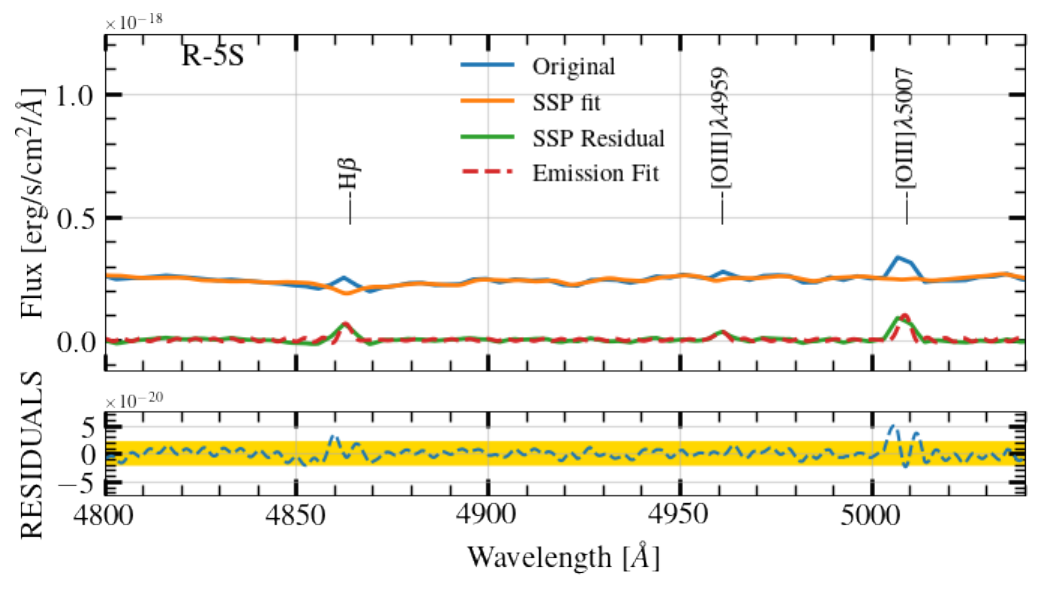}\includegraphics[width=\columnwidth]{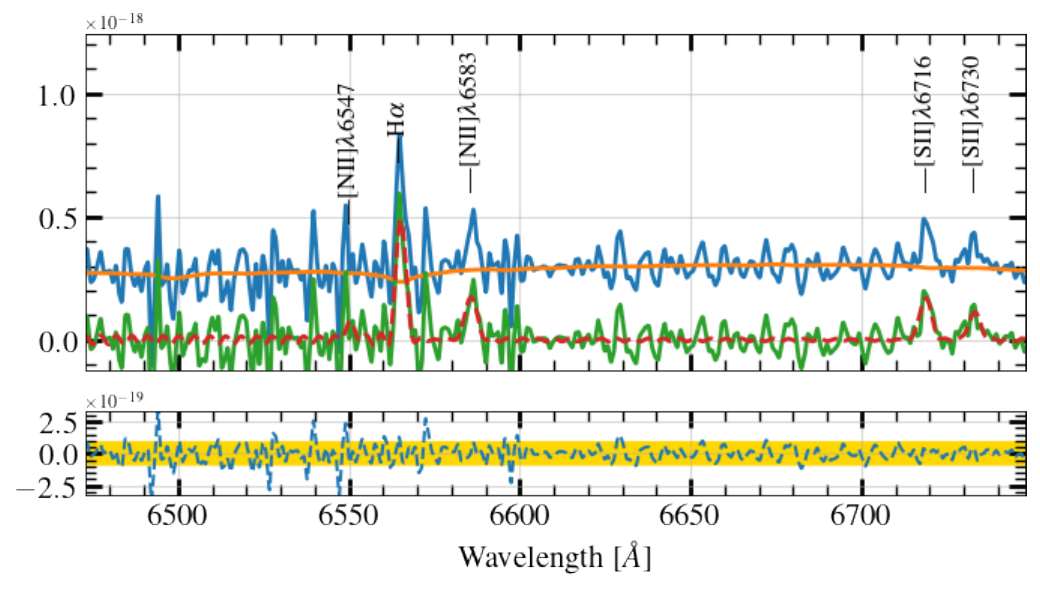}
    \contcaption{}
    \label{tab:_stellar_spec_fit_t}
\end{figure*}

\begin{figure*}
\centering
\includegraphics[scale=0.55]{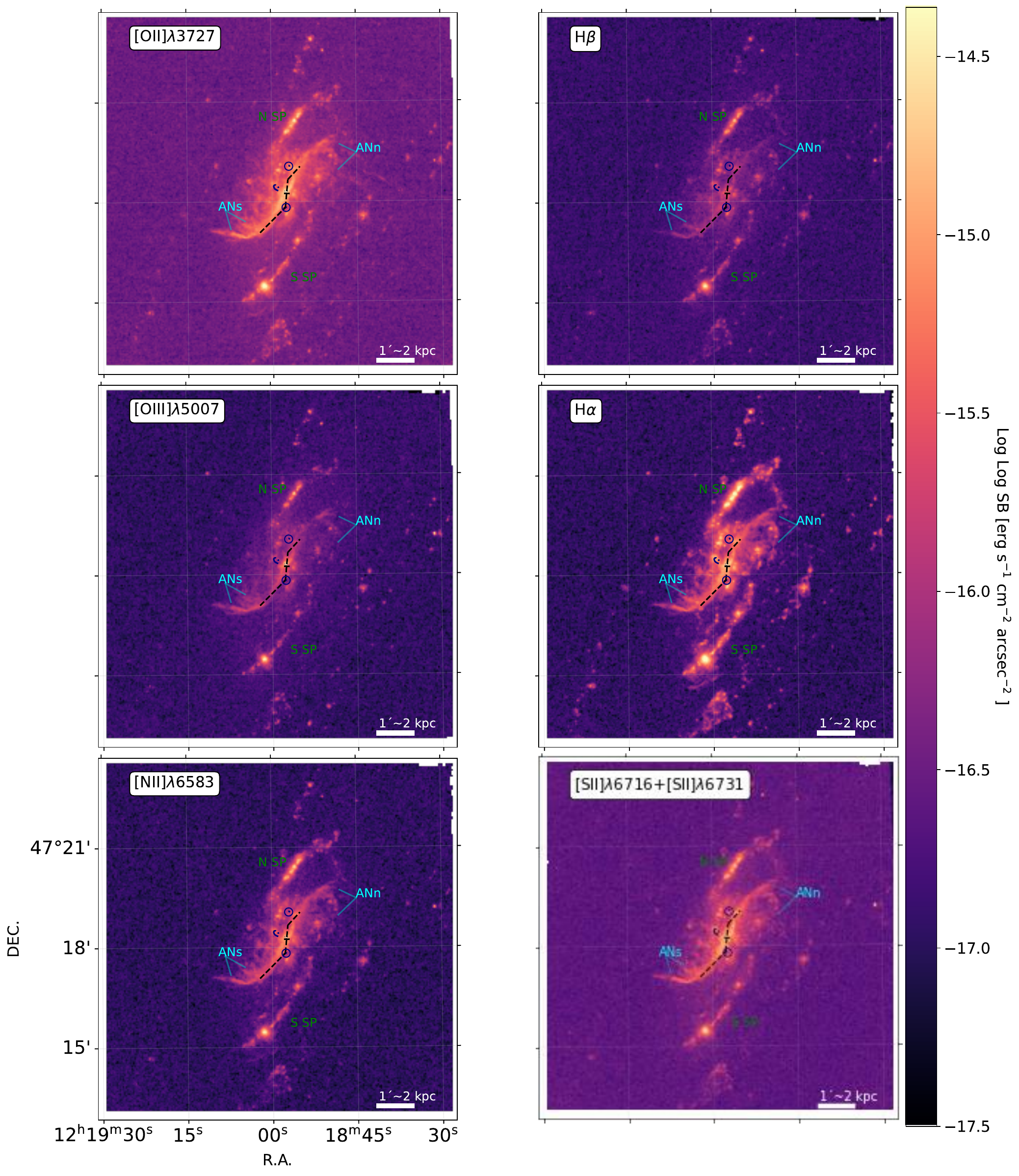}
   \caption{Surface brightness maps of NGC~4258 of \oii\lin3727 for SN1, \hbeta, \oiii\lin5007 for SN2 and \halpha, \nii\lin6583, \sii\lin6716,6731 for SN3. The fits have been obtained using \texttt{ORCS} routine after sky background subtraction and absorption correction of the stellar component.}
    \label{flux_maps}
\end{figure*}

\bsp	
\label{lastpage}
\end{document}